\documentclass[tight,twocolumn]{aastex62}
\usepackage{apjfonts}
\usepackage{psfig}
\usepackage{float}
\usepackage[caption=false]{subfig}

\usepackage{graphics}
\usepackage{amssymb,amsmath}
\usepackage{color}

\newcommand{\host}{\hbox{SDSS J075654.53+341543.6}}

\def\lsim{\lower0.3em\hbox{$\,\buildrel <\over\sim\,$}}
\def\gsim{\lower0.3em\hbox{$\,\buildrel >\over\sim\,$}}

\newcommand{\mstar}{\hbox{$m_{\star}$}}
\newcommand{\rstar}{\hbox{$r_{\star}$}}

\newcommand{\msun}{\hbox{M$_{\odot}$}}

\newcommand{\lsun}{\hbox{L$_{\odot}$}}

\newcommand{\halpha}{\hbox{H$\alpha$}}
\newcommand{\hbeta}{\hbox{H$\beta$}}

\newcommand{\swift}{\textit{Swift}}

\definecolor{agreen}{rgb}{0.05, 0.5, 0.06}
\definecolor{cobalt}{rgb}{0., 0.28, 0.67}

\begin{document}

\title{PS18kh: A New Tidal Disruption Event with a Non-Axisymmetric Accretion Disk}  
\shorttitle{Discovery of the TDE PS18kh} 
\shortauthors{Holoien, et al. }

\author[0000-0001-9206-3460]{T.~W.-S.~Holoien}
\affiliation{The Observatories of the Carnegie Institution for Science, 813 Santa Barbara St., Pasadena, CA 91101, USA}

\author{M.~E.~Huber}
\affiliation{Institute for Astronomy, University of Hawai'i, 2680 Woodlawn Drive, Honolulu, HI 96822, USA}

\author{B.~J.~Shappee}
\affiliation{Institute for Astronomy, University of Hawai'i, 2680 Woodlawn Drive, Honolulu, HI 96822, USA}

\author{M.~Eracleous}
\affiliation{Department of Astronomy \& Astrophysics and Institute for Gravitation and the Cosmos, The Pennsylvania State University, University Park, PA 16802, USA}

\author{K.~Auchettl}
\affiliation{Center for Cosmology and AstroParticle Physics (CCAPP), The Ohio State University, 191 W.\ Woodruff Ave., Columbus, OH 43210, USA}
\affiliation{Department of Physics, The Ohio State University, 191 W. Woodruff Avenue, Columbus, OH 43210, USA}

\author{J.~S.~Brown}
\affiliation{Department of Astronomy, The Ohio State University, 140 West 18th Avenue, Columbus, OH 43210, USA}

\author[0000-0002-2471-8442]{M.~A.~Tucker}
\affiliation{Institute for Astronomy, University of Hawai'i, 2680 Woodlawn Drive, Honolulu, HI 96822, USA}

\author{K.~C.~Chambers}
\affiliation{Institute for Astronomy, University of Hawai'i, 2680 Woodlawn Drive, Honolulu, HI 96822, USA}

\author{C.~S.~Kochanek}
\affiliation{Center for Cosmology and AstroParticle Physics (CCAPP), The Ohio State University, 191 W.\ Woodruff Ave., Columbus, OH 43210, USA}
\affiliation{Department of Astronomy, The Ohio State University, 140 West 18th Avenue, Columbus, OH 43210, USA}

\author{K.~Z.~Stanek}
\affiliation{Center for Cosmology and AstroParticle Physics (CCAPP), The Ohio State University, 191 W.\ Woodruff Ave., Columbus, OH 43210, USA}
\affiliation{Department of Astronomy, The Ohio State University, 140 West 18th Avenue, Columbus, OH 43210, USA}

\author{A.~Rest}
\affiliation{Space Telescope Science Institute, 3700 San Martin Drive, Baltimore, MD 21218, USA.}
\affiliation{Department of Physics and Astronomy, Johns Hopkins University, Baltimore, MD 21218, USA.}

\author{D.~Bersier}
\affiliation{Astrophysics Research Institute, Liverpool John Moores University, 146 Brownlow Hill, Liverpool L3 5RF, UK}

\author{R.~S.~Post}
\affiliation{Post Observatory, Lexington, MA 02421, USA}

\author{G.~Aldering}
\affiliation{Lawrence Berkeley National Laboratory, Physics Division, One Cyclotron Rd, Berkeley, CA 94720, USA}

\author{K.~A.~Ponder}
\affiliation{Berkeley Center for Cosmological Physics, 341 Campbell Hall, University of California Berkeley, Berkeley, CA 94720, USA}

\author{J.~D.~Simon}
\affiliation{The Observatories of the Carnegie Institution for Science, 813 Santa Barbara St., Pasadena, CA 91101, USA}

\author{E.~Kankare}
\affiliation{Astrophysics Research Centre, School of Mathematics and Physics, Queens University Belfast, Belfast BT7 1NN, UK}

\author{D.~Dong}
\affiliation{Division of Physics, Mathematics and Astronomy, California Institute of Technology, 1200 East California Boulevard, Pasadena, CA 91125, USA}

\author{G.~Hallinan}
\affiliation{Division of Physics, Mathematics and Astronomy, California Institute of Technology, 1200 East California Boulevard, Pasadena, CA 91125, USA}

\author{N.~A.~Reddy}
\affiliation{Department of Physics and Astronomy, University of California, Riverside, 900 University Avenue, Riverside, CA 92521, USA}

\author{R.~L.~Sanders}
\affiliation{Department of Physics, University of California, Davis, 1 Shields Avenue, Davis, CA 95616, USA}

\author{M.~W.~Topping}
\affiliation{Department of Physics and Astronomy, University of California, Los Angeles, 430 Portola Plaza, Los Angeles, CA 90095, USA}

\collaboration{}
\collaboration{\textit{Pan-STARRS}}

\author{J.~Bulger}
\affiliation{Institute for Astronomy, University of Hawai'i, 2680 Woodlawn Drive, Honolulu, HI 96822, USA}

\author{T.~B.~Lowe}
\affiliation{Institute for Astronomy, University of Hawai'i, 2680 Woodlawn Drive, Honolulu, HI 96822, USA}

\author{E.~A.~Magnier}
\affiliation{Institute for Astronomy, University of Hawai'i, 2680 Woodlawn Drive, Honolulu, HI 96822, USA}

\author{A.~S.~B.~Schultz}
\affiliation{Institute for Astronomy, University of Hawai'i, 2680 Woodlawn Drive, Honolulu, HI 96822, USA}

\author{C.~Z.~Waters}
\affiliation{Institute for Astronomy, University of Hawai'i, 2680 Woodlawn Drive, Honolulu, HI 96822, USA}

\author{M.~Willman}
\affiliation{Institute for Astronomy, University of Hawai'i, 2680 Woodlawn Drive, Honolulu, HI 96822, USA}

\author{D.~Wright}
\affiliation{Astrophysics Research Centre, School of Mathematics and Physics, Queens University Belfast, Belfast BT7 1NN, UK}

\author{D.~R.~Young}
\affiliation{Astrophysics Research Centre, School of Mathematics and Physics, Queens University Belfast, Belfast BT7 1NN, UK}

\collaboration{}
\collaboration{\textit{ASAS-SN}}

\author{Subo~Dong}
\affiliation{Kavli Institute for Astronomy and Astrophysics, Peking University, Yi He Yuan Road 5, Hai Dian District, Beijing 100871, China}

\author{J.~L.~Prieto}
\affiliation{N\'ucleo de Astronom\'ia de la Facultad de Ingenier\'ia y Ciencias, Universidad Diego Portales, Av. Ej\'ercito 441, Santiago, Chile}
\affiliation{Millennium Institute of Astrophysics, Santiago, Chile}

\author{Todd~A.~Thompson}
\affiliation{Center for Cosmology and AstroParticle Physics (CCAPP), The Ohio State University, 191 W.\ Woodruff Ave., Columbus, OH 43210, USA}
\affiliation{Department of Astronomy, The Ohio State University, 140 West 18th Avenue, Columbus, OH 43210, USA}

\collaboration{}
\collaboration{\textit{ATLAS}}

\author{L.~Denneau}
\affiliation{Institute for Astronomy, University of Hawai'i, 2680 Woodlawn Drive, Honolulu, HI 96822, USA}

\author{H.~Flewelling}
\affiliation{Institute for Astronomy, University of Hawai'i, 2680 Woodlawn Drive, Honolulu, HI 96822, USA}

\author{A.~N.~Heinze}
\affiliation{Institute for Astronomy, University of Hawai'i, 2680 Woodlawn Drive, Honolulu, HI 96822, USA}

\author{S.~J.~Smartt}
\affiliation{Astrophysics Research Centre, School of Mathematics and Physics, Queens University Belfast, Belfast BT7 1NN, UK}

\author{K.~W.~Smith}
\affiliation{Astrophysics Research Centre, School of Mathematics and Physics, Queens University Belfast, Belfast BT7 1NN, UK}

\author{B.~Stalder}
\affiliation{LSST, 950 North Cherry Avenue, Tucson, AZ 85719, USA}

\author{J.~L.~Tonry}
\affiliation{Institute for Astronomy, University of Hawai'i, 2680 Woodlawn Drive, Honolulu, HI 96822, USA}

\author{H.~Weiland}
\affiliation{Institute for Astronomy, University of Hawai'i, 2680 Woodlawn Drive, Honolulu, HI 96822, USA}

\correspondingauthor{T.~W.-S.~Holoien}
\email{tholoien@carnegiescience.edu}

\date{\today}

\begin{abstract}

We present the discovery of PS18kh, a tidal disruption event (TDE) discovered at the center of \host{} ($d\simeq322$~Mpc) by the Pan-STARRS Survey for Transients. Our dataset includes pre-discovery survey data from Pan-STARRS, the All-Sky Automated Survey for Supernovae (ASAS-SN), and the Asteroid Terrestrial-impact Last Alert System (ATLAS) as well as high-cadence, multi-wavelength follow-up data from ground-based telescopes and {\swift}, spanning from 56 days before peak light until 75 days after. The optical/UV emission from PS18kh is well-fit as a blackbody with temperatures ranging from $T\simeq12000$~K to $T\simeq25000$~K and it peaked at a luminosity of $L\simeq8.8\times10^{43}$~ergs~s$^{-1}$. PS18kh radiated $E=(3.45\pm0.22)\times10^{50}$~ergs over the period of observation, with $(1.42\pm0.20)\times10^{50}$~ergs being released during the rise to peak. Spectra of PS18kh show a changing, boxy/double-peaked \halpha{} emission feature, which becomes more prominent over time. We use models of non-axisymmetric accretion disks to describe the profile of the H$\alpha$ line and its evolution. We find that at early times the high accretion rate leads the disk to emit a wind which modifies the shape of the line profile and makes it bell-shaped. At late times, the wind becomes optically thin, allowing the non-axisymmetric perturbations to show up in the line profile. The line-emitting portion of the disk extends from $r_{\rm in}\sim60r_{\rm g}$ to an outer radius of $r_{\rm out}\sim1400r_{\rm g}$ and the perturbations can be represented either as an eccentricity in the outer rings of the disk or as a spiral arm in the inner disk.
\end{abstract}
\keywords{accretion, accretion disks --- black hole physics --- galaxies: nuclei}


\section{Introduction}
\label{sec:intro}

Tidal disruption events (TDEs) occur when a star crosses the tidal radius of a supermassive black hole (SMBH) and the tidal shear forces of the SMBH are able to overcome the self-gravity of the star. For main-sequence stars, approximately half of the stellar material is ejected from the system, while the other half remains bound to the SMBH. The bound material falls back to pericenter at a rate proportional to $t^{-5/3}$ and a fraction of it is accreted onto the black hole, resulting in a short-lived, luminous flare \citep[e.g.,][]{lacy82,rees88,evans89,phinney89}. 

Initially, it was commonly assumed that the flare emission would peak at soft X-ray energies and that the luminosity would be proportional to the $t^{-5/3}$ rate of return of the stellar material to pericenter. However, in recent years a number of well-studied TDEs have been discovered that exhibit a wide range of observational properties \citep[e.g.,][]{velzen11,cenko12a,gezari12b,arcavi14, chornock14,holoien14b,gezari15,vinko15,holoien16a,holoien16b,brown16a,auchettl17,blagorodnova17,brown17a,brown17b,gezari17,holoien18a}. It is now known that the emission depends on many factors, including the physical properties of the disrupted star \citep[e.g.,][]{macleod12,kochanek16}, the evolution of the accretion stream after disruption \citep[e.g.,][] {kochanek94,strubbe09,guillochon13,hayasaki13,hayasaki16,piran15,shiokawa15}, and radiative transfer effects \citep[e.g.,][]{gaskell14,strubbe15,roth16,roth18}. However, there have been few TDEs monitored in sufficient detail to directly infer these properties. In particular, most TDE candidates have been discovered after peak light, making it difficult to study the formation of the accretion disk and the evolution of the stellar debris.

Here we present the discovery of PS18kh, a TDE candidate discovered by the Pan-STARRS Survey for Transients\footnote{\url{https://star.pst.qub.ac.uk/ps1threepi/psdb/}} \citep[PSST;][]{chambers16} on 2018 March 02 in the spectroscopically unobserved galaxy SDSS J075654.53+341543.6. The discovery was announced publicly on 2018 March 04 on the Transient Name Server (TNS) and given the designation AT 2018zr\footnote{\url{https://wis-tns.weizmann.ac.il/object/2018zr}}. The discovery image indicated that the position of the transient was consistent with the nucleus of the host, with the Pan-STARRS coordinates lying within 0\farcs{1} of the measured center of the host in SDSS.

The transient was first spectroscopically observed by the Spectral Classification of Astronomical Transients \citep[SCAT;][]{SCATref} survey, which uses the SuperNova Integral Field Spectrograph \citep[SNIFS;][]{lantz04} on the University of Hawaii 88-inch telescope. The initial spectrum obtained on 2018 March 07 showed a blue continuum with no obvious emission or absorption features, and a second spectrum obtained on 2018 March 18 was very similar, with a strong blue continuum, but with the possible addition of broad Balmer emission lines \citep{ps18kh_spec_atel}. Based on these spectra, we obtained two additional low-resolution optical spectra on 2018 March 20 with the Wide Field Reimaging CCD Camera (WFCCD) mounted on the Las Campanas Observatory du Pont 2.5-m telescope ($3700-9600$~\AA, $\rm R\sim 7$~\AA) and the Fast Spectrograph \citep[FAST;][]{fabricant98} mounted on the Fred L. Whipple Observatory Tillinghast 1.5-m telescope ($3700-9000$~\AA, $\rm R\sim 3$~\AA). Both of these spectra also suggested the presence of broad Balmer emission lines with a strong blue continuum, both features of TDEs \citep[e.g.,][]{arcavi14}, and \citet{ps18kh_spec_atel} publicly announced that PS18kh was a TDE candidate on 2018 March 24. Based on  \ion{Ca}{2} H\&K absorption lines visible in the spectra, PS18kh has a redshift of $z=0.071$, corresponding to a luminosity distance of 322 Mpc ($H_0=69.6$~km~s$^{-1}$~Mpc$^{-1}$, $\Omega_M=0.29$, $\Omega_{\Lambda}=0.71$; see Section~\ref{sec:params}).

Based on the preliminary classification, we requested and were awarded target-of-opportunity (TOO) observations from the \textit{Neil Gehrels Swift Gamma-ray Burst Mission} \citep[\swift;][]{gehrels04} UltraViolet and Optical Telescope \citep[UVOT;][]{roming05} and X-ray Telescope \citep[XRT;][]{burrows05}. These observations confirmed that the transient was bright in the UV and appeared to have weak soft X-ray emission, so we began an extended multi-wavelength monitoring campaign to characterize PS18kh. With a peak $g$-band magnitude of $m_g\simeq17.3$, PS18kh was also well-observed by a number of ground-based optical surveys, and we include in our analysis multiwavelength pre- and post-discovery light curves from Pan-STARRS, the All-Sky Automated Survey for Supernovae \citep[ASAS-SN;][]{shappee14}, and the Asteroid Terrestrial-impact Last Alert System \citep[ATLAS;][]{tonry18} spanning from 56 days before the peak of the light curve until it became Sun-constrained 75 days after peak, making this one of the best-sampled early light curves for a TDE candidate to-date. 

In Section~\ref{sec:obs} we describe the available pre-outburst data for the host galaxy and fit the physical properties of the host. We also describe the new observations of the transient that were obtained by the Pan-STARRS, ASAS-SN, and ATLAS surveys and our follow-up campaign. In Section~\ref{sec:params} we perform detailed measurements of the position of PS18kh within its host, its redshift, and the time of peak light. In Section~\ref{sec:phot_anal} we analyze the photometric data and model the luminosity and temperature evolution of PS18kh. In Section~\ref{sec:spec_anal} we analyze the spectroscopic evolution of PS18kh and model the boxy, double-peaked emission line profiles in an attempt to determine the physical properties of the TDE-SMBH system. Finally, in Section~\ref{sec:disc} we compare the properties of PS18kh to those of supernovae and other TDEs and summarize our findings.


\section{Observations and Survey Data}
\label{sec:obs}


\subsection{Archival Data and Host Fits}
\label{sec:archival}

We retrieved archival optical $ugriz$ model magnitudes of {\host} from SDSS Data Release 14 \citep[DR14;][]{abolfathi18} and infrared $W1$ and $W2$ magnitudes from the Wide-field Infrared Survey Explorer \citep[WISE;][]{wright10} AllWISE catalog. The host is not detected in archival data from, or was not previously observed by, the Two Micron All-Sky Survey (2MASS), Spitzer, Herschel, the Hubble Space Telescope (HST), the Chandra X-ray Observatory, the X-ray Multi-Mirror Mission (XMM-Newton), or the Very Large Array Faint Images of the Radio Sky at Twenty-cm (VLA FIRST) survey. It is also not detected in Galaxy Evolution Explorer (GALEX) UV data, but we obtain 3-sigma 6\farcs{0} upper limits on the UV magnitudes of $NUV > 23.65$ and $FUV>23.69$ using single-epoch data obtained on 2008 January 19. The archival host magnitudes and limits are listed in Table~\ref{tab:host_mags}.

To place constraints on any X-ray emission prior to the flare that could be indicative of an AGN, we take advantage of data from the ROSAT All-sky Survey \citep{voges99}. We do not detect X-ray emission associated with the position of the host galaxy with a 3-sigma upperlimit on the count rate of $8\times 10^{-3}$ counts s$^{-1}$. Assuming an absorbed power law redshifted to the distance of the host galaxy and a photon index similar to that of known AGN ($\Gamma=1.75$: e.g., \citealt[][]{tozzi06, marchesi16, liu17, ricci17}), we derive a limit on the absorbed (unabsorbed) flux of $2.3~(2.6) \times 10^{-13}$ ergs~cm$^{-2}$~s$^{-1}$ in the 0.3-10.0 keV energy band. At the distance of PS18kh this flux limit corresponds to an X-ray luminosity of $3.2\times10^{42}$ ergs~s$^{-1}$. This is lower than the average luminosity of known AGN \citep[e.g.,][]{ricci17}, suggesting that the host galaxy of PS18kh does not harbor a strong AGN.

We fit the spectral energy distribution (SED) of the host galaxy to the archival limits and magnitudes from GALEX, SDSS, and WISE using the publicly available Fitting and Assessment of Synthetic Templates \citep[\textsc{fast};~][]{kriek09}. For the fit we assumed a \citet{cardelli89} extinction law with $R_V=3.1$ and a Galactic extinction of $A_V = 0.128$ mag \citep{schlafly11} and we adopted an exponentially declining star-formation history, a Salpeter initial mass function, and the \citet{bruzual03} stellar population models. In order to make a more robust estimate of the host SED and the uncertainties on its physical parameters, we generated 1000 realizations of the archival fluxes, perturbed by their respective uncertainties assuming Gaussian errors. Each realization was then modeled with \textsc{fast}. The median and 68\% confidence intervals on the host parameters from these 1000 realizations are: $M_{\star}=1.4^{+0.4}_{-0.4} \times 10^{10}$ M$_{\odot}$, age $=5.0^{+2.1}_{-1.9}$ Gyr, and a star formation rate $\textrm{SFR}= 6.8^{+4.0}_{-4.9} \times 10^{-3}$ M$_{\odot}$~yr$^{-1}$. We scaled the stellar mass of {\host} using the average stellar-mass-to-bulge-mass ratio from the hosts of ASASSN-14ae, ASASSN-14li, and ASASSN-15oi \citep{holoien14b,holoien16a,holoien16b}, to get a bulge mass estimate of $M_B\simeq10^{9.5}$~{\msun}. Using the $M_B-M_{BH}$ relation from \citet{mcconnell13}, we obtain a black hole mass of $M_{BH}=10^{6.9}$~{\msun}, comparable to what has been found for other optical TDE host galaxies \citep[e.g.,][]{holoien14b,holoien16a,holoien16b,brown17a,wevers17,mockler18}.


\begin{deluxetable}{ccc}
\tabletypesize{\footnotesize}
\tablecaption{Archival Photometry of \host}
\tablehead{
\colhead{Filter} &
\colhead{Magnitude} &
\colhead{Magnitude Uncertainty} }
\startdata
$FUV$ & $>23.69$ & --- \\
$NUV$ & $>23.65$ & --- \\
$u$ & 20.97 & 0.12 \\
$g$ & 18.93 & 0.01 \\
$r$ & 18.17 & 0.01 \\
$i$ & 17.76 & 0.01 \\
$z$ & 17.46 & 0.01 \\
$W1$ & 15.19 & 0.94 \\
$W2$ & 15.32 & 0.11
\enddata 
\tablecomments{Archival model magnitudes of {\host} from SDSS DR14 ($ugriz$) and PSF photometry magnitudes from the AllWISE catalog ($W1$ and $W2$). The GALEX $NUV$ and $FUV$ upper limits are 3-sigma upper limits measured with a 6\farcs{0} aperture from a single epoch of data obtained on 2008 January 19.} 
\label{tab:host_mags} 
\end{deluxetable}

Our photometric follow-up campaign includes $ugri$ photometry, for which the archival SDSS data can be used to subtract the host flux and isolate the transient flux. For the {\swift} UVOT and Johnson-Cousins $BV$ data, there are no available archival images. To obtain 5\farcs{0} aperture host flux measurements to use for host subtraction in the $ugri$ filters, we measured 5\farcs{0} aperture magnitudes from the archival SDSS images using the IRAF {\tt apphot} package, with the magnitudes calibrated using several stars in the field with well-defined magnitudes in SDSS DR14. In order to estimate the host flux in the filters without archival data, we used the bootstrapped SED fits for the host galaxy to derive synthetic host magnitudes for each photometric band in our follow-up campaign. For each of the 1000 host SEDs, we computed synthetic 5\farcs{0} aperture magnitudes in each of our follow-up filters. This yields a distribution of synthetic magnitudes for each filter, and we report the median and 68\% confidence intervals on the host magnitudes, along with the measured $ugri$ magnitudes, in Table~\ref{tab:synth_host}. These host magnitudes were used to obtain host-subtracted transient magnitudes for the non-survey data in our analyses.

\subsection{Pan-STARRS light curve}
\label{sec:PS_LC}

The Pan-STARRS1 telescope, located at the summit of Haleakala on Maui, has a 1.8-m diameter primary mirror with a f/4.4 Cassegrain focus. The telescope uses a wide-field 1.4 gigapixel camera mounted at the Cassegrain focus, consisting of sixty Orthogonal Transfer Array devices, each of which has a detector area of 4846$\times$4868 pixels. The 10 micron pixels have a plate scale of 0\farcs{26}, giving a full field-of-view area of 7.06 square degrees, with an active region of roughly 5 square degrees. Pan-STARRS1 uses the $grizy_{P1}$ filters, which are similar to those of SDSS \citep{abazajian09}, with the redder $y$ filter replacing the bluer SDSS $u$ filter. The Pan-STARRS1 photometric system in discussed in detail in \citet{tonry12}.


\begin{deluxetable}{ccc}
\tabletypesize{\footnotesize}
\tablecaption{5\farcs{0} Host Galaxy Magnitudes}
\tablehead{
\colhead{Filter} &
\colhead{Magnitude} &
\colhead{Magnitude Uncertainty} }
\startdata
$UVW2$ & 24.81 & 0.60 \\
$UVM2$ & 24.64 & 0.43 \\
$UVW1$ & 23.19 & 0.14 \\
$U_{UVOT}$ & 20.95 & 0.07 \\
$u$ & 21.28 & 0.43 \\
$B$ & 19.48 & 0.04 \\
$g$ & 18.94 & 0.21 \\
$V$ & 18.45 & 0.02 \\
$r$ & 18.07 & 0.14 \\
$i$ & 17.76 & 0.12
\enddata 
\tablecomments{5\farcs{0} aperture magnitudes of {\host} synthesized in the {\swift} UV$+U$ filters and the Johnson-Cousins $BV$ filters and their 68\% confidence intervals, and measured from archival SDSS images in the $ugri$ filters. Magnitudes were synthesized and measured using the processes described in \S\ref{sec:archival} and are presented in the AB system.} 
\label{tab:synth_host} 
\end{deluxetable} 

Pan-STARRS1 images are processed with the Image Processing Pipeline \citep[IPP; see details in][]{magnier13}. The IPP runs new images through successive stages of processing, including device ``de-trending'', a flux-conserving warping to a sky-based image plane, masking and artefact location that involves bias and dark correction, flatfielding, and illumination correction obtained by rastering sources across the field of view \citep{waters16}. After determining an initial astrometric solution, corrected images are then warped onto the tangent plane of the sky using a flux-conserving algorithm, which involves mapping the camera pixels to a defined set of skycells. For nightly processing, the zeropoints of the camera chips are set using a catalog of photometric reference stars from the ``ubercal'' analysis of the first reprocessing of the PS1 3$\pi$ data \citep{schlafly12,magnier13}. The internal calibration of this catalog has a relative precision of roughly 1\%, but the automated zeropoint applied in difference imaging is an average full-field zeropoint, which can result in variations across skycells of up to $\pm0.15$ magnitudes.

Transient searching is aided by having pre-existing sky images from the Pan-STARRS1 Sky Surveys \citep{chambers16}. The IPP creates difference images by subtracting stacked reference images from the PS1 $3\pi$ from newly observed images, and transient sources are then identified by the IPP through analysis of the difference images \citep[e.g.,][]{huber15}. Catalog source files from the IPP are transferred from Hawaii to Belfast and ingested into a MySQL database. A series of quality cuts are implemented  \citep{mccrum15,smartt16} together with a machine learning algorithm that distinguishes real sources from bogus sources \citep{wright15}. Sources are  accumulated into unique objects and spatially cross-matched against all large catalogs, therefore providing both a real-bogus value and a classification of variable star, AGN, supernova, CV, or nuclear transient. The $grizy_{P1}$ lightcurve presented in this manuscript was produced from this Pan-STARRS transient processing pipeline as described in \citet{mccrum14,mccrum15} and \citet{smartt16}. The Pan-STARRS1 $griz$ photometry is presented in Table \ref{tab:phot} and is shown in Figure~\ref{fig:lc}; we do not present the $y$ photometry as PS18kh was only detected in one $y$-band epoch.

\subsection{ASAS-SN light curve}
\label{sec:ASASSN_LC}

ASAS-SN is an ongoing project that monitors the full visible sky on a rapid cadence to find bright, nearby transients \citep{shappee14,kochanek17}. ASAS-SN uses units of four 14-cm telescopes on a common mount located at multiple sites in both hemispheres and hosted by the Las Cumbres Observatory global telescope network \citep{brown13}. The ASAS-SN network was expanded in 2017 and now comprises five units located in Hawaii, Chile, Texas, and South Africa. With its current capacity, ASAS-SN observes the entire visible sky every $\sim20$ hours to a depth of $g\simeq18.5$~mag, weather permitting. ASAS-SN has proven to be a powerful tool for discovering TDEs, and it has discovered three of the four nearest and brightest TDEs to-date: ASASSN-14ae \citep{holoien14b,brown16a}, ASASSN-14li \citep{holoien16a,prieto16,romero16,brown17a}, and ASASSN-15oi \citep{holoien16b,holoien18a}. The three ASAS-SN TDEs have since become some of the most well-studied TDEs, with multiwavelength datasets spanning multiple years.

ASAS-SN processes new images using a fully automatic pipeline that incorporates the ISIS image subtraction package \citep{alard98, alard00}. After the discovery of PS18kh, a host-galaxy reference image was constructed for each ASAS-SN unit that could observe it. As the transient was still brightening, we only used images obtained at least 35 days before the discovery of PS18kh to ensure that no transient flux was present in the references. These reference images were then used to subtract the host galaxy's background emission from all science images. Aperture photometry was computed for each host-template subtracted science image using the IRAF {\tt apphot} package, with the magnitudes being calibrated using multiple stars in the field of the host galaxy with known magnitudes in the AAVSO Photometric All-Sky Survey \citep[APASS;][]{henden15}. For some of the pre-discovery epochs when PS18kh was still very faint, we stacked multiple science images in order to improve the signal-to-noise (S/N) of our detections. All ASAS-SN photometric measurements (detections and 3-sigma limits) are presented in Table~\ref{tab:phot} and shown in Figure~\ref{fig:lc}, with error bars on the X-axis used to denote the date ranges of epochs that were combined.

\subsection{ATLAS light curve}
\label{sec:ATLAS_LC}

ATLAS is an ongoing survey project with the primary goal of detecting small (10--140 m) asteroids that are on a collision course with Earth \citep{tonry18}. ATLAS uses fully robotic 0.5m f/2 Wright Schmidt telescopes located on the summit of Haleakal\=a and at Mauna Loa Observatory to monitor the entire sky visible from Hawaii every few days. During normal operations, each telescope obtains four 30-second exposures of 200--250 target fields per night, allowing the two telescopes to cover roughly a quarter of the visible sky each night. The four observations of a given field are typically obtained within less than an hour of each other. ATLAS uses two broad filters for its survey operations, with the `cyan' filter ($c$) covering 420--650 nm and the `orange' filter ($o$) covering 560--820 nm \citep{tonry18}.


\begin{deluxetable}{cccc}
\tabletypesize{\footnotesize}
\tablecaption{Host-Subtracted Photometry of PS18kh}
\tablehead{
\colhead{MJD} &
\colhead{Filter} &
\colhead{Magnitude} &
\colhead{Telescope/Observatory} }
\startdata
58220.29 & $z$ & $18.22\pm0.04$ & PS1 \\
58225.25 & $z$ & $18.31\pm0.07$ & PS1 \\
58260.26 & $z$ & $19.13\pm0.04$ & PS1 \\
... & & & \\
58261.12 & $UVW2$ & $18.71\pm0.07$ & \swift \\ 
58264.04 & $UVW2$ & $18.84\pm0.07$ & \swift \\
58267.82 & $UVW2$ & $18.66\pm0.07$ & \swift \\
\enddata 
\tablecomments{Host-subtracted magnitudes and 3-sigma upper limits in all photometric filters used for follow-up data. The Telescope/Observatory column indicates the source of the data in each epoch: ``PS1'', ``ASAS-SN'', and ``ATLAS'' are used for Pan-STARRS, ASAS-SN, and ATLAS survey data, respectively; ``CFHT'', ``PO'', and ``LT'' are used for Canada-France-Hawaii Telescope, Post Observatory, and Liverpool Telescope data, respectively; and ``\swift'' is used for {\swift} UVOT data. ``Syn'' indicates magnitudes synthesized from follow-up spectra, as described in Section~\ref{sec:spec}. These measurements are corrected for Galactic extinction, and all magnitudes are presented in the AB system. This Table is published in its entirity in a machine-readable format in the online journal; a portion is shown here for guidance regarding its form and content.} 
\label{tab:phot} 
\end{deluxetable}

Every ATLAS image is processed by a fully automated pipeline that performs flat fielding, astrometric calibration, and photometric calibration. A low-noise reference image of the host field was constructed by stacking multiple images taken under excellent conditions and this reference was then subtracted from each science image of PS18kh in order to isolate transient flux. We performed forced photometry on the subtracted ATLAS images of PS18kh as described in \citet{tonry18}, and then combined the intra-night photometric observations using a weighted average to get a single flux measurement for each epoch of observation. The ATLAS $o$-band photometry and 3-sigma limits are presented in Table \ref{tab:phot} and are shown in Figure~\ref{fig:lc}. We do not present the $c$ photometry as there were few $c$ observations during this period due to weather and the design of the ATLAS survey. Because of this, PS18kh was only detected in two $c$-band epochs.

\subsection{Swift Observations}
\label{sec:swift}

After PS18kh was classified as a TDE candidate, we were awarded 20 epochs of {\swift} TOO observations of PS18kh between 2018 March 27 and 2018 May 29, after which it became Sun-constrained. The UVOT observations were obtained in the $V$ (5468 \AA), $B$ (4392 \AA), $U$ (3465 \AA), $UVW1$ (2600 \AA), $UVM2$ (2246 \AA), and $UVW2$ (1928 \AA) filters \citep{poole08} for all epochs. As each epoch contained 2 observations in each filter, we first combined the two images in each filter using the HEAsoft software task {\tt uvotimsum}, and then extracted counts from the combined images in a 5\farcs{0} radius region using the software task {\tt uvotsource}, with a sky region of $\sim$~40\farcs{0} radius used to estimate and subtract the sky background. The UVOT count rates were converted into magnitudes and fluxes based on the most recent UVOT calibration \citep{poole08,breeveld10}. 

\begin{figure*}
\begin{minipage}{\textwidth}
\centering
{\includegraphics[width=0.95\textwidth]{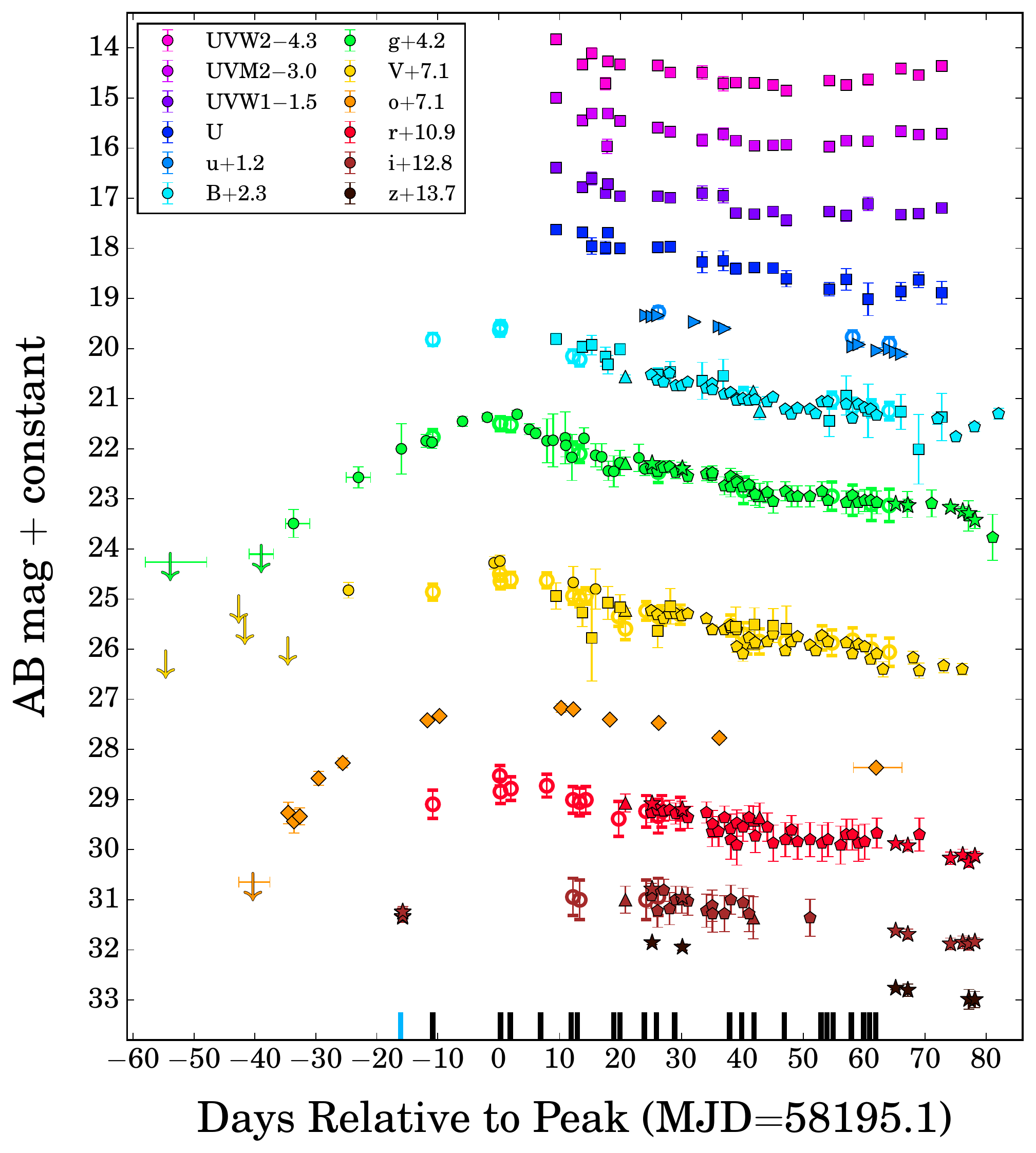}}
\caption{Host-subtracted UV and optical light curves of PS18kh spanning roughly 2 months before and 2.5 months after peak brightness (MJD$=$58195.1, measured from the ASAS-SN $g$ light curve; see Section~\ref{sec:params}). Pan-STARRS1 ($griz$), ASAS-SN ($gV$), and ATLAS ($o$) survey data are shown as stars, circles, and diamonds, respectively; follow-up {\swift} UVOT data are shown as squares; and follow-up ground data from LT ($BVgri$), Post Observatory ($BVgri$), and CFHT ($u$) are shown as triangles, pentagons, and right-facing triangles, respectively. Photometry synthesized from spectra are shown as open circles. 3-sigma upper limits are indicated with downward arrows. Error bars in time are used to denote the date range of observations that have been combined to obtain a single measurement. {\swift} $B$ and $V$ data have been converted to Johnson $B$ and $V$ magnitudes to enable direct comparison with ground-based follow-up data. The blue vertical bar on the X-axis shows the epoch of discovery, and the black bars show epochs of spectroscopic follow-up. All data have been corrected for Galactic extinction and are presented in the AB system.}
\label{fig:lc}
\end{minipage}
\end{figure*}

We corrected the UVOT magnitudes for Galactic extinction assuming a \citet{cardelli89} extinction law. Using the synthetic 5\farcs{0} host fluxes calculated from the \textsc{FAST} fits, we then subtracted the host flux from each UVOT observation to isolate the transient flux in each band. To enable direct comparison to ASAS-SN magnitudes and other ground-based follow-up photometry, we converted the UVOT $B$- and $V$-band data to Johnson $B$ and $V$ magnitudes using publicly available color corrections\footnote{\url{https://heasarc.gsfc.nasa.gov/docs/heasarc/caldb/swift/docs/uvot/uvot_caldb_coltrans_02b.pdf}}. The host-subtracted {\swift} UVOT photometry and 3-sigma limits are presented in Table \ref{tab:phot} and are shown in Figure~\ref{fig:lc}.


\begin{deluxetable}{ccc}
\tabletypesize{\footnotesize}
\tablecaption{\swift{} XRT photometry of PS18kh}
\tablehead{
\colhead{MJD Range} &
\colhead{Unabsorbed Flux} &
\colhead{Uncertainty} }
\startdata
$58204-58221$ & $3.44\times10^{-14}$ & $1.21\times10^{-14}$ \\
$58223-58240$ & $3.16\times10^{-14}$ & $1.21\times10^{-14}$ \\
$58242-58267$ & $<2.88\times10^{-14}$ & --- \\
\enddata 
\tablecomments{X-ray fluxes measured from merged observations from the \swift{} XRT. The first column gives the date range in MJD of the observations combined for each merged observation. Fluxes are given in ergs~cm$^{-2}$~s$^{-1}$. No X-ray emission was detected in the third merged observation, and the corresponding row gives a 3-sigma upper limit on the flux.} 
\label{tab:xray} 
\end{deluxetable}


PS18kh was also observed using the \swift{} XRT. All observations were taken in photon counting mode, and were reprocessed from level one XRT data using the \swift{} \textsc{XRTPIPELINE} version 0.13.2. As suggested in the \swift{} XRT data reduction guide\footnote{\url{https://swift.gsfc.nasa.gov/analysis/xrt_swguide_v1_2.pdf}}, standard filters and screening were applied, along with the most up-to-date calibration files. We used a source region centered on the position of PS18kh with a radius of 30\arcsec, and a source free background region centered at $(\alpha,\delta)=$(07:57:07.71, $+$34:20:59.97) with a radius of 150\farcs{0}. All extracted count rates were corrected for the encircled energy fraction (a 30\farcs{0} source radius contains only $\sim$90\% of the counts from a source at 1.5 keV; \citealt{moretti04}).

To increase the signal-to-noise of our observations, we combined the individual XRT observations using \textsc{XSELECT} version 2.4d. We combined our observations into three time-bins spanning the full \swift{} observing campaign and merged all observations together to extract an X-ray spectrum with the highest signal-to-noise possible. From these merged observations, we used the task \textsc{XRTPRODUCTS} to extract both source and background spectra. Ancillary response files were derived using \textsc{XRTMKARF} and merged exposure maps were created from the individual observations using \textsc{XIMAGE} version 4.5.1. We took advantage of the ready-made response matrix files (RMFs), which are obtained from the most up-to-date \swift{} CALDB. The XRT fluxes and 3-sigma upper limits measured from the merged observations are given in Table~\ref{tab:xray}.

The spectral data were analyzed using the X-ray spectral fitting package (XSPEC) version 12.9.1 and $\chi^2$ statistics. Each spectrum was grouped using \textsc{FTOOLS} command \textit{grppha} to have a minimum of 10 counts per energy bin. Due to the faintness of the X-ray emission from this source, the signal-to-noise of the resulting spectrum is quite low. As such, the spectrum is insufficient to constrain the column density ($N_{H}$) and so we fixed it to $N_{H}=4.42\times10^{20}$ cm$^{-2}$, which is the Galactic H\textsc{I} column density in the direction of PS18kh \citep{kalberla05}.

\subsection{Other Photometric Observations}
\label{sec:other_phot}

In addition to the survey data and {\swift} observations, we also obtained photometric observations from multiple ground observatories. $BVgri$ observations were obtained from the 2-m Liverpool Telescope \citep{steele04} and from the 24-inch Post Observatory robotic telescopes located in Mayhill, New Mexico, and Sierra Remote Observatory in California. Additional $u$-band data were obtained with MegaCam \citep{boulade98} on the Canada-France-Hawaii Telescope (CFHT). After flat-field corrections were applied to these follow-up data, we measured 5\farcs{0} aperture magnitudes using the IRAF {\tt apphot} package, with the magnitudes calibrated using several stars in the field with well-defined magnitudes in SDSS DR14. $B$ and $V$ reference star magnitudes were calculated from the SDSS $ugriz$ magnitudes using the corrections from \citet{lupton05}. 

As was done with the {\swift} UVOT magnitudes, after calculating the 5\farcs{0} aperture fluxes in each image, we corrected for Galactic extinction and subtracted the host flux using the synthetic host magnitudes calculated from the FAST fits. The host-subtracted ground-based follow-up photometry are presented in Table \ref{tab:phot} and are shown in Figure~\ref{fig:lc}.

\subsection{Spectroscopic Observations}
\label{sec:spec}

After classifying PS18kh as a TDE candidate, we began a program of spectroscopic follow-up to complement our photometric follow-up. The telescopes and instruments used to obtain follow-up spectra as part of this campaign included SNIFS on the University of Hawaii 88-inch telescope, the Inamori-Magellan Areal Camera and Spectrograph \citep[IMACS;][]{dressler11} on the 6.5-m Magellan-Baade telescope, the Gemini Multi-Object Spectrograph \citep[GMOS;][]{hook04} on the 8.2-m Gemini North telescope, the SPectrograph for the Rapid Acquisition of Transients (SPRAT) on the Liverpool Telescope, the Low-Resolution Imaging Spectrometer \citep[LRIS;][]{oke95} on the Keck I 10-m telescope, and the Multi-Object Double Spectrographs (MODS; \citealt{Pogge2010}) mounted on the dual 8.4-m Large Binocular Telescope (LBT).


\begin{figure*}
\begin{minipage}{\textwidth}
\centering
{\includegraphics[width=0.90\textwidth]{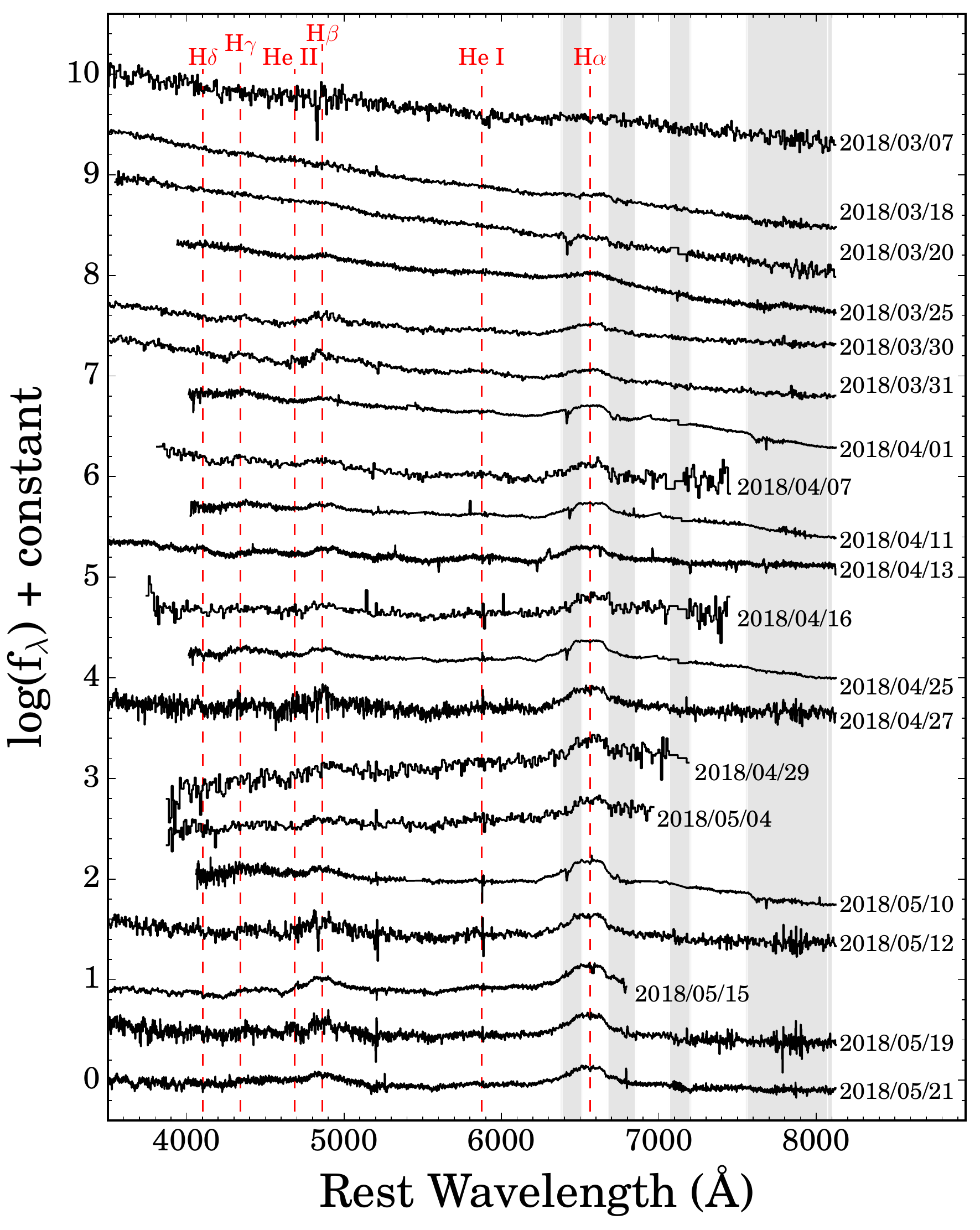}}
\caption{Spectroscopic evolution of PS18kh spanning from 11 days before peak (2018 March 18) through 64 days after peak. The spectra have been flux-calibrated to the photometry, as described in Section~\ref{sec:spec}. Hydrogen and helium emission features common to TDEs are indicated with red dashed lines and telluric bands are shown in light gray. For cases where the telluric features were not removed in calibration, the A-band telluric feature has been masked to facilitate plotting. The spectra labelled ``2018/05/12'' and ``2018/05/19'' are coadded spectra from SNIFS, combining data from 2018 May $11-12$ and 2018 May $17-19$, respectively.}
\label{fig:spec_evol}
\end{minipage}
\end{figure*}

We reduced and calibrated the majority of the spectra using IRAF following standard procedures, including bias subtraction, flat-fielding, 1-D spectral extraction, and wavelength calibration by comparison to an arc lamp. The MODS spectra were reduced using the MODS spectroscopic pipeline\footnote{\url{http://www.astronomy.ohio-state.edu/MODS/Software/modsIDL/}}. The observations were flux calibrated with spectroscopic standard star spectra obtained on the same nights as the science spectra. In some cases, we also performed telluric corrections using the standard star spectra, and in other cases we masked prominent telluric features. In order to increase the signal-to-noise of later observations from SNIFS, spectra taken within $2-3$ days of each other were co-added, with each spectrum weighted by its uncertainty. Details of all spectra obtained for PS18kh are presented in Table~\ref{tab:spec_details}.

We futher calibrated the spectra using the photometric measurements. We extracted synthetic photometric magnitudes for each filter that was completely contained in the wavelength range covered by the spectrum and for which we could either interpolate the photometric light curves or extrapolate them by 1 hour or less. We fit a line to the difference between the observed and synthetic flux as a function of central wavelength and scaled each spectrum by this fit. We corrected the observed spectra for Galactic reddening using a Milky Way extinction curve and assuming $R_V=3.1$ and $A_V=0.128$ \citep{schlafly11}.

The spectroscopic evolution of PS18kh is shown in Figure~\ref{fig:spec_evol}. For cases where multiple observations were obtained on a given night, only one spectrum is shown. The SNIFS spectra labelled ``2018/05/12'' and ``2018/05/19'' are coadded spectra combining data from 2018 May $11-12$ and 2018 May $17-19$, respectively. The SNIFS dichroic split falls very close to the \hbeta{} line, and some of the SNIFS spectra (2018 March 7, March 18, March 30, March 31, April 27, May 12, and May 19) show residual noise around \hbeta{} as a result.

After calibrating the spectra, we synthesized photometric magnitudes from each follow-up spectrum for each filter that was completely contained in the wavelength range covered by the spectrum. These magnitudes were corrected for Galactic extinction and host fluxes were subtracted using the synthetic and measured host 5\farcs{0} magnitudes, as was done with the {\swift} and ground follow-up data. The host-subtracted synthetic photometry are presented in Table \ref{tab:phot} and are shown in Figure~\ref{fig:lc}.

\section{Analysis}
\label{sec:anal}

\subsection{Position, Redshift, and $t_{Peak}$ Measurements}
\label{sec:params}

We used the discovery $i$-band image obtained by Pan-STARRS1 on 2018 March 02 and the corresponding Pan-STARRS1 $i$-band reference image to measure an accurate position of the transient. We first measured the centroid position of the transient in the host-subtracted discovery image and the centroid position of the host galaxy nucleus in the reference image using the \textsc{Iraf} task \texttt{imcentroid}, then calculated the offset between the two positions. From this method, we obtain a position of RA$=$07:56:54.53, Dec$=+$34:15:43.58 for PS18kh. We calculate an offset of $0.28\pm0.29$ arcseconds from the host nucleus, corresponding to a physical projected distance of $0.45\pm0.48$~kpc at the distance of the host.

We initially obtained a redshift of the transient using a Gaussian fit to the \halpha{} emission line in the Magellan IMACS spectrum obtained on 2018 March 25, as this spectrum had both high S/N and was obtained before the double-peaked feature started to appear in the emission lines. This preliminary redshift was $z=0.074$, but this was uncertain due to being measured from such a broad feature. We were later able to refine this measurement using \ion{Ca}{2} H \& K lines from the host galaxy that are visible in the LBT MODS spectrum obtained on 2018 May 21. From these narrow features, we obtain a redshift of $z=0.071$, corresponding to a distance of $d=322.4$~Mpc.

To estimate the time of peak light, we fit a parabolic function to the ASAS-SN $g$ and ATLAS $o$ light curves near peak. In order to estimate the uncertainty on the peak dates, we used a procedure similar to the one used to estimate the uncertainties on the host galaxy parameters: we generated 10000 realizations of the $g$ and $o$ light curves near peak, with each magnitude perturbed by their respective uncertainties and assuming Gaussian errors. We then fit a parabola to each of these light curves and calculated the 68\% confidence interval and median $t_{peak}$ values. For $g$-band, we obtain $t_{g,peak}=58195.1^{+0.8}_{-0.8}$ and $m_{g,peak}=17.4$, while for $o$-band we obtain $t_{o,peak}=58198.5^{+0.5}_{-0.6}$ and $m_{o,peak}=17.6$. This discrepancy between filters is not unexpected, as PS18kh was becoming redder in optical filters, which will result in later peak dates in redder filters. We adopt the median $g$-band peak of $t_{g,peak}=58195.1$, corresponding to 2018 March 18.1, when discussing data with respect to peak time throughout the manuscript.

\subsection{Light Curve Analysis and SED Fits}
\label{sec:phot_anal}

The ASAS-SN and ATLAS survey data make PS18kh one of the few TDE candidates with a well-sampled rising light curve. PS18kh brightened by roughly 2.1 magnitudes over 40 days in $g$-band, reaching a peak of $m_{g,peak}=17.3$. It brightened by a similar amount in the ATLAS $o$-band over the same time frame, but the rise is less dramatic in redder filters such as $i$ and $z$. After peak, PS18kh faded gradually in all optical filters redder than $U$, but was still brighter than the magnitude of first detection in $g$-band in the observations obtained 78 days after peak. At $i$-band, in contrast, the transient was fainter in later data than it was in the discovery epoch, and in some cases was consistent with the measured host magnitude. In the {\swift} UV$+U$ bands, the flux plateaus, or begins to re-brighten $\sim50$~days after peak, with the effect being more pronounced in bluer filters.


\begin{figure}
\centering
\subfloat{{\includegraphics[width=0.45\textwidth]{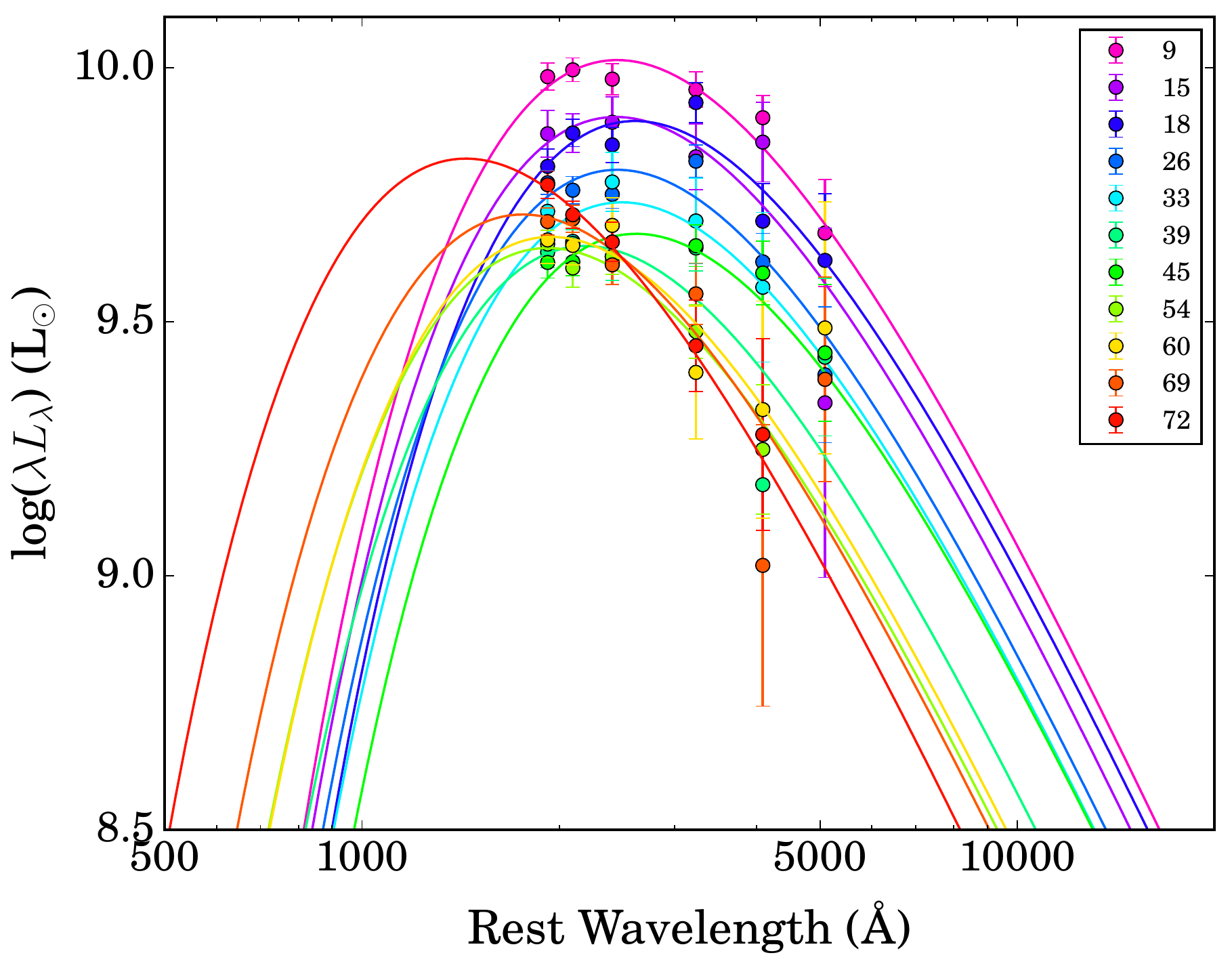}}}
\caption{Evolution of the blackbody SED fit to the {\swift} data, with rest-frame phase relative to peak light shown in the legend. Data from the individual {\swift} filters are shown for each epoch in matching colors. For ease of visibility, only every other epoch is shown in the figure.}
\label{fig:sed_evol}
\end{figure}

\begin{figure}
\centering
\subfloat{{\includegraphics[width=0.45\textwidth]{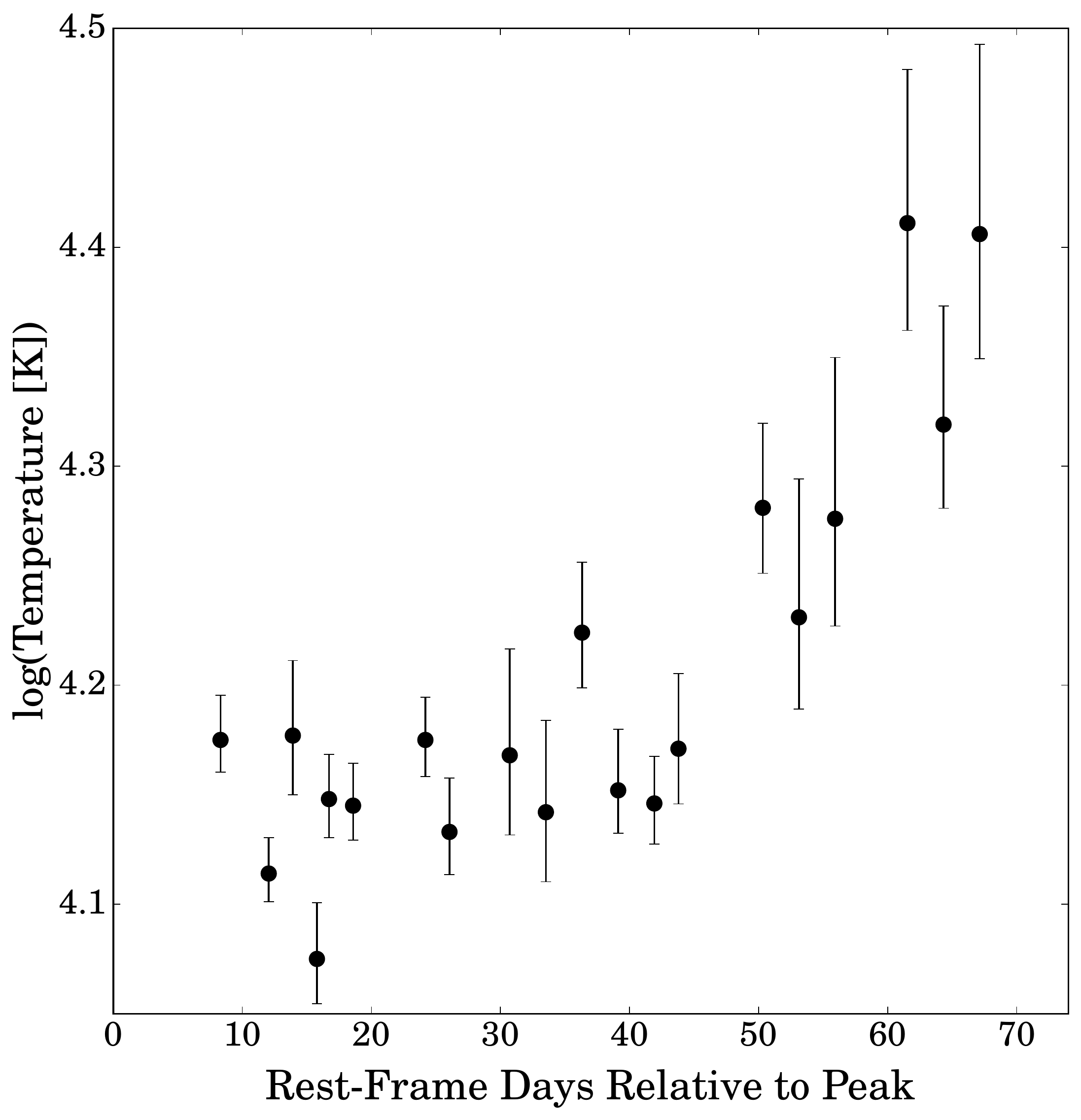}}}
\caption{Temperature evolution of PS18kh from blackbody fits to the UV/optical {\swift} SED. All fits were made with a flat prior of $4.00\leq \log{T} \leq 4.74$.}
\label{fig:temp_evol}
\end{figure}

To better quantify the physical parameters of the system, we modeled the UV and optical SED of PS18kh for epochs where {\swift} data were available as a blackbody using Markov Chain Monte Carlo methods, as was done for the previous ASAS-SN TDEs \citep[e.g.,][]{holoien14b,holoien16a,holoien16b,brown16a,brown17a,holoien18a}. So as not to overly influence the fits, we performed the blackbody fits using a flat prior of $10000$~K~$\leq T \leq55000$~K in all epochs. As can be seen in Figure~\ref{fig:sed_evol}, which shows the best-fit blackbody SED at various epochs compared to the \swift{} photometry, the blackbody fits provide good fits to the data. The resulting temperature evolution in rest-frame days relative to peak is shown in Figure~\ref{fig:temp_evol}, with time corrected to rest-frame days relative to peak.

The blackbody fits indicate that for the first $\sim45$ days after peak, the temperature of PS18kh held relatively constant around $T\simeq14000$~K. This temperature and flat evolution is not uncommon for TDEs \citep[e.g.,][]{holoien14b,holoien16a,brown16a,brown17a,holoien18a}. However, after the UV flux began to rise, the transient became hotter, with the temperature increasing to $T\simeq25000$~K over the following 3 weeks. This temperature is similar to that of other TDEs, but the rising behavior seen $\sim50$ days after peak is unusual. Unfortunately, it is unclear whether the temperature continued to increase further, as PS18kh became Sun-constrained for {\swift} not long after the source began to rebrighten in the UV.


\begin{figure}
\centering
\subfloat{{\includegraphics[width=0.45\textwidth]{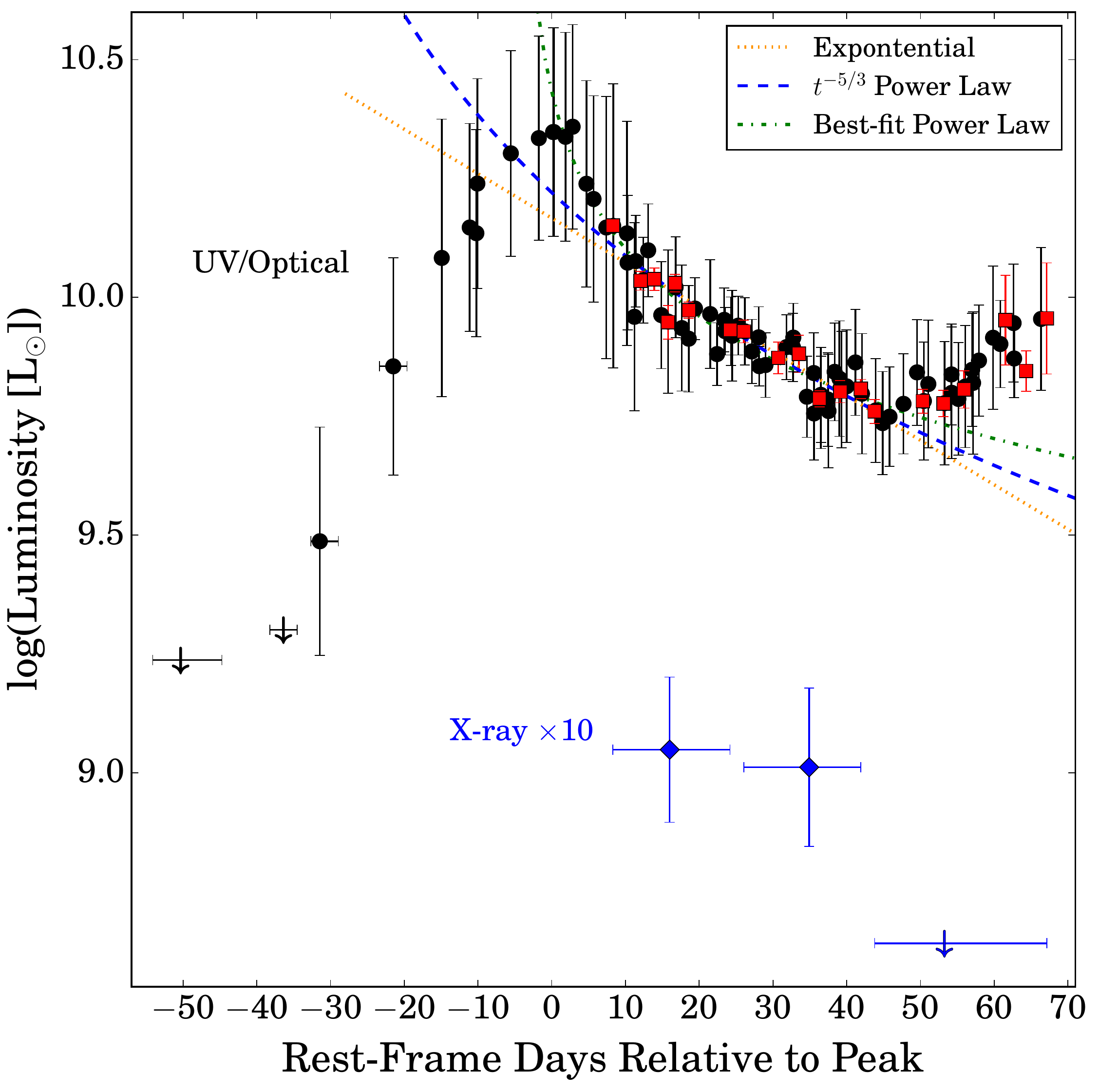}}}
\caption{Luminosity evolution of PS18kh from blackbody fits to the UV/optical {\swift} SED (red squares) and estimated from the $g$-band light curve after applying bolometric corrections based on the {\swift} fits (black circles). The dotted, dashed, and dash-dotted lines show exponential, $t^{-5/3}$ power-law, and best-fit power-law fits to the early fading luminosity curve, respectively. The blue diamonds show the {\swift} XRT luminosity evolution, multiplied by a factor of 10 to improve readability. Downward arrows indicate upper limits, and X-axis error bars indicate date ranges of data combined to obtain a single measurement.}
\label{fig:lum_evol}
\end{figure}

For those epochs with {\swift} data, we also estimated the bolometric luminosity of PS18kh from the blackbody fits. In order to better take advantage of the high-cadence light curve, we used the epochs with {\swift} blackbody fits to calculate bolometric corrections to the $g$-band data taken within 1 day of the {\swift} observations, or to $g$-band magnitudes interpolated between the previous and next $g$-band observations if there was no observation within 1 day of the {\swift} observation. We then used these bolometric corrections to estimate the bolometric luminosity of PS18kh from the $g$-band data for epochs when we did not have {\swift} data, linearly interpolating the bolometric corrections for each $g$-band epoch. For epochs prior to our first {\swift} observation, we used the bolometric correction from the first {\swift} SED fit. We do not correct the data taken after the last {\swift} observation, as the $g$-band continued to decline while the UV was re-brightening, and we do not want to extrapolate a rising or falling behavior beyond what our SED fits can tell us. The luminosity evolution calculated from the {\swift} SED fits and estimated from the $g$-band light curve is shown in Figure~\ref{fig:lum_evol}.

As suggested by the {\swift} light curves, while the luminosity initially drops after peak, it begins to rise again $\sim50$ (rest-frame) days after peak. As we did with previous TDEs, we fit the initial fading light curve ($0<t<50$~days) with an exponential profile $L= L_0e^{-(t-t_0)/\tau}$, a $L=L_0 (t-t_0)^{-5/3}$ power-law profile, and a power law where the power-law index is fit freely, $L\propto (t-t_0)^{-\alpha}$. Our best fit parameters for each model are as follows: for the exponential profile we obtain $L_0=10^{44.0}$~ergs~s$^{-1}$, $t_0=58163.3$, and $\tau=49.8$ days; for the $t^{-5/3}$ power law we obtain $L_0=10^{46.7}$~ergs~s$^{-1}$ and $t_0=58142.0$; and for the free power law we obtain $L_0=10^{44.4}$~ergs~s$^{-1}$, $t_0=58190.9$, and $\alpha=0.60$. We find that both power laws provide better fits than the exponential profile, with $\chi^2=31.0$, $\chi^2=43.4$, and $\chi^2=57.2$, for the free power law, the $t^{-5/3}$ power law, and the exponential fit, respectively. All three fits are shown in Figure~\ref{fig:lum_evol}. 


As can be seen in the Figure and from the $\chi^2$ value of the fit, the $t^{-5/3}$ profile is not a particularly good fit to the data, as the luminosity initially declines at a steeper rate, and then levels off sooner than such a profile would predict. However, it is expected that there should be some deviation from this profile near peak, as the luminosity is not expected to track the fallback rate until later in the flare, and the initial steeper decline after peak could be due to inefficient circularization of the stellar debris \citep[e.g.,][]{dai15,guillochon15}. The best-fit $t^{-0.60}$ power law profile is closest to the $t^{-5/12}$ power law expected for disk-dominated emission \citep[e.g.,][]{lodato11,auchettl17}, though the flare is not expected to exhibit this decline rate until later times after peak. It is clear that the luminosity evolution of PS18kh is more complicated than the simple $t^{-5/3}$ rate that would be observed if the luminosity tracked the mass fallback rate, as predicted in \citet{rees88} and \citet{phinney89}, and it is not a good match to any individual theory, implying that multiple physical processes may be contributing to the observed luminosity.

Figure~\ref{fig:lum_evol} also shows the X-ray luminosity calculated from the binned {\swift} XRT observations. While there is weak X-ray emission detected in the two earlier time bins, we do not detect any X-ray emission at later times, and the detected X-ray luminosity is 2 or more orders of magnitude weaker than the UV/optical emission in all epochs. The X-ray detections are below the archival limit from ROSAT, and we cannot definitively determine whether it is associated with the host or the transient based on the measured flux. Similar to what was seen with ASASSN-15oi and ASASSN-14li at early times \citep{holoien16a,holoien16b}, the X-ray emission does not show strong evolution during the period of observation. 

Modeling the X-ray spectrum obtained by combining all the XRT data, we find that the X-ray emission favors an absorbed power-law spectrum with a photon index of $\Gamma=3\pm1$. We also tested an absorbed blackbody model, but find that this produces a significantly worse fit (reduced $\chi_{r}^{2} \sim 2)$ compared to the simple powerlaw ($\chi_{r}^{2} \sim 1)$. \citet{auchettl17,auchettl18} showed that the X-ray emission of a non-jetted TDE can be well described by photon indices larger than $\sim3$, which is consistent with that obtained for PS18kh. These values are much softer than seen for AGN, which have photon indexes $\sim1.75$ \citep[e.g.,][]{auchettl17}, suggesting that the emission we see arises from the TDE, rather than an underlying AGN.

Integrating over the entire rest-frame bolometric light curve calculated from the $g$-band data and the {\swift} blackbody fits gives a total radiated energy of $E=(3.46\pm0.22)\times10^{50}$~ergs, with $(1.42\pm0.20)\times10^{50}$~ergs being released during the rise to peak. This shows that a significant fraction of energy radiated from TDEs can be emitting during the rise to peak, and highlights the need for early detection. The total radiated energy corresponds to an accreted mass of $M_{Acc}\simeq0.002\eta_{0.1}^{-1}$~\msun, where the accretion efficiency is $\eta=0.1\eta_{0.1}$. As with other TDEs, a negligible fraction of the bound stellar material appears to actually accrete onto the black hole, or the material is accreting with a very low radiative efficiency.

\subsection{Spectroscopic Analysis}
\label{sec:spec_anal}

The dominant spectral features of PS18kh are a strong blue continuum and broad hydrogen emission lines, similar to the features that have been seen in most TDEs discovered at optical wavelengths \citep[e.g.,][]{arcavi14}. PS18kh falls into the ``hydrogen-rich'' group of TDEs, with strong Balmer lines, particularly \halpha{} and \hbeta, visible in most epochs, but with weak or absent helium emission features. There is some suggestion of emission that is consistent with \ion{He}{1}~5875\AA~at the redshift of PS18kh, but the \ion{He}{2}~4686\AA~line seen in many TDEs is notably absent.

Our earliest spectroscopic follow-up was obtained prior to or within a few days of the $g$-band peak, and some interesting trends can be seen in the spectra. In particular, the spectral slope becomes steeper near peak before beginning to slowly flatten again over the course of our observations, which is unsurprising given that the TDE was optically brightest at peak. The emission lines become stronger as time progresses, and only become clearly visible shortly after peak light. Unfortunately our first spectrum, the classification spectrum obtained on 2018 March 7, was taken through clouds, making it difficult to determine whether there were emission lines prior to peak. As was seen with the optical photometry, there is little evidence of the UV re-brightening in the optical spectra---the continuum level remains relatively flat, and the lines show no significant evolution.

The spectra of PS18kh differ from the majority of other TDEs in one respect: the \halpha, and in some cases \hbeta, lines show evidence of an evolving, boxy shape that becomes more prominent over time, and in some later epochs there is a suggestion of double peaks in the \halpha{} profile. A similar double-peaked \halpha{} profile was seen in the TDE PTF09djl, though in that case, the peaks showed a much larger separation \citep{arcavi14,liu17b}.


\begin{figure*}
\begin{minipage}{\textwidth}
\centering
\subfloat{{\includegraphics[width=0.31\textwidth]{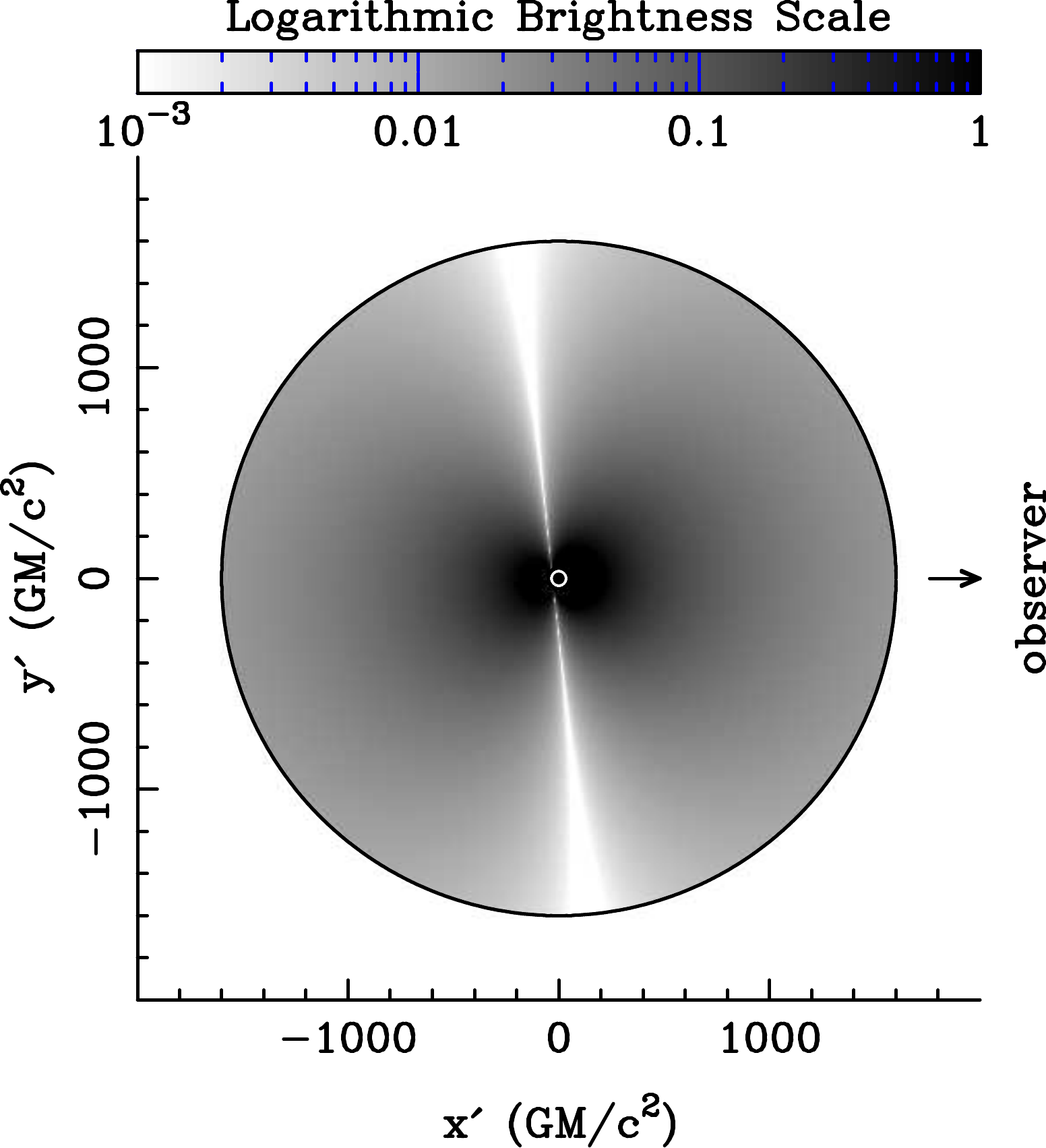}}}
\hfill
\subfloat{{\includegraphics[width=0.31\textwidth]{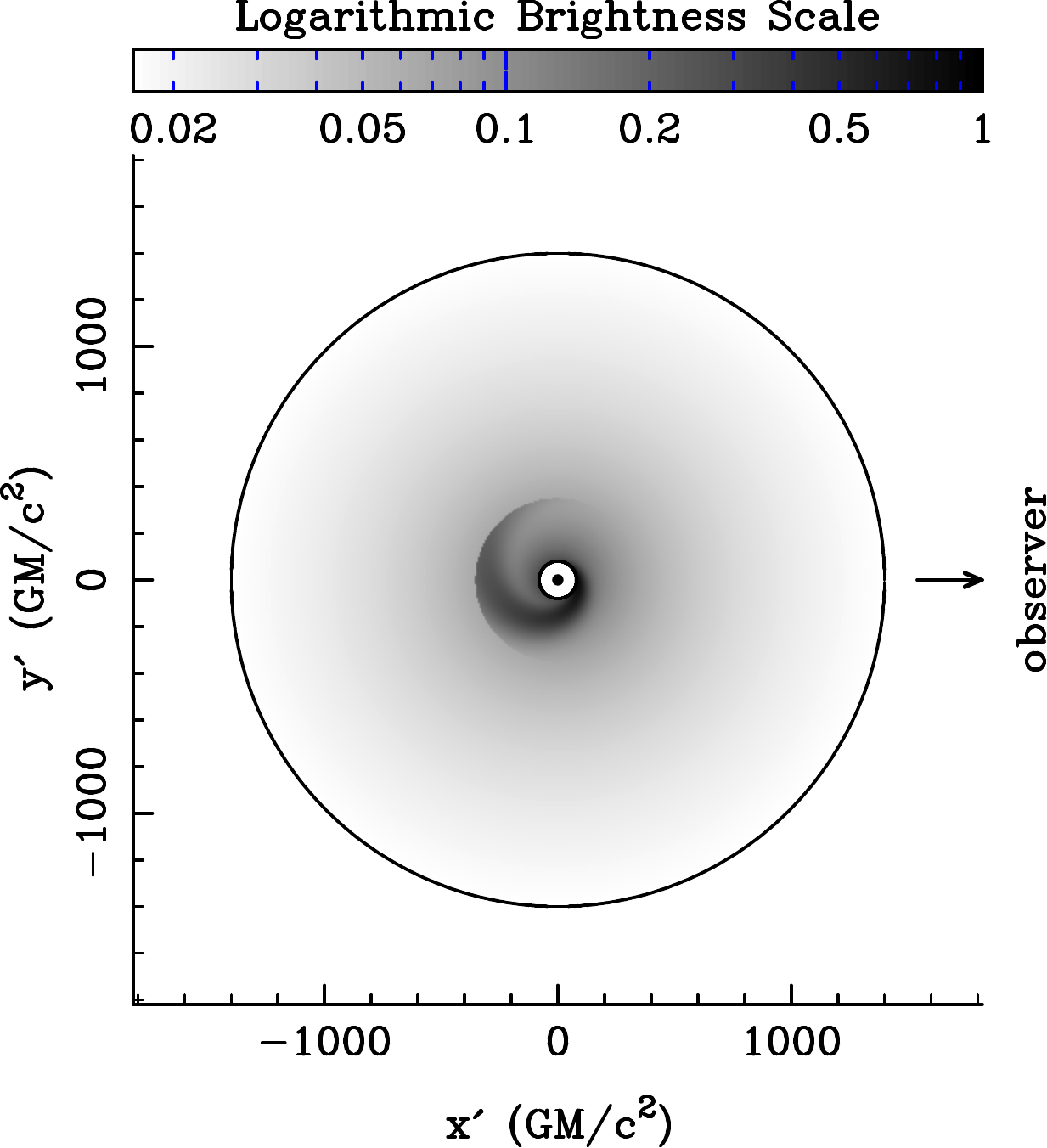}}}
\hfill
\subfloat{{\includegraphics[width=0.31\textwidth]{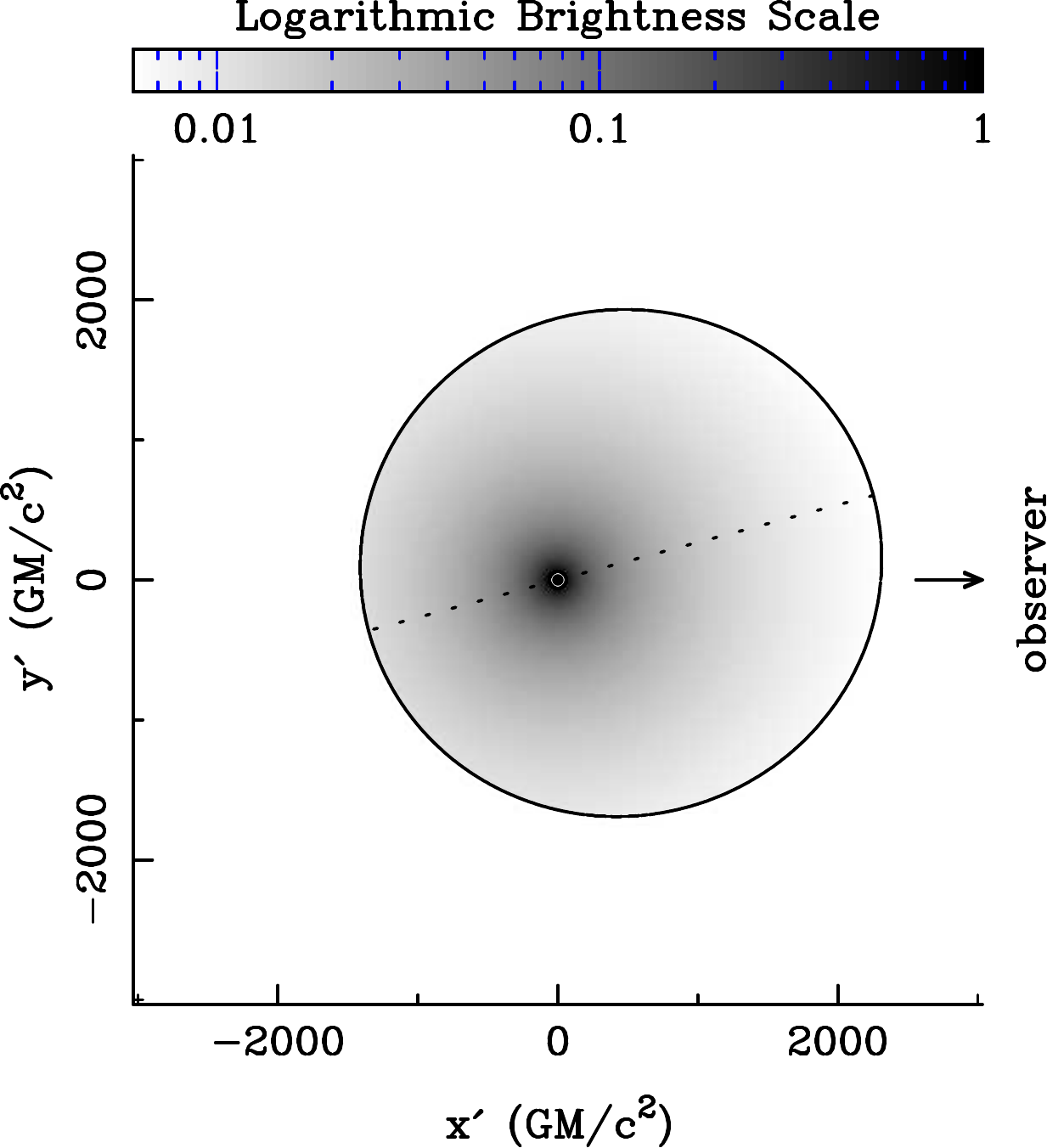}}}
\caption{Illustrations of the {\it relative} surface brightness distributions of some of the disk models used in this work. The shading is logarithmic with darker shades indicating higher intensities. See \S\ref{sec:spec_anal} for details of the models, their adjustable parameters, and other conventions. \emph{Left Panel}: Disk+wind model for the 2018 March 25 profile. The non-axisymmetric pattern is a result of the anisotropic escape probability of line photons caused by the non-negligible optical depth at the base of the wind. The lowest emissivity values are of order $10^{-6}$ of the maximum but they are not plotted here so that the overal pattern can be displayed more clearly. \emph{Middle Panel}: A disk spanning radii 80--1500$\;r_{\rm g}$ with a spiral arm, used to model the profiles after 2018 April 11. The arm extends up to 350$\;r_{\rm g}$, it has a pitch angle of $p=10^\circ$, and its azimuth at the inner disk is $\varphi_{in}=10^\circ$. The brightness of the arm is five times that of the underlying disk at all radii. The arm is superposed on an axisymmetric emissivity of the form $r^{-1}$. \emph{Right Panel}: An elliptical disk model for the 2018 April 1 profile. The disk spans a range of pericenter distances of 60--1400$\;r_{\rm g}$ and has an emissivity of the form $\epsilon\propto r^{-1.4}$. The eccentricity increases linearly with pericienter distance from 0 to 0.25. The dotted line marks the semi-major axis, which makes an angle of $\varphi_0=15^\circ$ with the line of sight.}
\label{fig:disk_models}
\end{minipage}
\end{figure*}

The possibility that TDEs could lead to the formation of line-emitting (elliptical) disks was discussed by \citet{eracleous95} and \citet{guillochon14}. In the cases of two recent TDEs, PTF09djl and ASASSN-14li, an elliptical disk model has been used to fit the emission line profiles and model the properties of the accretion disk \citep{liu17b,cao18}. Here we use similar models to infer the properties of the accretion disk, and potentially the stellar debris, of PS18kh.

We consider models for the profiles of the broad H$\alpha$ emission lines that attribute the emission to gas in a relativistic keplerian disk. We were motivated by the success of such models in describing the Balmer line profiles of active galaxies and quasars in general \citep[e.g.,][]{popovic04,bon09,lamura09,storchi17} and recent theoretical scenarios that associate the broad-line region with the accretion disk in quasars and active galaxies \citep[e.g.,][]{elitzur14}, as well as the studies of PTF09djl and ASASSN-14li mentioned above. An alternative family of models, which we do not consider here, attribute the emission lines to spherically expanding outflow \citep[see][]{roth16,roth18}. Those models employ more rigorous radiative transfer calculations than ours and incorporate electron scattering (our models adopt the Sobolev apprximation for radiative transfer in an accelerating medium). They can also produce asymmetric line profiles in the early stages of the evolution of the event with asymmetries resulting from radiative transfer effects. In contrast, in our models the asymmetries result from relativistic effects. In fact, an interpretation of the Balmer line profiles of PS18kh in terms of an outflow model is discussed in a recent paper by \citet{hung19}. The blueshifted broad absorption lines (BALs) found by \citet{hung19} in the UV spectra of PS18kh can be explained by both their model and ours, since the models are qualitatively similar: they invoke accretion-powered outflows \citep[our models are based on the accretion-disk wind calculations of][]{murray95a}. The models do differ, however, in the exact geometry and velocity field of the outflow, the layers taken to emit the Balmer lines (we attribute the Balmer lines to the base of the outflow, i.e. the accretion disk atmosphere), and methods they used to treat radiative transfer.


The model line profiles are obtained in the observer's frame by adopting the formalism detailed in \citet{chen89a}, \citet{chen89b}, \citet{eracleous95}, and \citet{flohic12} by computing the integral 
\begin{equation}
  f_\nu \propto \int d\varphi \int \xi\,d\xi\; 
  \,I_\nu(\xi,\varphi,\nu_e)\,D^3(\xi,\varphi)\, \Psi(\xi,\varphi)\; 
  \label{eqn:surfint}
\end{equation}
over the surface of the disk. The functions in the integrand are expressed in polar coordinates in the frame of the disk where $\varphi$ is the azimuthal angle in the plane of the disk, $\xi\equiv r/r_{\rm g}$ is the dimensionless radial coordinate, $r_{\rm g}\equiv GM_\bullet/c^2$ is the gravitational radius, and $M_\bullet$ is the mass of the black hole. The axis of the disk makes an angle $i$ with the line of sight to the observer (the ``inclination'' angle) and the line-emitting portion of the disk is enclosed between radii $\xi_{disk}^{in}$ and $\xi_{disk}^{out}$ \citep[see Fig.~1 of][]{chen89a}. 

The functions $D$ and $\Psi$ describe the gravitational and transverse redshifts and light bending, respectively, in the weak-field approximation. The function $I_\nu$ represents the {\it apparent} emissivity of the disk and includes terms that account for the intrinsic brightness distribution of the disk, the (potentially anisotropic) escape probability of line photons in the direction of the observer, and local line broadening \citep[see equation~2 of][and the associated discussion]{flohic12}. The local profile of the line is assumed to be a Gaussian of standard deviation $\sigma$ that includes contributions from local turbulence, electron scattering, and blurring resulting from the finite cells used in the numerical integration. The intrinsic brightness profile of the disk is parameterized by a power-law of the form $\xi^{-q}$ where $q$ takes values between 1 and 3, inspired by the results of photoionization  calculations by \citet{dumont90iv,dumont90v}. This axisymmetric emissivity pattern can be perturbed either by making the disk elliptical or by superposing a logarithmic spiral, as we explain below. 

At early times, the observed profile of the H$\alpha$ line in PS18kh appears bell-shaped and somewhat asymmetric with an extended red wing. At late times, the profile evolves to a flat-topped or, sometimes, double-peaked shape. It maintains its red wing and it sometimes shows a blue shoulder. We interpret this sequence of line profiles as indicating a progressive decline in the optical depth of the line emitting region of the disk. This interpretation is based on the behavior of the theoretical line profiles with optical depth and on the expected evolution of the accretion rate through the disk and onto the black hole. At early times the high accretion rate is likely to lead to the emission of a wind from the surface of the accretion disk, consisting of stellar debris from the disruption, whose dense base layers will provide a substantial optical depth to the line photons. As the accretion rate drops and the debris moves outward from the black hole, the density of the wind and the optical depth of the surface layers of the disk decline accordingly. We also note that the blue shoulder in the observed late-time H$\alpha$ profiles cannot be reproduced by a model of an axisymmetric disk. Therefore, we postulate that a non-axisymmetric perturbation is present and we explore whether an elliptical disk or a disk with a spiral arm can describe this perturbation successfully. 

The spectra obtained prior to 2018 March 25 show a weak \halpha{} emission line, suggesting that the optical depth of the material surrounding the accretion disk is too large to obtain a model fit to the data. To represent the observed early-time H$\alpha$ profiles (those between 2018 March 25 and 2018 April 1) we adopt the wind model discussed by \citet[][see also \citealt{flohic12}, \citealt{chiang96}, \citealt{murray97}, and \citealt{chajet13}]{murray95a}. In these models, the {\it apparent} brightness profile of the disk is non-axisymmetric, as shown, for example, in Figure~4 of \citet{flohic12}, because of the large optical depth and anisotropic escape probability of photons through the emission layer. The resulting line profiles have round or somewhat flat tops and an extended red wing because of relativistic effects \citep[see examples in Fig.~5 of][]{flohic12}. The free parameters of the model are the inner and outer radii of the line-emitting portion of the disk, $\xi_{disk}^{in}$ and $\xi_{disk}^{out}$, the local line width, $\sigma$ (in km~s$^{-1}$), the emissivity power-law index, $q$, the disk inclination angle, $i$, the angle of the wind streamlines relative to the plane of the disk, $\lambda$ \citep[see Fig.~1 of][]{murray97}, and the normalization of the position-dependent optical depth pattern, given in terms of $\tau$, the optical depth in the direction of the observer at a fiducial position in the disk of $(\xi,\varphi)=(\xi_{disk}^{in},0)$.\footnote{In the current implementation of this model we do not allow the optical depth normalization to vary with radius; in the notation of \citet{flohic12} we set $\eta=0$.}
\footnote{The speed of the wind is not a parameter of the model since the optical depth depends on the velocity {\it gradient} rather than the velocity itself.}

For a circular disk with an axisymmetric emissivity pattern, the line profile has a net redshift and a red wing that is more pronounced than the blue wing because of a combination of gravitational and transverse redshifts. If $\xi_{disk}^{out}/\xi_{disk}^{in}\lsim 10$, there are two well-separated peaks and the blue peak is stronger than the red peak because of Doppler boosting. As this ratio increases, the two peaks get closer together, eventually blending to form a profile with a flat or round top. If the emitting layer of the disk is accelerating to form a wind, radiative transfer effects make the apparent emissivity  non-axisymmetric, as illustrated in the left panel of Figure~\ref{fig:disk_models}. The emissivity is enhanced at low projected velocities and depressed at high projected velocities, which enhances the core and depresses the two peaks of the profile, making it flat- or round-topped.


\begin{figure*}
\begin{minipage}{\textwidth}
\centering
{\includegraphics[width=0.95\textwidth]{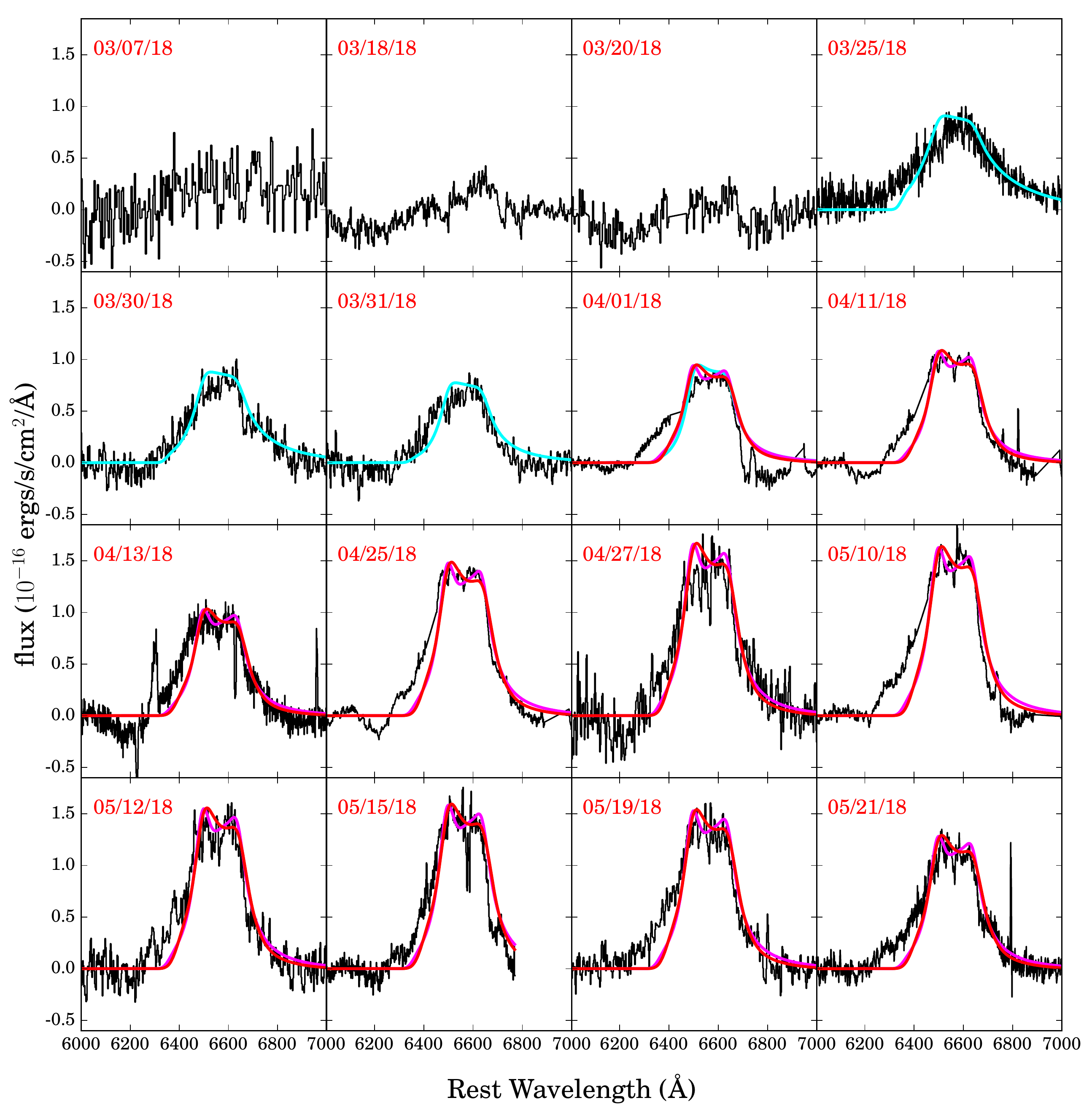}}
\caption{Evolution of the \halpha{} profile of PS18kh. A linear estimate of the continuum emission was subtracted from each epoch and the date of each spectrum is shown in the upper-left corner of each panel. The cyan lines show disk+wind model fits to the spectra taken between 2018 March 25 and 2018 April 1, the magenta lines show elliptical disk model fits to the 2018 April 1 and later spectra, and the red lines show disk+spiral arm model fits to the 2018 April 1 and later spectra. The models shown in all epochs after 2018 April 1 are the same models, which have been scaled by a factor of $1.15-1.8$ to fit the line profiles. All models shown are described in Section~\ref{sec:spec_anal}. The spectra from 2018 April 1 and 2018 May 10 have prominent telluric water vapor absorption bands in the red wing of the line (6700--6800~\AA) that have not been corrected.}
\label{fig:halpha}
\end{minipage}
\end{figure*}

To fit the profiles at late times, we tried two different models, an elliptical disk \citep[see][]{eracleous95}, and a circular disk with a single spiral arm \citep[see][]{gilbert99,storchi03}. Figure~\ref{fig:disk_models} shows illustrations of these models, which are described in more detail below.


\begin{deluxetable}{cccccccc}
\tabletypesize{\footnotesize}
\tablecaption{Fixed Model Parameters}
\tablehead{
\colhead{}&
\colhead{Parameter} &
\colhead{}&
\colhead{}&
\colhead{}&
\colhead{}&
\colhead{Value}&
\colhead{}}
\startdata
\multicolumn{8}{c}{Underlying Disk} \\
& $i$ & & & & & $26^\circ$ & \\
& $\sigma$ & & & & & 800 km~s$^{-1}$ & \\
\hline
\multicolumn{8}{c}{Wind} \\
& $\lambda$ & & & & & $15^\circ$ & \\
& $\eta$ & & & & & 0 & \\
\hline
\multicolumn{8}{c}{Spiral Arm} \\
& $A$ & & & & & 5 & \\
& $\varphi_{in}$ & & & & & $10^\circ$ & \\
& $p$ & & & & & $10^\circ$ & \\
& $w$ & & & & & $80^\circ$ & \\
\enddata 
\tablecomments{Fixed parameters for the disk+wind+spiral~arm model, as defined in \S\ref{sec:spec_anal} of the text. The values listed in this table to not change with time. The parameters that do change with time are given in Table~\ref{table:var_params}.} 
\label{table:fixed_params} 
\end{deluxetable}

\begin{description}

\item[\it Disk With Spiral Arm] The axisymmetric emissivity of a circular disk is perturbed by a logarithmic spiral, as described in equation (2) of \citet{storchi03}. In addition to the five free parameters that describe a circular disk, there are five free parameters that describe the spiral pattern: the  pitch angle, width, and azimuth of the spiral arm at the inner disk, $p$, $w$, and $\varphi_{in}$, respectively, its brightness contrast relative to the underlying axisymmetric disk, $A$, and its outer radius, $\xi_{spiral}^{out}$. We take its inner radius to be the same as the inner radius of the line emitting portion of the disk, i.e., $\xi_{spiral}^{in}=\xi_{disk}^{in}$. The disk may also emit a wind of modest optical depth that modifies the line profiles because of radiative transfer effects, as discussed earlier in this section. 
  
\item[\it Elliptical Disk] The disk streamlines are nested ellipses with aligned semi-major axes whose eccentricity increases linearly with distance from the center (from 0 to a maximum value of $e$). The emitting gas is optically thin to the line photons. There are two more free parameters in addition to those of an axxisymmetric circular disk (without a wind), the outer eccentricity and orientation of the semi-major axis relative to the observer, $e$ and $\varphi_0$. A model of this type was considered by \citet{guillochon14} in their discussion of the evolution of the tidal disruption event PS1-10jh and applied to PTF09djl and ASASSN-14li by \citet{liu17b} and \citet{cao18}. Moreover, a structure resembling an elliptical disk is discernible in the simulations of \citet{shiokawa15} that follow the evolution of the post-disruption debris to late times.
  
\end{description}

Introducing non-axisymmetric perturbations, such as a spiral arm or eccentric orbits, enhances the disk emissivity at specific projected velocities. Thus the asymmetries present in axisymmetric disk models can be changed (e.g., reversed or eliminated) or the two peaks can become less pronounced because the valley between them is filled in. In the case of a spiral arm, the profile modifications are determined largely by the shape, orientation, and contrast of the spiral arm. In the case of an eccentric disk of the type employed in this work, the modifications of the line profile are controlled largely by the combination of eccentricity and orientation of the major axis.

While a complete exploration of the model parameter space is beyond the scope of this work, we carried out a limited, qualitative exploration where the goodness of all fits was assessed by eye. We focused our attention on disks with low inclination (i.e., closer to face-on) so as to obtain fits with models that have small disk radii. We took this approach in order to reduce the angular momentum of the debris in the disk so that it does not exceed the initial angular momentum of the approaching star; we discuss this issue in detail in Section~\ref{sec:angmom}. We note  that in models of line profiles from a non-relativistic disk, the line profiles are symmetric and their widths depend on the combination $(M_\bullet/R)^{1/2}\sin i \propto \xi^{-1/2}\sin i$,
making the inclination angle and disk radius degenerate. Once special and general relativistic effects are included, the line profiles become progressively more asymmetric and redshifted as $\xi$ decrases, and the degeneracy between $i$ and $\xi$ is broken. In the models that we explore here, we look for the minimum inclination angle that can reproduce the shape/asymmetry of the line profiles.

In practice, we first fitted the 2018 March 25 spectrum with a wind model (including a spiral arm) and then adjusted the optical depth and disk radii to reproduce the March 30, March 31, and April 1 spectra. We found that the minimum inclination angle for which the model can match the red wing of the H$\alpha$ profile well and the blue wing approximately is $26^\circ$. At smaller inclination angles the model profiles are too asymmetric to fit the data well. We then fitted the 2018 April 25 spectrum with a disk and spiral model and an elliptical disk model of the same inclinaiton and compared that model with the other observed spectra obtained after 2018 April 1 to check whether they could adequately describe those spectra as well. We estimated the uncertainties in the model parameters by perturbing them about their best-fit values and adjusting the other parameters to get a good fit until no good fit was possible. Thus, we found that the inner disk radius can be determined to $\pm 20\; r_{\rm g}$, the outer radius to $\pm 200\; r_{\rm g}$, the emissivity power law index to $\pm 0.2$ and the broadening parameter to $\pm 300$~km~s$^{-1}$. The wind optical depth could be determined to a factor of 3 while the wind opening angle was held fixed at $15^\circ$ based on the physical considerations discussed in \cite{murray97}. The orientation of the spiral pattern could be determined to $\pm 5^\circ$, its pitch angle to $\pm 10^\circ$, its angular width to $\pm20^\circ$, its outer radius to $\pm 200\; r_{\rm g}$, and its contrast to $\pm 2$. The eccentricity of the elliptical disk could be determined to $\pm 0.2$ and the orientation of its major axis to $\pm 10^\circ$. The best-fit parameters for the disk$+$wind$+$spiral arm models are given in Tables~\ref{table:fixed_params} and~\ref{table:var_params}. The former table gives the values of the model parameters that wer held fixed while the latter gives tha values of the parameters that were allowed vary with time in order to reproduce the evolution of the line profiles. 


\begin{deluxetable}{ccccc}
\tabletypesize{\footnotesize}
\tablecaption{Variable Parameters for Circular Disk Models}
\tablehead{
\colhead{Date} &
\colhead{$\tau(\xi_{disk}^{in},0)$}  &
\colhead{$\xi_{disk}$} &
\colhead{$\xi_{spiral}$} &
\colhead{$q$}}
\startdata
3/25            & $>$3.0 & 50--1600 & \dots & 1.8 \\
3/30            & $>$3.0 & 50--1600 & \dots & 1.5 \\
3/31            &    0.9 & 50--1600 & \dots & 1.3 \\
4/01            &    0.3 & 50--1600 & \dots & 1.2 \\
4/11 and later  &    0.1 & 80--1500 & 80--350 & 1.0 \\
\enddata 
\tablecomments{Parameters for the disk+wind+spiral model used to model the \halpha{} emission line profile that change with time. All optical depths are in the inner disk the direction of the observer, i.e., at $\xi=\xi_{disk}^{in}$ and $\varphi=0$. Radii are the outer radius of the disk and the outer radius of the spiral arm in the disk (the inner radius of the disk and the spiral arm are fixed). The parameter $q$ is the index of the power law that describes the axisymmetric portion of the disk emissivity.} 
\label{table:var_params} 
\end{deluxetable} 

Figure~\ref{fig:halpha} shows the evolution of the \halpha{} emission line and the model fits for each epoch after the line emerges. As described above, the profiles between 2018 March 25 and 2018 April 1 are well-fit by a disk$+$wind model, shown with a cyan line in the Figure. As the optical depth of the wind drops, the emission from the underlying disk becomes apparent. We show three fits to the 2018 April 1 spectrum, a disk$+$wind model, an elliptical disk model (shown in magenta), and a disk with spiral arm model (shown in red). The parameters of the elliptical disk model are  $i=26^\circ$, $\xi_{disk}=60-1400$ (pericenter distances), $e=0.25$, $\varphi_0=15^\circ$, $q=1.4$, and $\sigma=800\;{\rm km\; s^{-1}}$ (see illustration in Fig.~\ref{fig:disk_models}).

All models can provide a reasonable fit to the top of the 2018 April 1 line profile and its red wing but none can reproduce the blue wing very well. The same is true for the models in most of the later epochs, although the blue wing of the line becomes weaker in some epochs in late April and early May and the models can approximate it better. The model parameters that do not change with time are summarized in Table~\ref{table:fixed_params} while the parameters that vary with time are in Table~\ref{table:var_params}. Each of the later epochs has been fit with the same models, with the only difference between epochs being a flux scaling factor ranging from $1.15-1.8$. The flux scaling factor increases with time until 15 May 2018 after which it begins to fall again. The changing scaling factor and and varying difference between the models and the blue wing of the line profile, as well as small changes in the peak and wings of the emission line from epoch-to-epoch, are likely a result of short-term variability in the disk structure on spatial scales too small to properly capture with the models used here.

Finally, we note that we also experimented with disk models with higher inclination angles, up to $i=60^\circ$. These models have correspondingly larger radii and the other parameters have to be adjusted somewhat to obtain a good fit. For example, the models with $i=60^\circ$ have $\xi_{disk}\sim 500-15,000$ and $\tau(\xi_{disk}^{in},0)$ of order a few. We do not favor such models because they imply that the debris carries too much angular momentum, as we discuss in detail in Section~\ref{sec:angmom}.


\begin{figure}
\centering
\subfloat{{\includegraphics[width=0.47\textwidth]{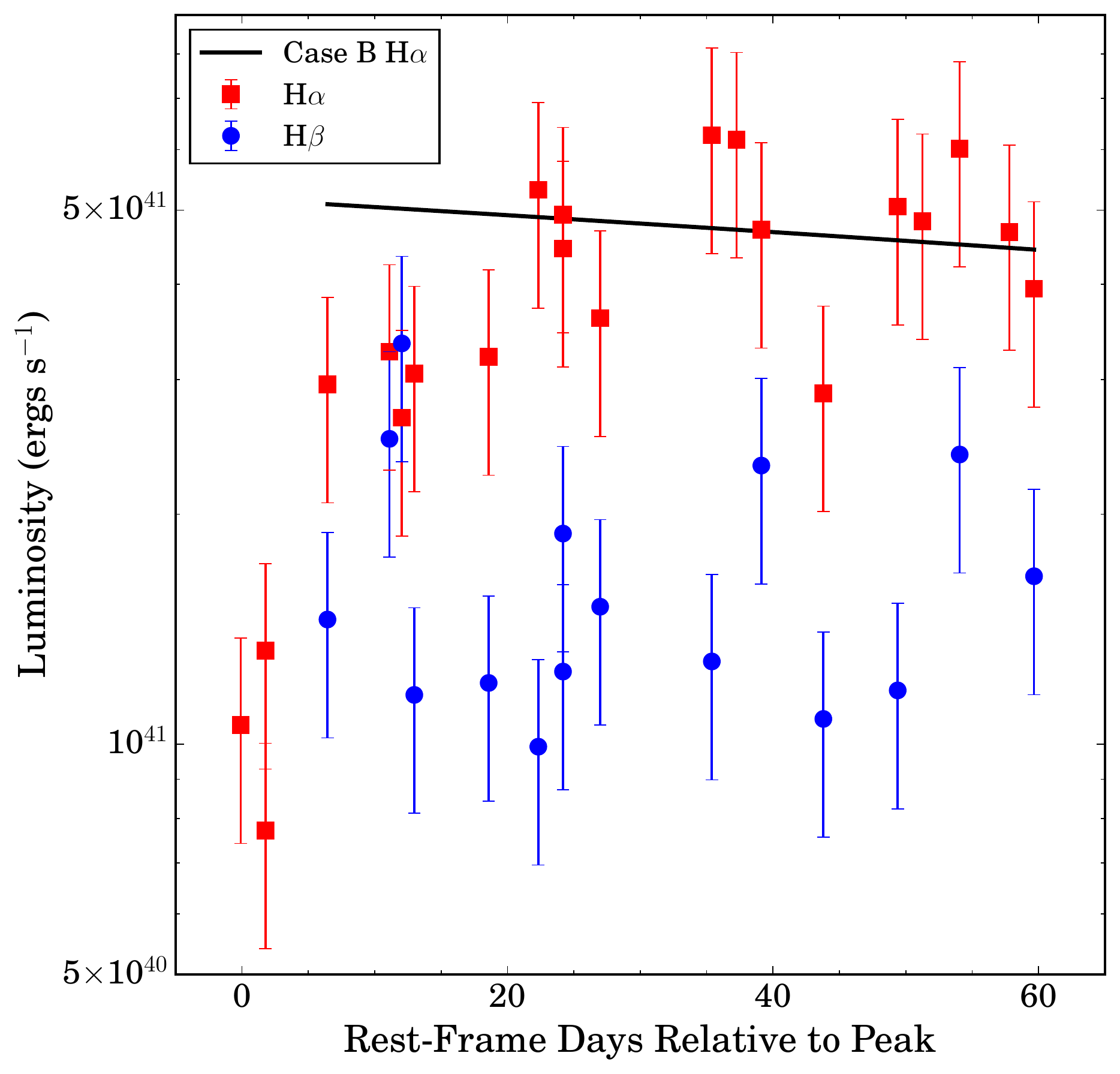}}}
\caption{Evolution of the \halpha{} (red squares) and \hbeta{} (blue circles) luminosities of PS18kh. Errorbars show 30\% errors on the line fluxes. The black line shows the \halpha{} emission that would be expected from case B recombination, given the \hbeta{} emission.}
\label{fig:line_lum}
\end{figure}

Taken as a whole, the evolution of the \halpha{} profile in PS18kh shows that as the TDE is brightening towards its peak, the disk is obscured by optically thick material, likely debris from the disruption. Between 2018 March 25 and 2018 April 1, this material becomes progressively more optically thin while the line-emitting portion of the disk grows in size, and the emission lines are well-fit by a disk$+$wind model. After 2018 April 1, the wind become feeble, the emission from the disk is clearly seen, and the double-peaked/boxy profile is well-fit by a disk with a spiral arm model. The scale of the disk is similar to that seen in PTF09ge and ASASSN-14li \citep{liu17b,cao18}, indicating that this is likely a common feature of TDEs. The disk has non-axisymmetric perturbations that are approximated by the spiral arm. Slight variation in the scaling of the models to the line profiles after 2018 April 1 suggests that the perturbations are changing with time.

We also measured the luminosities of the \halpha{} and \hbeta{} emission lines from our follow-up spectra for all epochs where the lines were pronounced enough to measure the flux. In Figure~\ref{fig:line_lum} we show the luminosity evolution of these two emission features from the spectra of PS18kh. As estimating the true error on the line fluxes is difficult given their complex shape, we assume 30\% errors on the emission fluxes calculated in each epoch. From the spectra taken at and shortly after peak, we can see the \halpha{} line becoming more luminous and more pronounced, peaking at a luminosity of $L_{\rm H\alpha}\sim6\times10^{41}$~ergs~s$^{-1}$ roughly 20 rest-frame days after the continuum peaks. After peaking, the \halpha{} luminosity remains relatively constant for the rest of the period of observations. Similarly, though it is not measurable prior to peak, the \hbeta{} luminosity remains at roughly $L_{\rm H\beta}\sim1-2\times10^{41}$~ergs~s$^{-1}$ in all epochs where it is measurable. This roughly constant evolution of the line luminosities differs from that of ASASSN-14li, which showed declining line luminosities following discovery \citep{holoien16a,brown17a}.

Figure~\ref{fig:line_lum} also shows the \halpha{} emission that would be expected given the measured \hbeta{} emission, assuming the emission is driven by case B recombination. The \halpha/\hbeta{} ratio is largely consistent with what would be expected from recombination, within noise, similar to what was seen in ASASSN-14li \citep{holoien16a}. This also indicates there is little additional extinction from the host galaxy. The measured luminosities for both lines are given in Table~\ref{tab:line_lum}.

The excellent spectroscopic coverage of PS18kh before, during, and after peak light may also allow us to trace the evolution of the accretion state in the TDE. Few TDEs prior to PS18kh have exhibited such a strong evolution in emission line profiles as we see here with H$_\alpha$, and this evolution may be due to changes in the accretion state. For example, some theories predict that shortly after disruption, the accretion in a TDE is expected to be super-Eddington, launching a wind where strong optical reprocessing can occur \citep[e.g.,][]{roth16,dai18}. The accretion rate is expected to drop to roughly Eddington shortly after peak for a 10$^{6.9}$~{\msun} black hole, when the H$_\alpha$ line began to emerge, and later as the accretion becomes sub-Eddington the wind is expected to become optically thin, allowing the disk to be observed. While there doesn't seem to be significant evolution in the H$_\alpha$ profile of PS18kh after it emerges, indicating this theoretical picture may not be exactly correct, there have been few TDEs observed at such early times and even fewer that are observed before the lines appear, indicating more objects with similarly early spectra are needed to refine these models.

\subsection{Angular Momentum of the Line-Emitting Gas}
\label{sec:angmom}

To further evaluate the plausibility of the models for the H$\alpha$ line profiles, we compare the angular momentum of the line-emitting gas to the angular momentum of the star prior to disruption. We assume that the star is on a parabolic orbit around the black hole with a pericenter distance equal to the tidal disruption radius, $r_{\rm t}$, and a pericenter speed $\upsilon_{\rm p}(r_t)=(2GM_\bullet/r_{\rm t})^{1/2}$. The {\it specific} angular momentum of the star is, then, $j_\star=r_t\, \upsilon_{\rm p}(r_ t) = (2GM_\bullet r_{\rm t})^{1/2}$. If the post-disruption debris conserves specific angular momentum and settles down into a circular disk (or ring), the radius of that disk should be $r_{\rm d} = 2r_{\rm t}$, which is two orders of magnitude smaller than the disk radius inferred from fitting the line profiles (see illustration below). Thus, in order to reconcile the size of the line emitting disk with the dynamics of the debris we must adopt a picture in which the line-emitting gas represents a small fraction of the mass of the debris and as much larger specific angular momentum than the initial star. This implies that angular momentum is transported very quickly, perhaps with the help of shocks, \citep[e.g.,][]{shiokawa15,bonnerot16,hayasaki16}, and redistributed so that a small fraction of the mass of the debris ends up at a large distance from the center and emits lines. We assess this scenario below by comparing the {\it total} angular momentum of the star to that of the line-emitting disk.

The {\it total} angular momentum of the star at the tidal disruption radius, $r_t$, is $J_\star=\mstar\, r_t\, \upsilon_{\rm p}(r_ t)\;$, which we re-write as $J_\star=\mstar\, (GM_\bullet/c)\, (2\xi_t)^{1/2}$, where $\xi_t \equiv r_t/r_g$. We use the expression for $r_t$ from \citet{phinney89} and re-cast it in terms of the average density of the star. We then take advantage of the mass-radius relation for zero-age main-sequence stars ($\rstar\propto\mstar^{0.57}$ for $\mstar > 1.66\; {\rm M_\odot}$; e.g., \citealt{demircan91}) to relate the average density of the star to its mass and obtain $\xi_t = 7.1 (K/0.7)\; M_7^{-2/3}\; (\mstar/{\rm M}_\odot)^{0.237}$, where $M_7 = M_\bullet/10^7\;{\rm M_\odot}$. The parameter $K$ is equivalent to the combination $(k/f)^{1/6}$ used by \citet{phinney89} to parameterize the structure of the star and takes values between 0.52 and 0.82. Substituting this expression for $\xi_t$ into the equation for $J_\star$ and carrying our some algebra we obtain $J_\star = 3.3\times 10^{56}\, \left(K / 0.7\right)^{1/2} \left(\mstar / {\rm M}_\odot\right)^{1.12} M_7^{1/3} ~{\rm g\;cm^2\;s^{-1}}$.

To compute the angular momentum of the line-emitting debris, $J_d$, we assume that the debris is concentrated at the outer boundary of the disk adopted in the models for the line profiles. Using the same approach as for the angular momentum of the star, and assuming that the disk is generally elliptical, we write $J_d = m_d\, (GM_\bullet/c)\, \left[\xi_{disk}^{out} (1+e)\right]^{1/2}$, where $m_d$ is the mass of the debris and $\xi_{disk}^{out}$ and $e$ are, respectively, the outer pericenter distance and eccentricity of the disk (inferred from the fits to the line profiles). To get the mass of the debris, we use the H$\alpha$ luminosity, assuming that it is produced by case-B recombination in an ionized, thermal plasma of uniform density. The luminosity per unit volume is, then, $ L_{\rm H\alpha} / V = n_e\, n_p\, \epsilon^{\rm eff}_{\rm H\alpha}(T,n_e)$, where $n_e$ is the electron density, $n_{\rm H}$ and $n_p$ are the electron and proton densities, and $\epsilon^{\rm eff}_{\rm H\alpha}(T,n_e)$ is the case-B effective recombination coefficient (a function of the temperature, $T$, and $n_e$). Expressing $n_p$ in terms of $m_d$ and $V$ and re-arranging, we obtain the following expression for the mass of the debris:
$m_d = 0.084\; (L_{42} / n_{e,11}\,\epsilon_{-25})\; {\rm M}_\odot$, where $L_{42} = L_{\rm H\alpha}/10^{42}\;{\rm erg\;s^{-1}}$, $n_{e,11} = n_e/10^{11}\;{\rm cm}^{-3}$, and $\epsilon_{-25}=\epsilon^{\rm eff}_{\rm H\alpha}(T,n_e)/10^{-25}\;{\rm erg\;s^{-1}\;cm^{-3}}$. For the range of temperatures and densities of interest here, $\epsilon_{-25}$ has values of order a few \citep[see][]{storey95}. The electron density is unknown, but we can draw an analogy with the broad-line region of Seyfert galaxies and quasars where recent work modelling the \ion{Fe}{2} UV line complex \citep{bruhweiler08, hryniewicz14}, the optical and UV intermediate-width lines \citep{adhikari16}, and the far-UV resonance lines \citep{moloney14} suggests densities in excess of $10^{11}\;{\rm cm}^{-3}$. Thus, we scale $n_e$ by $10^{11}\;{\rm cm}^{-3}$ in the above expression. The angular momentum of the line emitting debris is, then, $J_d = 2.3\times 10^{56} \, L_{42} \, M_7\, n_{e,11}^{-1}\,\epsilon_{-25}^{-1}\,\left[\xi_{disk}^{out} (1+e)/1000\right]^{1/2}~{\rm g\;cm^2\;s^{-1}}$.

Combining the above estimates, the ratio of the angular momentum of the debris to that of the star is 
\begin{equation}
  {J_d\over J_\star} =
  0.7\; {L_{42}\, M_7^{2/3}\over n_{e,11}\, \epsilon_{-25}}
  \,\left[\xi_{disk}^{out}(1+e)\over 1000\right]^{1/2}
  \left(K\over 0.7\right)^{-1/2} \left(\mstar\over M_\odot\right)^{-1.12}
    \label{eq:Jratio}
\end{equation}
Inserting values for the quantities that we can constrain observationally, $\left[\xi_{disk}^{out} (1+e)/1000\right]^{1/2} \lsim 1.3$, $L_{42}\approx 0.5$, $M_7^{2/3}=0.86$, and $\epsilon_{-25}\approx 2$ (from \citealt{storey95} for $T=1-3\times 10^4\;$K and $n_e=10^9-10^{13}\;{\rm cm}^{-3}$), we obtain $J_d/J_\star\lsim 0.2$. The value of $K$, although unknown, does not change the result by more than 10\%. The distance of closest approach of the star to the black hole is unknown but it cannot be less than $0.14$ of $\xi_t$, otherwise the star would enter the event horizon even for a maximally spinning black hole; since $J_\star \propto \xi_t^{1/2}$ a closer encounter could increase $J_d/J_\star$ to $\lsim 0.5$, at most. The mass of the star is also unknown but it is clear from the above analysis that a more massive star could be disrupted more easily because of its lower average density and that would also lead to a smaller value of the angular momentum ratio. For example, assuming $\mstar=4\;{\rm M}_\odot$ would lead to $J_d/J_\star\sim0.06$. Moreover, if the mass of the black hole is lower than the value we have estimated, the fraction of the angular momentum carried by the line-emitting gas is correspondingly lower. Taken at face value, these estimates suggest that the angular momentum of the line-emitting debris can be considerably smaller than the angular momentum of the star, reinforcing the plausibility of the model fits to the H$\alpha$ line profiles. However, two significant uncertainties must be borne in mind: ($i$) the density of the debris is unknown, and ($ii$) there may be a substantial amount of neutral gas associated with the line-emitting gas that would contribute to the mass, hence the angular momentum, of the debris.


\begin{figure*}
\begin{minipage}{\textwidth}
\centering
\subfloat{{\includegraphics[width=0.48\textwidth]{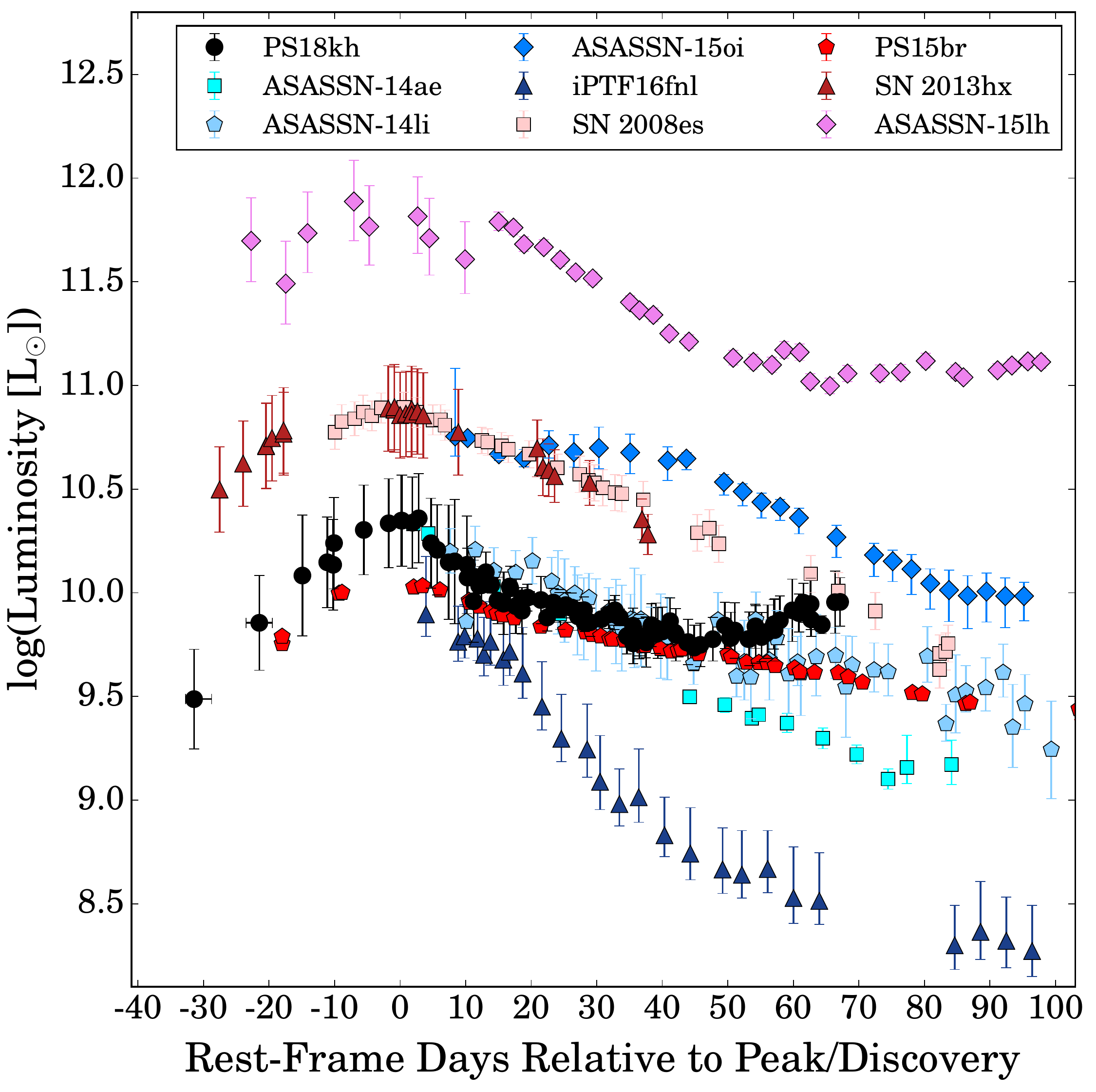}}}
\subfloat{{\includegraphics[width=0.48\textwidth]{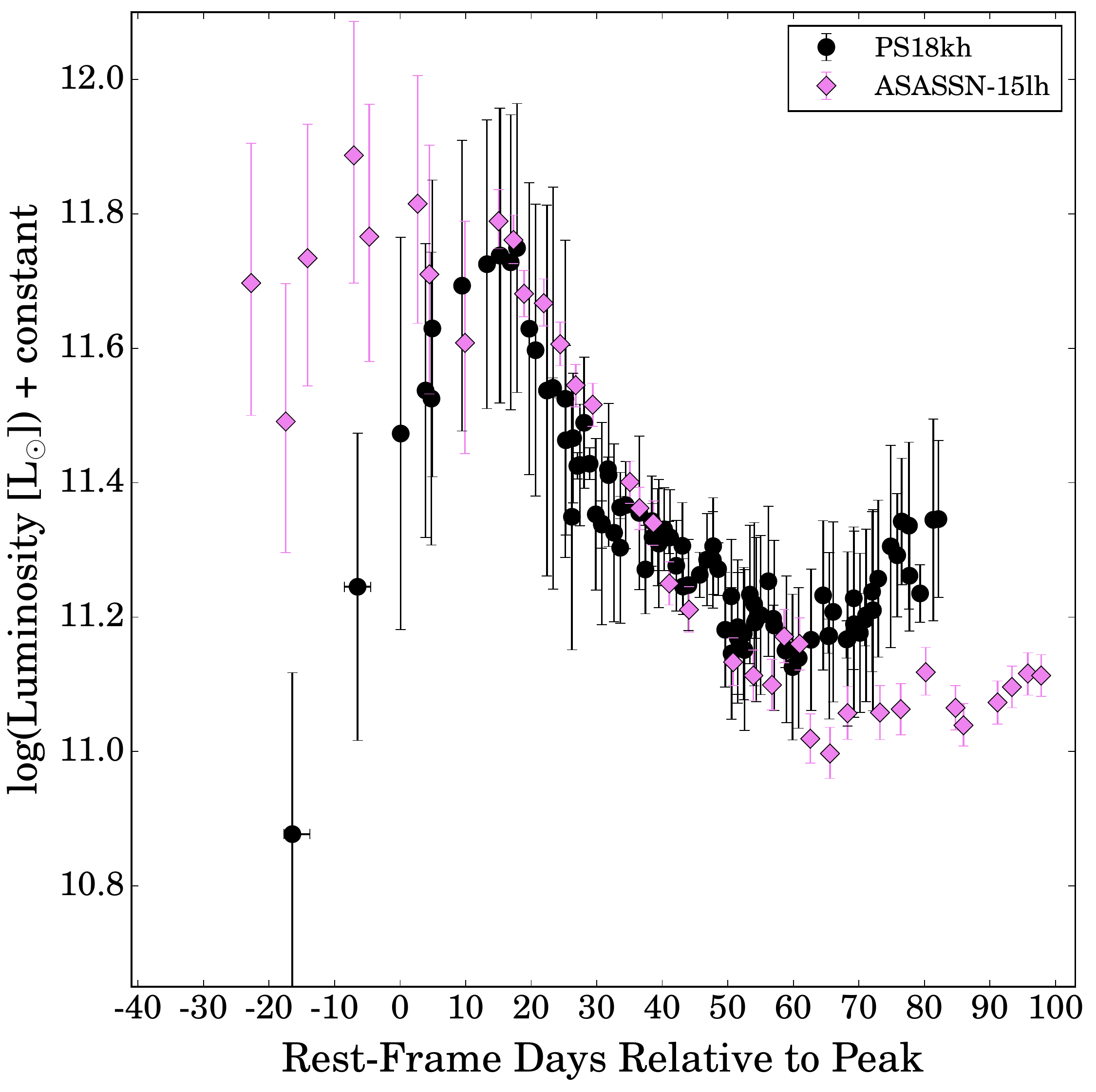}}}
\caption{\emph{Left Panel}: Luminosity evolution of PS18kh compared to that of the TDEs ASASSN-14ae \citep[cyan squares;][]{holoien14b}, ASASSN-14li \citep[light blue penatgons;][]{holoien16a}, ASASSN-15oi \citep[blue diamonds;][]{holoien16b}, and iPTF16fnl \citep[navy triangles;][]{brown17b}, the hydrogen-rich superluminous supernovae SN 2008es \citep[light red squares;][]{miller09,gezari09b}, SN 2013hx \citep[dark red triangles;][]{inserra18}, and PS15br \citep[red pentagons;][]{inserra18}, and the extremely luminous transient ASASSN-15lh \citep[magenta diamonds;][]{dong16,godoy-rivera17}. The full luminosity curve, including both the luminosities calculated from blackbody fits to the {\swift} data and the luminosities estimated from the $g$-band light curve, is shown for PS18kh. Time is shown in rest-frame days relative to peak for those objects which have observations spanning the peak of the light curve (PS18kh, SN 2008es, SN 2013hx, PS15br, and ASASSN-15lh) and in days relative to discovery for those objects which do not (ASASSN-14ae, ASASSN-14li, ASASSN-15oi, and iPTF16fnl). {Right Panel}: The luminosity evolution of PS18kh scaled by a factor of 24.5 and shifted by 15 days compared with that of ASASSN-15lh. These are the only two objects in the sample to exhibit a re-brightening in their UV light curves.}
\label{fig:lum_comp}
\end{minipage}
\end{figure*}

\section{Discussion}
\label{sec:disc}

The temperature, luminosity, radius and spectroscopic evolution of PS18kh are all consistent with other TDEs. However, many of these features are also common to type II superluminous supernovae (SLSNe II), and some of the observational characteristics of PS18kh (e.g., the UV re-brightening and the double-peaked line profiles) are not common to most (or any) other TDEs. In this Section we compare its luminosity, temperature, radius, and spectroscopic evolution to those of TDEs and SLSNe in literature to further investigate the nature of PS18kh.

Our sample of comparison objects includes the TDEs ASASSN-14ae \citep{holoien14b}, ASASSN-14li \citep{holoien16a}, ASASSN-15oi \citep{holoien16b}, and iPTF16fnl \citep{brown17b}, and the supernovae SN 2008es \citep{miller09,gezari09b}, SN 2013hx \citep{inserra18}, and PS15br \citep{inserra18}. The SN sample was chosen because these are the only three SLSNe that show both a broad \halpha{} feature and no signs of strong interaction between fast moving ejecta and circumstellar shells in their early spectra \citep{inserra18}, making them spectroscopically similar to PS18kh. Also included in our comparison sample is ASASSN-15lh, an extremely luminous transient whose nature has been debated, but which is likely either the most luminous SLSN ever discovered \citep{dong16,godoy-rivera17} or an extreme TDE around a maximally spinning black hole \citep{leloudas16}. ASASSN-15lh also exhibited a UV re-brightening, similar to PS18kh \citep{godoy-rivera17}, making it an interesting comparison object.

The left panel of Figure~\ref{fig:lum_comp} shows the rest-frame luminosity evolution of PS18kh and the transients in our comparison sample, with TDEs and SNe differentiated by color. The TDE sample has peak luminosities in the range $10^{9.8}$\lsun$\lesssim L \lesssim 10^{10.8}${\lsun} while the SN sample ranges from $10^{10}\lsun\lesssim L \lesssim 10^{10.8}\lsun$, meaning the luminosity of PS18kh is consistent with both types of object. ASASSN-15lh is clearly an outlier in peak luminosity from all the other objects in the sample, including PS18kh. While none of the TDEs in the sample were discovered prior to peak, preventing a comparison of the rising phase of the light curve, the rise time of PS18kh seems to be roughly consistent with that of the SNe in the sample.


\begin{figure}
\centering
\includegraphics[width=0.425\textwidth]{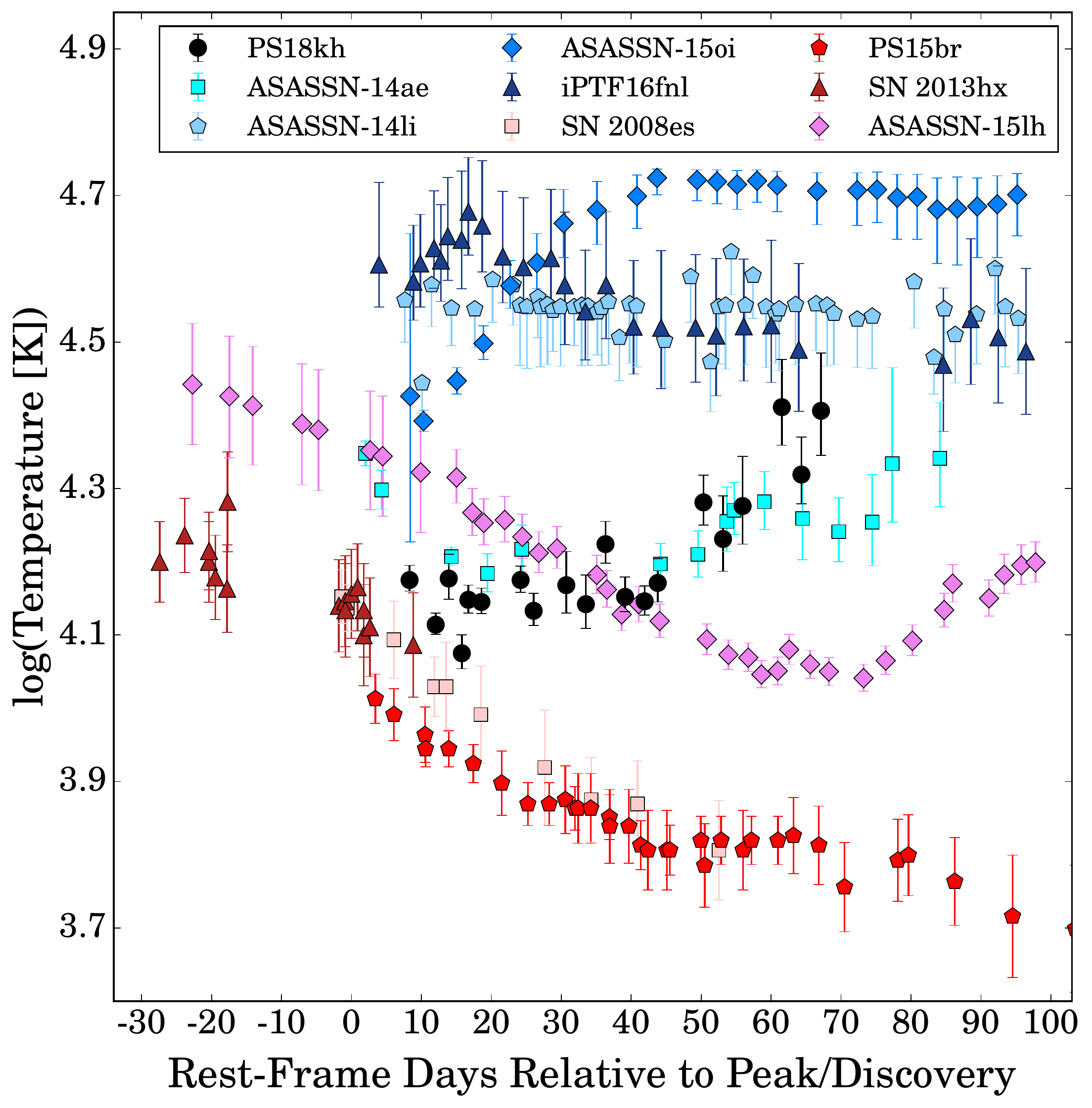}
\caption{Temperature evolution of PS18kh taken from blackbody fits to epochs with {\swift} observations compared with the temperature evolution of the objects in our comparison sample. Symbols and colors match those of Figure~\ref{fig:lum_comp} and all times are plotted in days relative to peak or discovery, as outlined in the caption of Figure~\ref{fig:lum_comp}.}
\label{fig:temp_comp}
\end{figure}

\begin{figure}
\centering
\subfloat{{\includegraphics[width=0.47\textwidth]{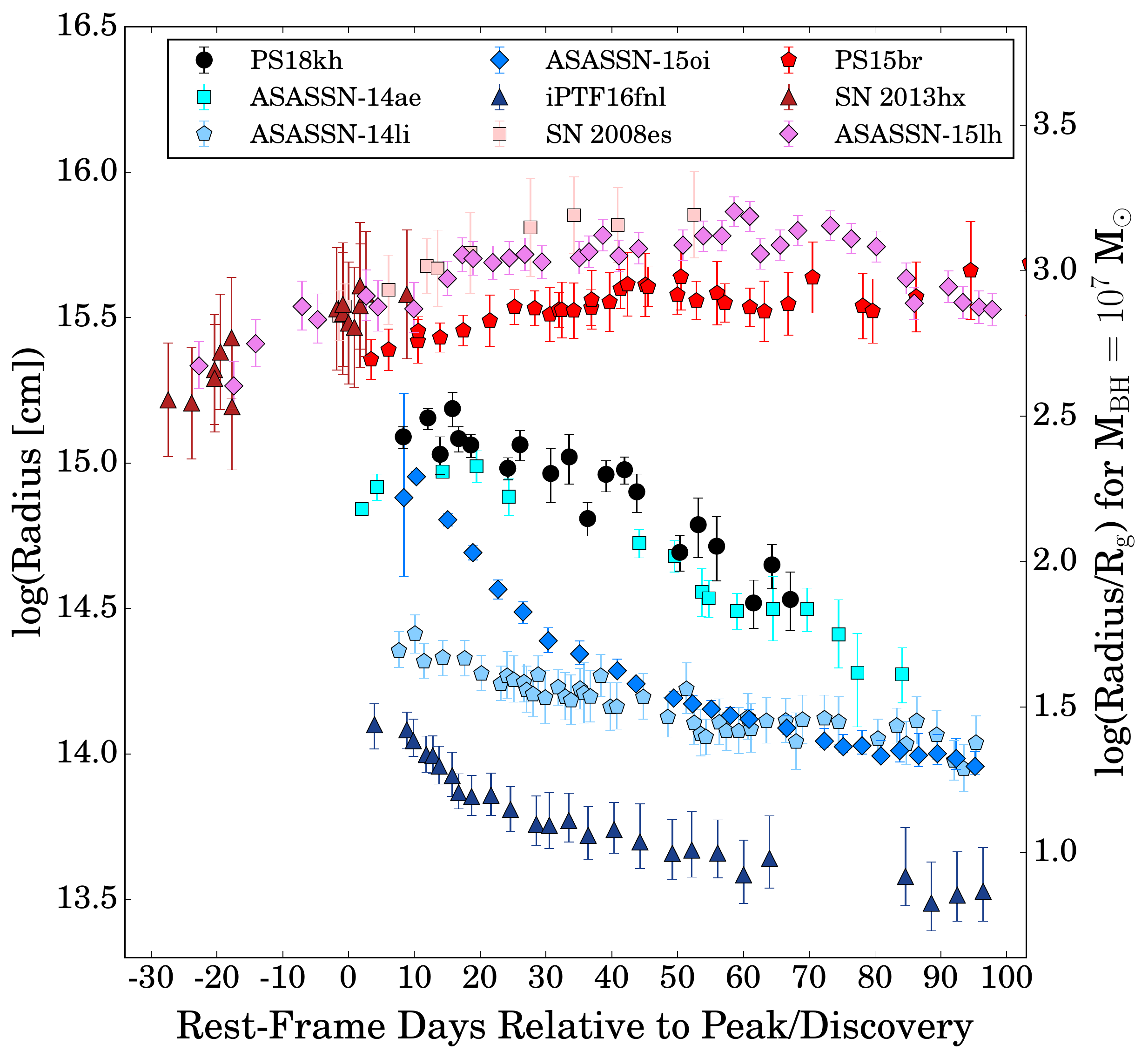}}}
\caption{Radius evolution of PS18kh taken from blackbody fits to epochs with {\swift} observations compared with the radius evolution of the objects in our comparison sample. Symbols and colors match those of Figure~\ref{fig:lum_comp} and all times are plotted in days relative to peak or discovery, as outlined in the caption of Figure~\ref{fig:lum_comp}. The left scale shows the radius in units of cm, while the right scale gives the corresponding radius in units of the gravitational radius for a $10^7$~\msun~black hole.}
\label{fig:rad_comp}
\end{figure}

\begin{figure*}
\begin{minipage}{\textwidth}
\centering
\subfloat{{\includegraphics[width=0.48\textwidth]{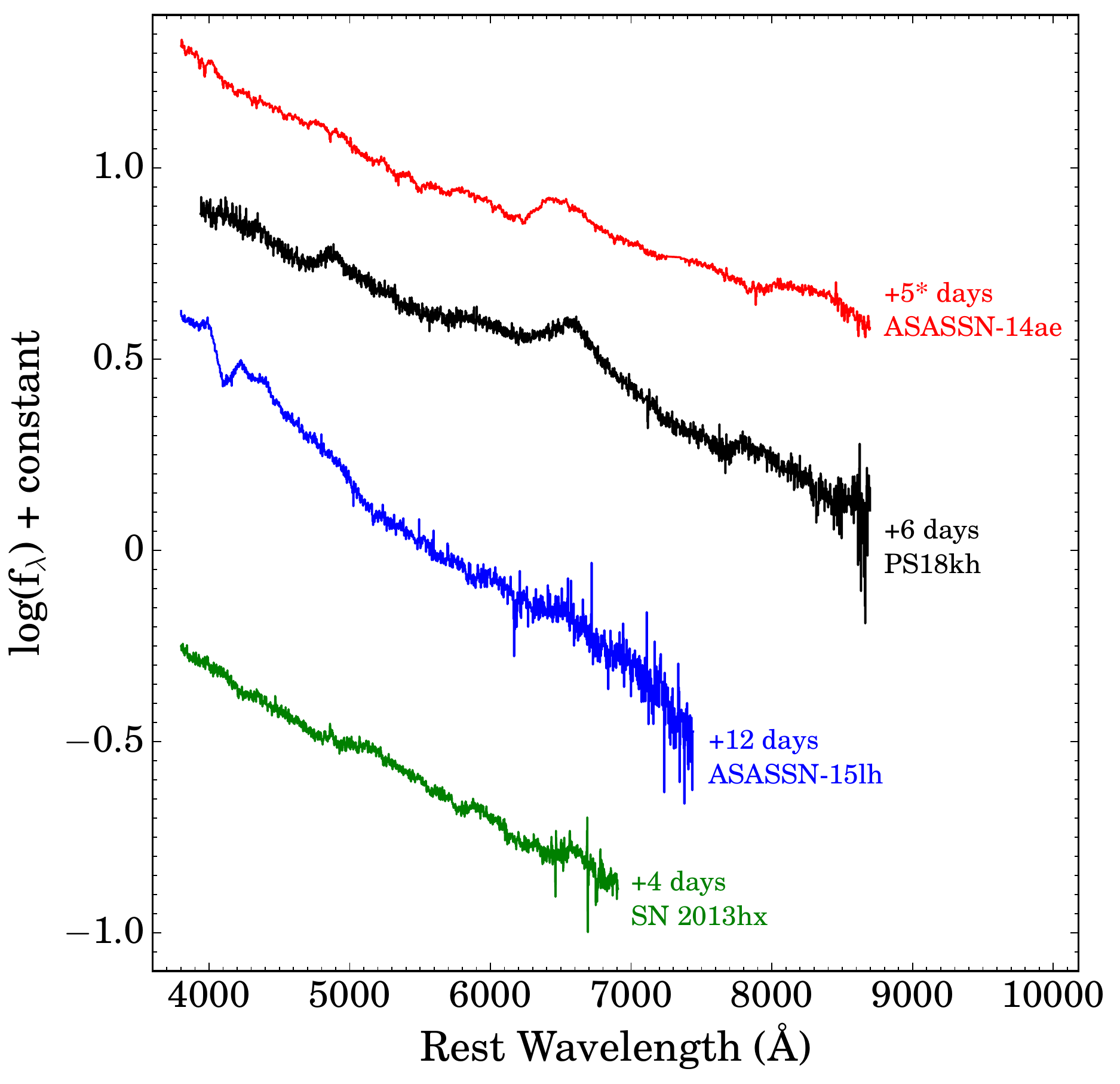}}}
\subfloat{{\includegraphics[width=0.48\textwidth]{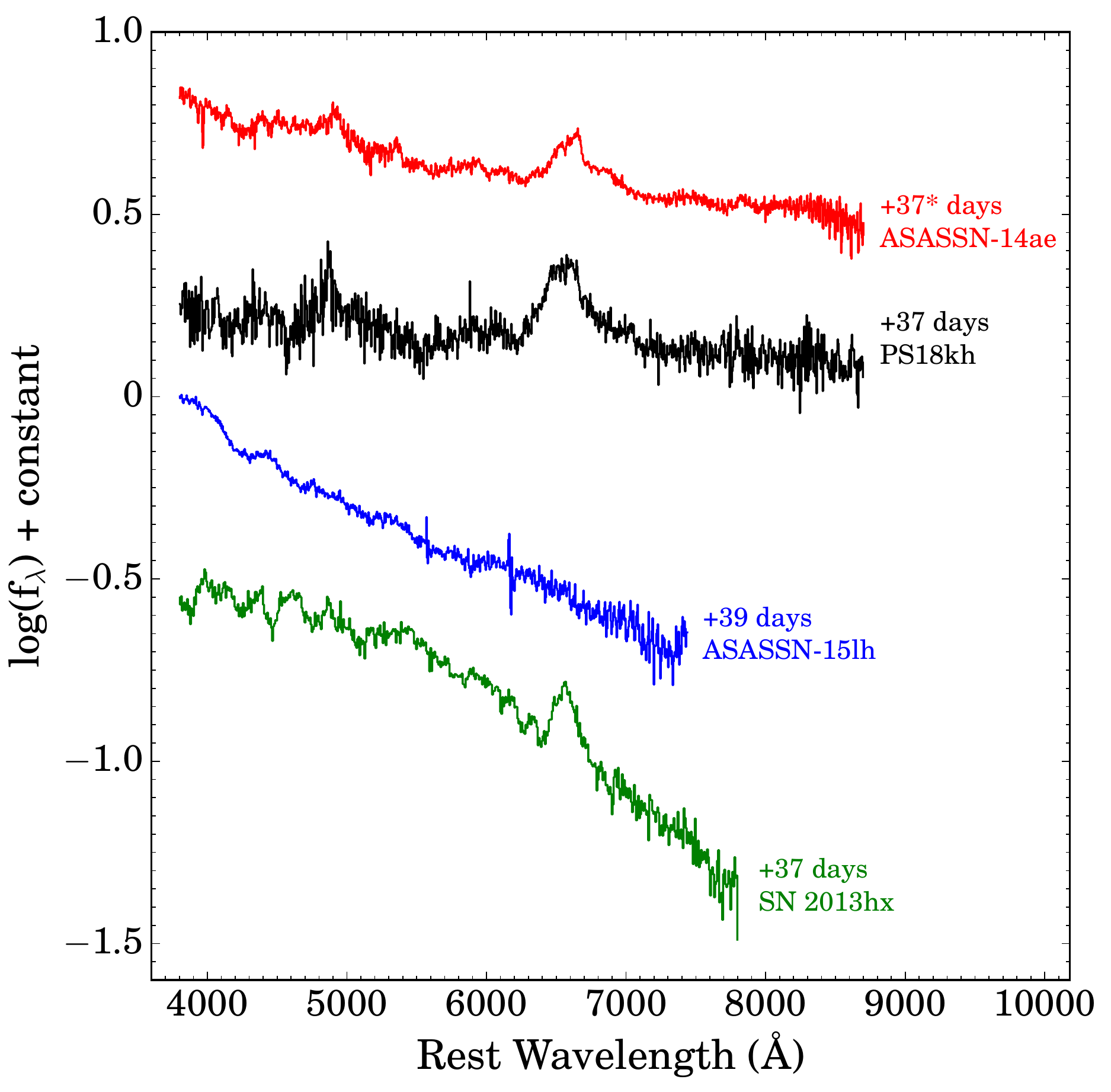}}}
\caption{\emph{Left Panel}: Spectra of PS18kh (black), ASASSN-14ae \citep[red;][]{holoien16a}, ASASSN-15lh \citep[blue;][]{dong16}, and SN 2013hx \citep[green;][]{inserra18} taken at similar phase shortly after rest-frame peak. (Phase for ASASSN-14ae is in days relative to discovery, as it was discovered after peak light.) Spectra have been offset for clarity and the phase is indicated to the right of each spectrum. {Right Panel}: Spectra of the same four objects taken 37$-$39 days after rest-frame peak/discovery.}
\label{fig:spec_comp}
\end{minipage}
\end{figure*}

To examine the similarity of the re-brightening seen in the light curves of PS18kh and ASASSN-15lh, we scaled the peak luminosity of PS18kh by a factor of 24.5 to match the peak of ASASSN-15lh, and shifted the light curve of PS18kh by 15 rest-frame days so that the peak of the PS18kh light curve aligns with the highest measured luminosity of ASASSN-15lh. The resulting comparison is shown in the right panel of Figure~\ref{fig:lum_comp}. PS18kh rises a bit more steeply than ASASSN-15lh does, but after peak the rate of decline is very similar between the two objects. PS18kh begins to re-brighten sooner, with the rise beginning at $t\simeq59$~rest-frame days, while ASASSN-15lh begins to re-brighten at $t\simeq73$ rest-frame days, but the shape of the two light curves is very similar. Assuming PS18kh is a TDE, this perhaps lends credence to the interpretation that ASASSN-15lh was the result of a TDE. However, the two objects differ in other respects, such as their temperature and radius evolution and their spectroscopic features (see Figures~\ref{fig:temp_comp}, \ref{fig:rad_comp}, and \ref{fig:spec_comp}), which indicates that the physical mechanisms responsible for the re-brightening likely differ between the two transients. 

Figure~\ref{fig:temp_comp} shows the evolution of the temperature measured from the blackbody fits to the {\swift} observations of PS18kh compared to the temperature evolution of the other objects in our comparison sample. All three hydrogen-rich SLSNe show a very similar temperature evolution, with the temperature declining steadily from a peak of $T\sim10000$~K, while the TDEs all show either rising or constant temperature evolution, with temperatures in the range of $10000~\textrm{K}\lesssim T \lesssim 50000$~K. ASASSN-15lh clearly stands out from the other objects, showing both a decline similar in shape to that of the hydrogen-rich SLSNe, and a later rise similar to that of the TDEs. The temperature evolution of PS18kh very strongly resembles that of ASASSN-14ae in both shape and magnitude, including a rising temperature after $t\sim40$~days. This evolution strongly differentiates it from the SLSN sample and from ASASSN-15lh.

Figure~\ref{fig:rad_comp} shows the evolution of the radius measured from the blackbody fits to the {\swift} observations of PS18kh compared to the radius evolution of the other objects in our comparison sample. All three hydrogen-rich SLSNe and ASASSN-15lh  stand out very clearly from the TDEs and PS18kh. While the SNe show larger and relatively constant photospheric radii, all the TDEs show a declining radius. PS18kh again very closely resembles ASASSN-14ae in the shape and magnitude of its radius evolution, and is clearly differentiated from the SLSN sample and ASASSN-15lh.

Finally, in Figure~\ref{fig:spec_comp} we compare spectra of PS18kh to those of ASASSN-14ae, SN 2013hx, and ASASSN-15oi at two similar rest-frame phases (near peak/discovery and roughly 40 days after peak/discovery). In the early epoch, the spectra of PS18kh resembles both that of ASASSN-14ae and that of SN 2013hx, with a broad \halpha{} emission feature and strong, blue, relatively featureless continuum. However, the later epoch clearly differentiates PS18kh from the SLSN, as both ASASSN-14ae and PS18kh continue to exhibit fairly strong continuum emission and broad hydrogen emission features, while the continuum shape of the spectra of SN 2013hx has started to change, reflecting its cooling temperature, and a number of absorption features have appeared. The spectra of ASASSN-15lh show almost no evolution at all between the two epochs, as it exhibits very blue spectra with broad absorption features at bluer wavelengths and no emission features, and it is clearly differentiated from the other three objects.


These comparisons show that luminosity evolution does not differentiate between SLSNe and TDEs at early times---while SLSNe tend to be more luminous, objects from both the TDE and SLSN samples show similar peak luminosities and decline rates. Conversely, TDEs and SLSNe quickly differentiate themselves in their temperature, radius, and spectroscopic evolution. SLSNe have smoothly declining temperatures, growing or relatively constant photospheric radii, and absorption features emerge in the spectra over time. TDEs exhibit constant or rising temperatures, shrinking photospheres, and consistently blue spectra with broad hydrogen and helium emission features. ASASSN-15lh is an outlier from both comparison groups in some respects, although its radius evolution very closely matches the SLSN sample and no TDE has shown similar spectra, while numerous SLSNe have similar spectroscopic evolution. While the shape of its luminosity evolution curve is somewhat similar to that of PS18kh, it is more luminous than any other object in the sample, it has a unique temperature evolution, and its spectra show little-to-no evolution between peak light and $\sim40$ days after peak light, with no evidence of the broad hydrogen emission features seen in the other objects' spectra. 

It is clear from these comparisons that despite the uniqueness of its light curve shape and the double-peaked line profiles, PS18kh bears a strong resemblance to other known TDEs, and this is the most likely origin for the emission we see during the outburst. Our early survey observations allow us to see the rise to peak light in multiple bands and to estimate its luminosity prior to peak, where we see that a significant fraction of the total early radiated energy is emitted during the rise to peak. UV observations obtained prior to peak will allow us to fit the blackbody SED and better quantify the fraction of energy emitted early for future TDE discoveries.

\begin{figure*}
\begin{minipage}{\textwidth}
\centering
\subfloat{{\includegraphics[width=0.48\textwidth]{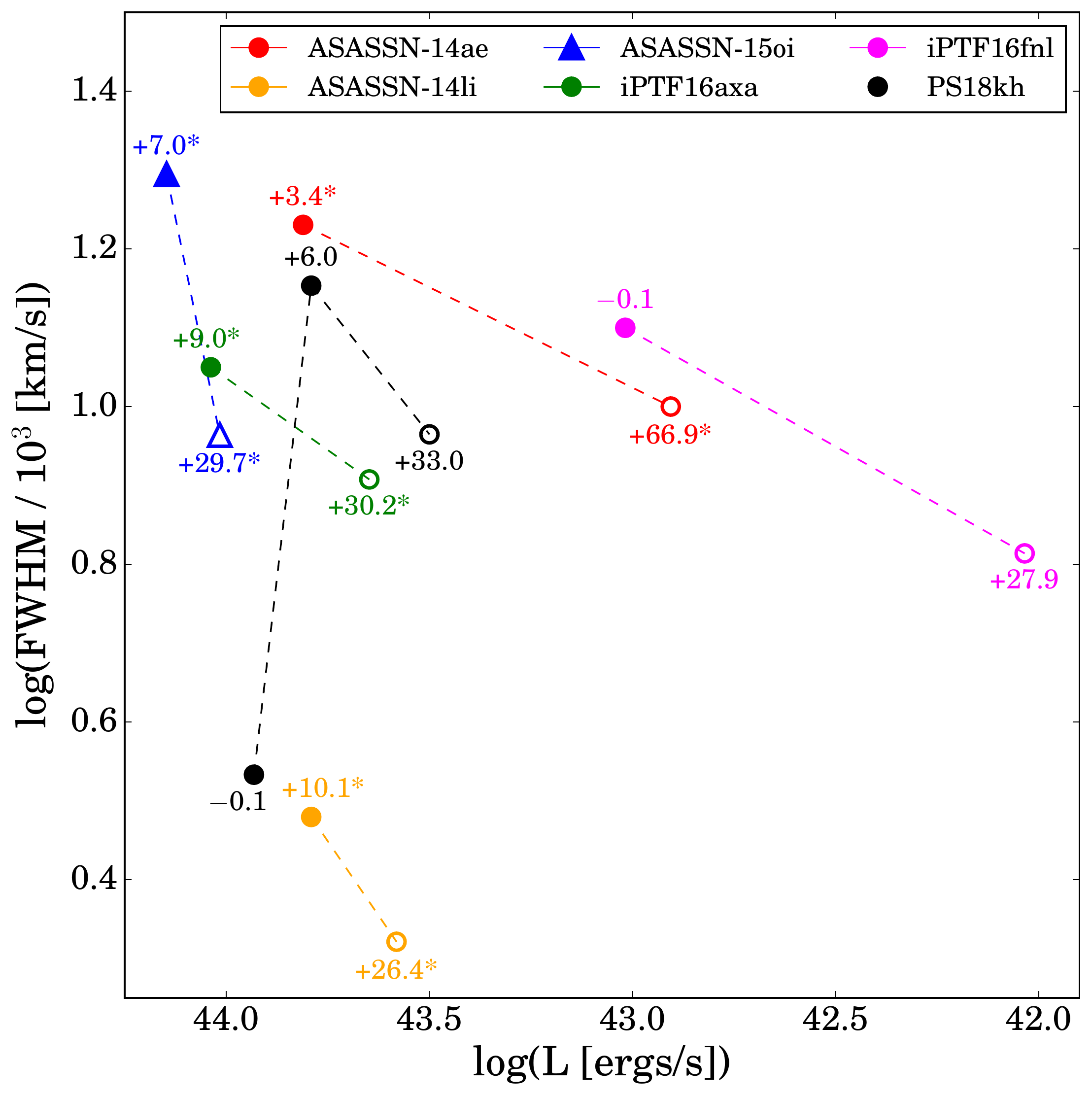}}}
\subfloat{{\includegraphics[width=0.48\textwidth]{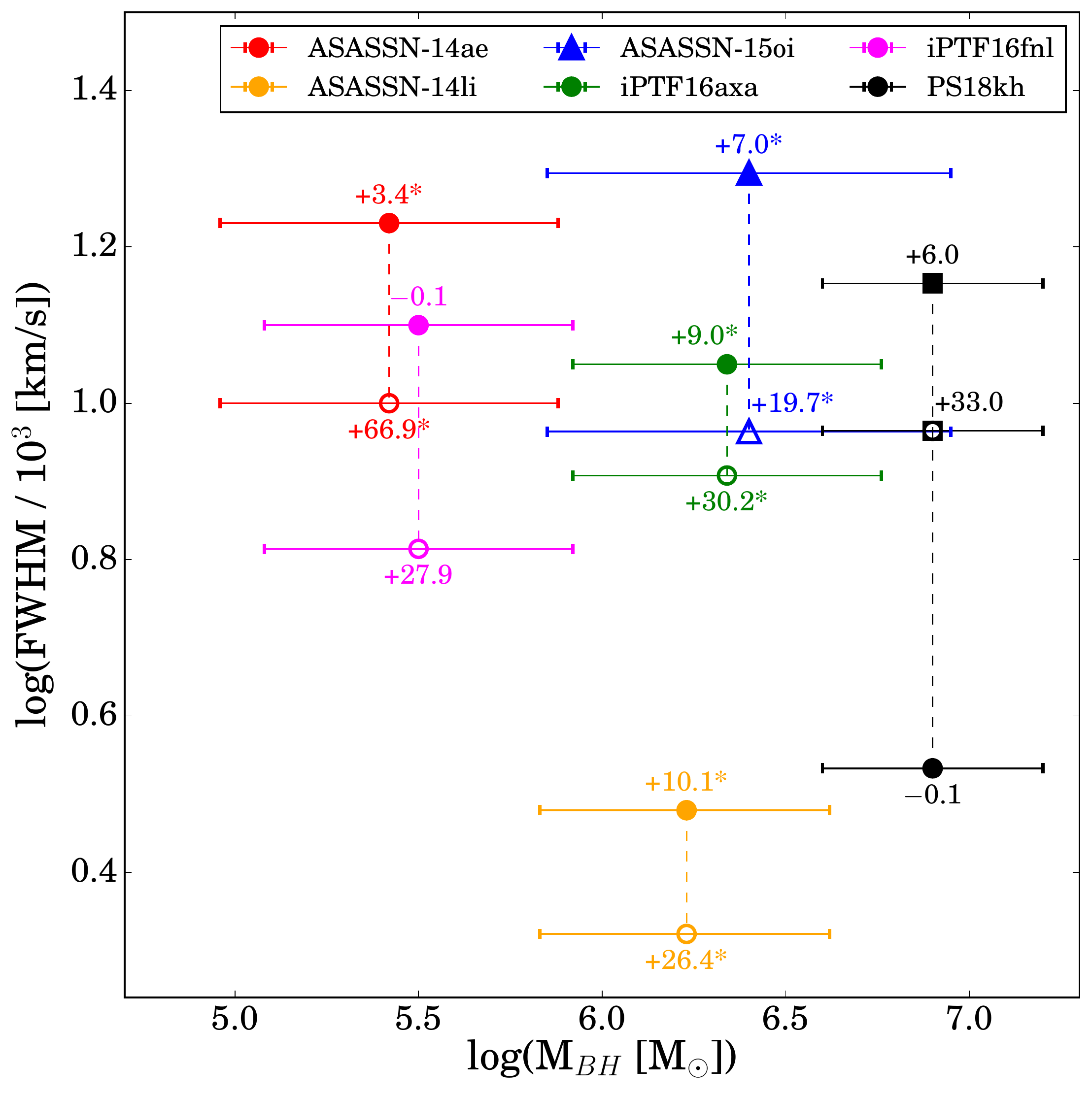}}}
\caption{\emph{Left Panel}: FWHM of the H$_\alpha$ line (circles) or \ion{He}{2} line (triangles) compared to luminosity for PS18kh and several TDEs from \citet{hung17} for epochs close to peak/discovery (filled points) and epochs 20-30 rest-frame days later (open points). Phase relative to peak/discovery is shown for each point, with an asterisk noting phase relative to discovery (as opposed to phase relative to peak). For PS18kh we show both the epoch at peak and the epoch 6 days later, as the FWHM initially increases before beginning to decline, as seen in the other TDES. {Right Panel}: Comparison of the FWHM of the same lines and epochs to the black hole mass for the same TDEs.}
\label{fig:fwhm_comp}
\end{minipage}
\end{figure*}

Having concluded that PS18kh is likely a TDE, we present a final comparison between it and other TDEs with similar spectroscopic coverage in Figure~\ref{fig:fwhm_comp}. In the Figure we show the FWHM of the most prominent spectroscopic emission line in PS18kh and a sample of TDEs from ASAS-SN and iPTF near peak brightness or near discovery and 20$-$30 days later compared to the luminosity of the TDE at similar times and the mass of the black hole. Data for comparison objects are taken from \citet{hung17}.

Comparing the emission line FWHM to luminosity (left panel of Figure~\ref{fig:fwhm_comp}, we see that in all cases the FWHM of the line decreases as the luminosity decreases, with no particular correlation between decline rate or absolute luminosity and FWHM. The comparison between line FWHM and black hole mass (right panel of the Figure) also indicates that there seems to be little correlation between these two properties, with the TDEs in the sample exhibiting a range of FWHM values and decline rates despite spanning roughly 1.5 orders of magnitude in black hole mass.


The early spectroscopic coverage of PS18kh also lets us look at the FWHM of the line at peak, compared to the evolution a few days later, and for the initial few days after peak the FWHM of the H$\alpha$ line increases. The only other TDE in the sample with similarly early coverage, iPTF16fnl, does not show the same behavior, so while it is clear that after an initial period the lines become narrower as the luminosity decreases, it is not clear whether the initial broadening seen in PS18kh is common or not. This highlights the need for more TDEs with spectra before, during, and shortly after peak brightness, as these times are largely unobserved for most TDEs in literature, and thus we cannot draw strong conclusions about possible correlations between the spectroscopic features and the TDE flare or black hole at these times.

PS18kh is the third TDE, after PTF09ge and ASASSN-14li \citep{arcavi14,liu17b,cao18}, to exhibit emission lines that can be fit by an elliptical disk model, and the first to have spectroscopic coverage prior to and throughout the peak of the light curve. Our modeling allows us to see the likely origin of the broad emission features that are ubiquitous in optically discovered TDEs, and to develop a physical picture for how these lines form in the early stages after the star is disrupted. Similarly detailed datasets will allow us to perform similar analysis on future TDEs, and will be able to tell us whether the model parameters seen in PS18kh are common to all TDEs, or whether there is a range of physical properties that can produce the observations we see. Real-time, high-cadence sky surveys like Pan-STARRS, ASAS-SN, and ATLAS will be able to provide early detection and long-term monitoring of future TDEs, providing us with a population of objects to study to further develop our physical understanding of these highly energetic events.

\acknowledgments

The authors thank Cosimo Inserra and Tiara Hung for providing comparison data used in Section~\ref{sec:disc} and Mark Seibert for assistence with calculating the GALEX host flux limit.

CSK and KZS are supported by NSF grants AST-1515876 and AST-1515927. SD acknowledges Project 11573003 supported by NSFC. Support for JLP is provided in part by FONDECYT through the grant 1151445 and by the Ministry of Economy, Development, and Tourism's Millennium Science Initiative through grant IC120009, awarded to The Millennium Institute of Astrophysics, MAS. TAT is supported in part by Scialog Scholar grant 24215 from the Research Corporation. SJS’s group  acknowledges funding from the European Research Council under the European Union's Seventh Framework Programme (FP7/2007-2013)/ERC Grant agreement n$^{\rm o}$ [291222] and STFC grant Grant Ref: ST/P000312/1 and  ST/N002520/1.

The Pan-STARRS1 Surveys (PS1) and the PS1 public science archive have been made possible through contributions by the Institute for Astronomy, the University of Hawaii, the Pan-STARRS Project Office, the Max-Planck Society and its participating institutes, the Max Planck Institute for Astronomy, Heidelberg and the Max Planck Institute for Extraterrestrial Physics, Garching, The Johns Hopkins University, Durham University, the University of Edinburgh, the Queen's University Belfast, the Harvard-Smithsonian Center for Astrophysics, the Las Cumbres Observatory Global Telescope Network Incorporated, the National Central University of Taiwan, the Space Telescope Science Institute, the National Aeronautics and Space Administration under Grant No. NNX08AR22G issued through the Planetary Science Division of the NASA Science Mission Directorate, the National Science Foundation Grant No. AST-1238877, the University of Maryland, Eotvos Lorand University (ELTE), the Los Alamos National Laboratory, and the Gordon and Betty Moore Foundation.

We thank the Las Cumbres Observatory and its staff for its continuing support of the ASAS-SN project. ASAS-SN is supported by the Gordon and Betty Moore Foundation through grant GBMF5490 to the Ohio State University and NSF grant AST-1515927. Development of ASAS-SN has been supported by NSF grant AST-0908816, the Mt. Cuba Astronomical Foundation, the Center for Cosmology and AstroParticle Physics at the Ohio State University, the Chinese Academy of Sciences South America Center for Astronomy (CASSACA), the Villum Foundation, and George Skestos.

This research has made use of the NASA/IPAC Extragalactic Database (NED) which is operated by the Jet Propulsion Laboratory, California Institute of Technology, under contract with the National Aeronautics and Space Administration.  This research has made use of NASA's Astrophysics Data System Bibliographic Services.  IRAF is distributed by the National Optical Astronomy Observatory, which is operated by the Association of Universities for Research in Astronomy (AURA) under a cooperative agreement with the National Science Foundation. 

This manuscript uses data obtained from the Keck telescopes, and we wish to extend special thanks to those of Hawaiian ancestry on whose sacred mountain we are privileged to be guests.

Based on data acquired using the Large Binocular Telescope (LBT). The LBT is an international collaboration among institutions in the United States, Italy, and Germany. LBT Corporation partners are: The University of Arizona on behalf of the Arizona university system; Istituto Nazionale di Astrofisica, Italy; LBT Beteiligungsgesellschaft, Germany, representing the Max-Planck Society, the Astrophysical Institute Potsdam, and Heidelberg University; The Ohio State University, and The Research Corporation, on behalf of The University of Notre Dame, University of Minnesota and University of Virginia.

The Liverpool Telescope is operated on the island of La Palma by Liverpool John Moores University in the Spanish Observatorio del Roque de los Muchachos of the Instituto de Astrofisica de Canarias with financial support from the UK Science and Technology Facilities Council.

This publication makes use of data products from the Wide-field Infrared Survey Explorer, which is a joint project of the University of California, Los Angeles, and the Jet Propulsion Laboratory/California Institute of Technology, funded by the National Aeronautics and Space Administration.

Funding for the SDSS and SDSS-II has been provided by the Alfred P. Sloan Foundation, the Participating Institutions, the National Science Foundation, the U.S. Department of Energy, the National Aeronautics and Space Administration, the Japanese Monbukagakusho, the Max Planck Society, and the Higher Education Funding Council for England. The SDSS Web Site is http://www.sdss.org/.

The SDSS is managed by the Astrophysical Research Consortium for the Participating Institutions. The Participating Institutions are the American Museum of Natural History, Astrophysical Institute Potsdam, University of Basel, University of Cambridge, Case Western Reserve University, University of Chicago, Drexel University, Fermilab, the Institute for Advanced Study, the Japan Participation Group, Johns Hopkins University, the Joint Institute for Nuclear Astrophysics, the Kavli Institute for Particle Astrophysics and Cosmology, the Korean Scientist Group, the Chinese Academy of Sciences (LAMOST), Los Alamos National Laboratory, the Max-Planck-Institute for Astronomy (MPIA), the Max-Planck-Institute for Astrophysics (MPA), New Mexico State University, Ohio State University, University of Pittsburgh, University of Portsmouth, Princeton University, the United States Naval Observatory, and the University of Washington.

\software{FAST (Kriek et al. 2009), IRAF (Tody 1986, Tody 1993), IPP (Magnier et al. 2013), HEAsoft (Arnaud 1996), XSPEC (v12.9.1; Arnaud 1996)}

\bibliography{bibliography.bib}

\begin{thebibliography}{126}
\expandafter\ifx\csname natexlab\endcsname\relax\def\natexlab#1{#1}\fi

\bibitem[{{Abazajian} {et~al.}(2009){Abazajian}, {Adelman-McCarthy},
  {Ag{\"u}eros}, {Allam}, {Allende Prieto}, {An}, {Anderson}, {Anderson},
  {Annis}, {Bahcall}, \& et~al.}]{abazajian09}
{Abazajian}, K.~N., {et~al.} 2009, \apjs, 182, 543

\bibitem[{{Abolfathi} {et~al.}(2018){Abolfathi}, {Aguado}, {Aguilar}, {Allende
  Prieto}, {Almeida}, {Tasnim Ananna}, {Anders}, {Anderson}, {Andrews},
  {Anguiano}, \& et~al.}]{abolfathi18}
{Abolfathi}, B., {et~al.} 2018, \apjs, 235, 42

\bibitem[{{Adhikari} {et~al.}(2016){Adhikari}, {R{\'o}{\.z}a{\'n}ska},
  {Czerny}, {Hryniewicz}, \& {Ferland}}]{adhikari16}
{Adhikari}, T.~P., {R{\'o}{\.z}a{\'n}ska}, A., {Czerny}, B., {Hryniewicz}, K.,
  \& {Ferland}, G.~J. 2016, \apj, 831, 68

\bibitem[{{Alard}(2000)}]{alard00}
{Alard}, C. 2000, AAPS, 144, 363

\bibitem[{{Alard} \& {Lupton}(1998)}]{alard98}
{Alard}, C., \& {Lupton}, R.~H. 1998, \apj, 503, 325

\bibitem[{{Arcavi} {et~al.}(2014){Arcavi}, {Gal-Yam}, {Sullivan}, {Pan},
  {Cenko}, {Horesh}, {Ofek}, {De Cia}, {Yan}, {Yang}, {Howell}, {Tal},
  {Kulkarni}, {Tendulkar}, {Tang}, {Xu}, {Sternberg}, {Cohen}, {Bloom},
  {Nugent}, {Kasliwal}, {Perley}, {Quimby}, {Miller}, {Theissen}, \&
  {Laher}}]{arcavi14}
{Arcavi}, I., {et~al.} 2014, \apj, 793, 38

\bibitem[{{Auchettl} {et~al.}(2017){Auchettl}, {Guillochon}, \&
  {Ramirez-Ruiz}}]{auchettl17}
{Auchettl}, K., {Guillochon}, J., \& {Ramirez-Ruiz}, E. 2017, \apj, 838, 149

\bibitem[{{Auchettl} {et~al.}(2018){Auchettl}, {Ramirez-Ruiz}, \&
  {Guillochon}}]{auchettl18}
{Auchettl}, K., {Ramirez-Ruiz}, E., \& {Guillochon}, J. 2018, \apj, 852, 37

\bibitem[{{Blagorodnova} {et~al.}(2017){Blagorodnova}, {Gezari}, {Hung},
  {Kulkarni}, {Cenko}, {Pasham}, {Yan}, {Arcavi}, {Ben-Ami}, {Bue}, {Cantwell},
  {Cao}, {Castro-Tirado}, {Fender}, {Fremling}, {Gal-Yam}, {Ho}, {Horesh},
  {Hosseinzadeh}, {Kasliwal}, {Kong}, {Laher}, {Leloudas}, {Lunnan}, {Masci},
  {Mooley}, {Neill}, {Nugent}, {Powell}, {Valeev}, {Vreeswijk}, {Walters}, \&
  {Wozniak}}]{blagorodnova17}
{Blagorodnova}, N., {et~al.} 2017, \apj, 844, 46

\bibitem[{{Bon} {et~al.}(2009){Bon}, {Popovi\'{c}}, {Gavrilovi\'{c}}, {Mura},
  \& {Mediavilla}}]{bon09}
{Bon}, E., {Popovi\'{c}}, L.~{\v{C}}., {Gavrilovi\'{c}}, N., {Mura}, G.~L., \&
  {Mediavilla}, E. 2009, \mnras, 400, 924

\bibitem[{{Bonnerot} {et~al.}(2016){Bonnerot}, {Rossi}, {Lodato}, \&
  {Price}}]{bonnerot16}
{Bonnerot}, C., {Rossi}, E.~M., {Lodato}, G., \& {Price}, D.~J. 2016, \mnras,
  455, 2253

\bibitem[{{Boulade} {et~al.}(1998){Boulade}, {Vigroux}, {Charlot}, {Borgeaud},
  {Carton}, {de Kat}, {Rousse}, {Mellier}, {Gigan}, {Crampton}, \&
  {Morbey}}]{boulade98}
{Boulade}, O., {et~al.} 1998, in \procspie, Vol. 3355, Optical Astronomical
  Instrumentation, ed. S.~{D'Odorico}, 614--625

\bibitem[{{Breeveld} {et~al.}(2010){Breeveld}, {Curran}, {Hoversten}, {Koch},
  {Landsman}, {Marshall}, {Page}, {Poole}, {Roming}, {Smith}, {Still},
  {Yershov}, {Blustin}, {Brown}, {Gronwall}, {Holland}, {Kuin}, {McGowan},
  {Rosen}, {Boyd}, {Broos}, {Carter}, {Chester}, {Hancock}, {Huckle}, {Immler},
  {Ivanushkina}, {Kennedy}, {Mason}, {Morgan}, {Oates}, {de Pasquale},
  {Schady}, {Siegel}, \& {vanden Berk}}]{breeveld10}
{Breeveld}, A.~A., {et~al.} 2010, \mnras, 406, 1687

\bibitem[{{Brown} {et~al.}(2017){Brown}, {Holoien}, {Auchettl}, {Stanek},
  {Kochanek}, {Shappee}, {Prieto}, \& {Grupe}}]{brown17a}
{Brown}, J.~S., {Holoien}, T.~W.-S., {Auchettl}, K., {Stanek}, K.~Z.,
  {Kochanek}, C.~S., {Shappee}, B.~J., {Prieto}, J.~L., \& {Grupe}, D. 2017,
  \mnras, 466, 4904

\bibitem[{{Brown} {et~al.}(2016){Brown}, {Shappee}, {Holoien}, {Stanek},
  {Kochanek}, \& {Prieto}}]{brown16a}
{Brown}, J.~S., {Shappee}, B.~J., {Holoien}, T.~W.-S., {Stanek}, K.~Z.,
  {Kochanek}, C.~S., \& {Prieto}, J.~L. 2016, \mnras, 462, 3993

\bibitem[{{Brown} {et~al.}(2018){Brown}, {Kochanek}, {Holoien}, {Stanek},
  {Auchettl}, {Shappee}, {Prieto}, {Morrell}, {Falco}, {Strader}, {Chomiuk},
  {Post}, {Villanueva}, {Mathur}, {Dong}, {Chen}, \& {Bose}}]{brown17b}
{Brown}, J.~S., {et~al.} 2018, \mnras, 473, 1130

\bibitem[{{Brown} {et~al.}(2013){Brown}, {Baliber}, {Bianco}, {Bowman},
  {Burleson}, {Conway}, {Crellin}, {Depagne}, {De Vera}, {Dilday}, {Dragomir},
  {Dubberley}, {Eastman}, {Elphick}, {Falarski}, {Foale}, {Ford}, {Fulton},
  {Garza}, {Gomez}, {Graham}, {Greene}, {Haldeman}, {Hawkins}, {Haworth},
  {Haynes}, {Hidas}, {Hjelstrom}, {Howell}, {Hygelund}, {Lister}, {Lobdill},
  {Martinez}, {Mullins}, {Norbury}, {Parrent}, {Paulson}, {Petry}, {Pickles},
  {Posner}, {Rosing}, {Ross}, {Sand}, {Saunders}, {Shobbrook}, {Shporer},
  {Street}, {Thomas}, {Tsapras}, {Tufts}, {Valenti}, {Vander Horst}, {Walker},
  {White}, \& {Willis}}]{brown13}
{Brown}, T.~M., {et~al.} 2013, \pasp, 125, 1031

\bibitem[{{Bruhweiler} \& {Verner}(2008)}]{bruhweiler08}
{Bruhweiler}, F., \& {Verner}, E. 2008, \apj, 675, 83

\bibitem[{{Bruzual} \& {Charlot}(2003)}]{bruzual03}
{Bruzual}, G., \& {Charlot}, S. 2003, \mnras, 344, 1000

\bibitem[{{Burrows} {et~al.}(2005){Burrows}, {Hill}, {Nousek}, {Kennea},
  {Wells}, {Osborne}, {Abbey}, {Beardmore}, {Mukerjee}, {Short}, {Chincarini},
  {Campana}, {Citterio}, {Moretti}, {Pagani}, {Tagliaferri}, {Giommi},
  {Capalbi}, {Tamburelli}, {Angelini}, {Cusumano}, {Br{\"a}uninger}, {Burkert},
  \& {Hartner}}]{burrows05}
{Burrows}, D.~N., {et~al.} 2005, SSR, 120, 165

\bibitem[{{Cao} {et~al.}(2018){Cao}, {Liu}, {Zhou}, {Komossa}, \& {Ho}}]{cao18}
{Cao}, R., {Liu}, F.~K., {Zhou}, Z.~Q., {Komossa}, S., \& {Ho}, L.~C. 2018,
  ArXiv e-prints

\bibitem[{{Cardelli} {et~al.}(1989){Cardelli}, {Clayton}, \&
  {Mathis}}]{cardelli89}
{Cardelli}, J.~A., {Clayton}, G.~C., \& {Mathis}, J.~S. 1989, \apj, 345, 245

\bibitem[{{Cenko} {et~al.}(2012){Cenko}, {Bloom}, {Kulkarni}, {Strubbe},
  {Miller}, {Butler}, {Quimby}, {Gal-Yam}, {Ofek}, {Quataert}, {Bildsten},
  {Poznanski}, {Perley}, {Morgan}, {Filippenko}, {Frail}, {Arcavi}, {Ben-Ami},
  {Cucchiara}, {Fassnacht}, {Green}, {Hook}, {Howell}, {Lagattuta}, {Law},
  {Kasliwal}, {Nugent}, {Silverman}, {Sullivan}, {Tendulkar}, \&
  {Yaron}}]{cenko12a}
{Cenko}, S.~B., {et~al.} 2012, \mnras, 420, 2684

\bibitem[{{Chajet} \& {Hall}(2013)}]{chajet13}
{Chajet}, L.~S., \& {Hall}, P.~B. 2013, \mnras, 429, 3214

\bibitem[{{Chambers} {et~al.}(2016){Chambers}, {Magnier}, {Metcalfe},
  {Flewelling}, {Huber}, {Waters}, {Denneau}, {Draper}, {Farrow}, {Finkbeiner},
  {Holmberg}, {Koppenhoefer}, {Price}, {Saglia}, {Schlafly}, {Smartt},
  {Sweeney}, {Wainscoat}, {Burgett}, {Grav}, {Heasley}, {Hodapp}, {Jedicke},
  {Kaiser}, {Kudritzki}, {Luppino}, {Lupton}, {Monet}, {Morgan}, {Onaka},
  {Stubbs}, {Tonry}, {Banados}, {Bell}, {Bender}, {Bernard}, {Botticella},
  {Casertano}, {Chastel}, {Chen}, {Chen}, {Cole}, {Deacon}, {Frenk},
  {Fitzsimmons}, {Gezari}, {Goessl}, {Goggia}, {Goldman}, {Grebel}, {Hambly},
  {Hasinger}, {Heavens}, {Heckman}, {Henderson}, {Henning}, {Holman}, {Hopp},
  {Ip}, {Isani}, {Keyes}, {Koekemoer}, {Kotak}, {Long}, {Lucey}, {Liu},
  {Martin}, {McLean}, {Morganson}, {Murphy}, {Nieto-Santisteban}, {Norberg},
  {Peacock}, {Pier}, {Postman}, {Primak}, {Rae}, {Rest}, {Riess}, {Riffeser},
  {Rix}, {Roser}, {Schilbach}, {Schultz}, {Scolnic}, {Szalay}, {Seitz},
  {Shiao}, {Small}, {Smith}, {Soderblom}, {Taylor}, {Thakar}, {Thiel},
  {Thilker}, {Urata}, {Valenti}, {Walter}, {Watters}, {Werner}, {White},
  {Wood-Vasey}, \& {Wyse}}]{chambers16}
{Chambers}, K.~C., {et~al.} 2016, ArXiv e-prints

\bibitem[{{Chen} \& {Halpern}(1989)}]{chen89b}
{Chen}, K., \& {Halpern}, J.~P. 1989, \apj, 344, 115

\bibitem[{{Chen} {et~al.}(1989){Chen}, {Halpern}, \& {Filippenko}}]{chen89a}
{Chen}, K., {Halpern}, J.~P., \& {Filippenko}, A.~V. 1989, \apj, 339, 742

\bibitem[{{Chiang} \& {Murray}(1996)}]{chiang96}
{Chiang}, J., \& {Murray}, N. 1996, \apj, 466, 704

\bibitem[{{Chornock} {et~al.}(2014){Chornock}, {Berger}, {Gezari}, {Zauderer},
  {Rest}, {Chomiuk}, {Kamble}, {Soderberg}, {Czekala}, {Dittmann}, {Drout},
  {Foley}, {Fong}, {Huber}, {Kirshner}, {Lawrence}, {Lunnan}, {Marion},
  {Narayan}, {Riess}, {Roth}, {Sanders}, {Scolnic}, {Smartt}, {Smith},
  {Stubbs}, {Tonry}, {Burgett}, {Chambers}, {Flewelling}, {Hodapp}, {Kaiser},
  {Magnier}, {Martin}, {Neill}, {Price}, \& {Wainscoat}}]{chornock14}
{Chornock}, R., {et~al.} 2014, \apj, 780, 44

\bibitem[{{Dai} {et~al.}(2015){Dai}, {McKinney}, \& {Miller}}]{dai15}
{Dai}, L., {McKinney}, J.~C., \& {Miller}, M.~C. 2015, \apjl, 812, L39

\bibitem[{{Dai} {et~al.}(2018){Dai}, {McKinney}, {Roth}, {Ramirez-Ruiz}, \&
  {Miller}}]{dai18}
{Dai}, L., {McKinney}, J.~C., {Roth}, N., {Ramirez-Ruiz}, E., \& {Miller},
  M.~C. 2018, \apjl, 859, L20

\bibitem[{{Demircan} \& {Kahraman}(1991)}]{demircan91}
{Demircan}, O., \& {Kahraman}, G. 1991, \apss, 181, 313

\bibitem[{{Dong} {et~al.}(2016){Dong}, {Shappee}, {Prieto}, {Jha}, {Stanek},
  {Holoien}, {Kochanek}, {Thompson}, {Morrell}, {Thompson}, {Basu}, {Beacom},
  {Bersier}, {Brimacombe}, {Brown}, {Chen}, {Conseil}, {Danilet}, {Falco},
  {Grupe}, {Kiyota}, {Masi}, {Nicholls}, {Olivares}, {Pignata}, {Pojmanski},
  {Simonian}, {Szczygiel}, \& {Wozniak}}]{dong16}
{Dong}, S., {et~al.} 2016, Science, 351, 257

\bibitem[{{Dressler} {et~al.}(2011){Dressler}, {Bigelow}, {Hare}, {Sutin},
  {Thompson}, {Burley}, {Epps}, {Oemler}, {Bagish}, {Birk}, {Clardy},
  {Gunnels}, {Kelson}, {Shectman}, \& {Osip}}]{dressler11}
{Dressler}, A., {et~al.} 2011, \pasp, 123, 288

\bibitem[{{Dumont} \& {Collin-Souffrin}(1990{\natexlab{a}})}]{dumont90iv}
{Dumont}, A.~M., \& {Collin-Souffrin}, S. 1990{\natexlab{a}}, \aap, 229, 313

\bibitem[{{Dumont} \& {Collin-Souffrin}(1990{\natexlab{b}})}]{dumont90v}
---. 1990{\natexlab{b}}, \aaps, 83, 71

\bibitem[{{Elitzur} {et~al.}(2014){Elitzur}, {Ho}, \& {Trump}}]{elitzur14}
{Elitzur}, M., {Ho}, L.~C., \& {Trump}, J.~R. 2014, \mnras, 438, 3340

\bibitem[{{Eracleous} {et~al.}(1995){Eracleous}, {Livio}, {Halpern}, \&
  {Storchi-Bergmann}}]{eracleous95}
{Eracleous}, M., {Livio}, M., {Halpern}, J.~P., \& {Storchi-Bergmann}, T. 1995,
  \apj, 438, 610

\bibitem[{{Evans} \& {Kochanek}(1989)}]{evans89}
{Evans}, C.~R., \& {Kochanek}, C.~S. 1989, \apjl, 346, L13

\bibitem[{{Fabricant} {et~al.}(1998){Fabricant}, {Cheimets}, {Caldwell}, \&
  {Geary}}]{fabricant98}
{Fabricant}, D., {Cheimets}, P., {Caldwell}, N., \& {Geary}, J. 1998, \pasp,
  110, 79

\bibitem[{{Flohic} {et~al.}(2012){Flohic}, {Eracleous}, \&
  {Bogdanovi\'{c}}}]{flohic12}
{Flohic}, H.~M.~L.~G., {Eracleous}, M., \& {Bogdanovi\'{c}}, T. 2012, \apj,
  753, 133

\bibitem[{{Gaskell} \& {Rojas Lobos}(2014)}]{gaskell14}
{Gaskell}, C.~M., \& {Rojas Lobos}, P.~A. 2014, \mnras, 438, L36

\bibitem[{{Gehrels} {et~al.}(2004){Gehrels}, {Chincarini}, {Giommi}, {Mason},
  {Nousek}, {Wells}, {White}, {Barthelmy}, {Burrows}, {Cominsky}, {Hurley},
  {Marshall}, {M{\'e}sz{\'a}ros}, {Roming}, {Angelini}, {Barbier}, {Belloni},
  {Campana}, {Caraveo}, {Chester}, {Citterio}, {Cline}, {Cropper}, {Cummings},
  {Dean}, {Feigelson}, {Fenimore}, {Frail}, {Fruchter}, {Garmire}, {Gendreau},
  {Ghisellini}, {Greiner}, {Hill}, {Hunsberger}, {Krimm}, {Kulkarni}, {Kumar},
  {Lebrun}, {Lloyd-Ronning}, {Markwardt}, {Mattson}, {Mushotzky}, {Norris},
  {Osborne}, {Paczynski}, {Palmer}, {Park}, {Parsons}, {Paul}, {Rees},
  {Reynolds}, {Rhoads}, {Sasseen}, {Schaefer}, {Short}, {Smale}, {Smith},
  {Stella}, {Tagliaferri}, {Takahashi}, {Tashiro}, {Townsley}, {Tueller},
  {Turner}, {Vietri}, {Voges}, {Ward}, {Willingale}, {Zerbi}, \&
  {Zhang}}]{gehrels04}
{Gehrels}, N., {et~al.} 2004, \apj, 611, 1005

\bibitem[{{Gezari} {et~al.}(2017){Gezari}, {Cenko}, \& {Arcavi}}]{gezari17}
{Gezari}, S., {Cenko}, S.~B., \& {Arcavi}, I. 2017, \apjl, 851, L47

\bibitem[{{Gezari} {et~al.}(2015){Gezari}, {Chornock}, {Lawrence}, {Rest},
  {Jones}, {Berger}, {Challis}, \& {Narayan}}]{gezari15}
{Gezari}, S., {Chornock}, R., {Lawrence}, A., {Rest}, A., {Jones}, D.~O.,
  {Berger}, E., {Challis}, P.~M., \& {Narayan}, G. 2015, \apjl, 815, L5

\bibitem[{{Gezari} {et~al.}(2009){Gezari}, {Halpern}, {Grupe}, {Yuan},
  {Quimby}, {McKay}, {Chamarro}, {Sisson}, {Akerlof}, {Wheeler}, {Brown},
  {Cenko}, {Rau}, {Djordjevic}, \& {Terndrup}}]{gezari09b}
{Gezari}, S., {et~al.} 2009, \apj, 690, 1313

\bibitem[{{Gezari} {et~al.}(2012){Gezari}, {Chornock}, {Rest}, {Huber},
  {Forster}, {Berger}, {Challis}, {Neill}, {Martin}, {Heckman}, {Lawrence},
  {Norman}, {Narayan}, {Foley}, {Marion}, {Scolnic}, {Chomiuk}, {Soderberg},
  {Smith}, {Kirshner}, {Riess}, {Smartt}, {Stubbs}, {Tonry}, {Wood-Vasey},
  {Burgett}, {Chambers}, {Grav}, {Heasley}, {Kaiser}, {Kudritzki}, {Magnier},
  {Morgan}, \& {Price}}]{gezari12b}
---. 2012, \nat, 485, 217

\bibitem[{{Gilbert} {et~al.}(1999){Gilbert}, {Eracleous}, {Filippenko}, \&
  {Halpern}}]{gilbert99}
{Gilbert}, A.~M., {Eracleous}, M., {Filippenko}, A.~V., \& {Halpern}, J.~P.
  1999, in Astronomical Society of the Pacific Conference Series, Vol. 175,
  Structure and Kinematics of Quasar Broad Line Regions, ed. C.~M. {Gaskell},
  W.~N. {Brandt}, M.~{Dietrich}, D.~{Dultzin-Hacyan}, \& M.~{Eracleous}, 189

\bibitem[{{Godoy-Rivera} {et~al.}(2017){Godoy-Rivera}, {Stanek}, {Kochanek},
  {Chen}, {Dong}, {Prieto}, {Shappee}, {Jha}, {Foley}, {Pan}, {Holoien},
  {Thompson}, {Grupe}, \& {Beacom}}]{godoy-rivera17}
{Godoy-Rivera}, D., {et~al.} 2017, \mnras, 466, 1428

\bibitem[{{Guillochon} {et~al.}(2014){Guillochon}, {Manukian}, \&
  {Ramirez-Ruiz}}]{guillochon14}
{Guillochon}, J., {Manukian}, H., \& {Ramirez-Ruiz}, E. 2014, \apj, 783, 23

\bibitem[{{Guillochon} \& {Ramirez-Ruiz}(2013)}]{guillochon13}
{Guillochon}, J., \& {Ramirez-Ruiz}, E. 2013, \apj, 767, 25

\bibitem[{{Guillochon} \& {Ramirez-Ruiz}(2015)}]{guillochon15}
---. 2015, \apj, 809, 166

\bibitem[{{Hayasaki} {et~al.}(2013){Hayasaki}, {Stone}, \& {Loeb}}]{hayasaki13}
{Hayasaki}, K., {Stone}, N., \& {Loeb}, A. 2013, \mnras, 434, 909

\bibitem[{{Hayasaki} {et~al.}(2016){Hayasaki}, {Stone}, \& {Loeb}}]{hayasaki16}
---. 2016, \mnras, 461, 3760

\bibitem[{{Henden} {et~al.}(2015){Henden}, {Levine}, {Terrell}, \&
  {Welch}}]{henden15}
{Henden}, A.~A., {Levine}, S., {Terrell}, D., \& {Welch}, D.~L. 2015, in
  American Astronomical Society Meeting Abstracts, Vol. 225, American
  Astronomical Society Meeting Abstracts \#225, 336.16

\bibitem[{{Holoien} {et~al.}(2018){Holoien}, {Brown}, {Auchettl}, {Kochanek},
  {Prieto}, {Shappee}, \& {Van Saders}}]{holoien18a}
{Holoien}, T.~W.-S., {Brown}, J.~S., {Auchettl}, K., {Kochanek}, C.~S.,
  {Prieto}, J.~L., {Shappee}, B.~J., \& {Van Saders}, J. 2018, \mnras, 480,
  5689

\bibitem[{{Holoien} {et~al.}(2014){Holoien}, {Prieto}, {Bersier}, {Kochanek},
  {Stanek}, {Shappee}, {Grupe}, {Basu}, {Beacom}, {Brimacombe}, {Brown},
  {Davis}, {Jencson}, {Pojmanski}, \& {Szczygie{\l}}}]{holoien14b}
{Holoien}, T.~W.-S., {et~al.} 2014, \mnras, 445, 3263

\bibitem[{{Holoien} {et~al.}(2016{\natexlab{a}}){Holoien}, {Kochanek},
  {Prieto}, {Grupe}, {Chen}, {Godoy-Rivera}, {Stanek}, {Shappee}, {Dong},
  {Brown}, {Basu}, {Beacom}, {Bersier}, {Brimacombe}, {Carlson}, {Falco},
  {Johnston}, {Madore}, {Pojmanski}, \& {Seibert}}]{holoien16b}
---. 2016{\natexlab{a}}, \mnras, 463, 3813

\bibitem[{{Holoien} {et~al.}(2016{\natexlab{b}}){Holoien}, {Kochanek},
  {Prieto}, {Stanek}, {Dong}, {Shappee}, {Grupe}, {Brown}, {Basu}, {Beacom},
  {Bersier}, {Brimacombe}, {Danilet}, {Falco}, {Guo}, {Jose}, {Herczeg},
  {Long}, {Pojmanski}, {Simonian}, {Szczygie{\l}}, {Thompson}, {Thorstensen},
  {Wagner}, \& {Wo{\'z}niak}}]{holoien16a}
---. 2016{\natexlab{b}}, \mnras, 455, 2918

\bibitem[{{Hook} {et~al.}(2004){Hook}, {J{\o}rgensen}, {Allington-Smith},
  {Davies}, {Metcalfe}, {Murowinski}, \& {Crampton}}]{hook04}
{Hook}, I.~M., {J{\o}rgensen}, I., {Allington-Smith}, J.~R., {Davies}, R.~L.,
  {Metcalfe}, N., {Murowinski}, R.~G., \& {Crampton}, D. 2004, \pasp, 116, 425

\bibitem[{{Hryniewicz} {et~al.}(2014){Hryniewicz}, {Czerny}, {Pych}, {Udalski},
  {Krupa}, {{\'S}wi{\c e}to{\'n}}, \& {Kaluzny}}]{hryniewicz14}
{Hryniewicz}, K., {Czerny}, B., {Pych}, W., {Udalski}, A., {Krupa}, M.,
  {{\'S}wi{\c e}to{\'n}}, A., \& {Kaluzny}, J. 2014, \aap, 562, A34

\bibitem[{{Huber} {et~al.}(2015){Huber}, {Carter Chambers}, {Flewelling},
  {Smartt}, {Smith}, \& {Wright}}]{huber15}
{Huber}, M., {Carter Chambers}, K., {Flewelling}, H., {Smartt}, S.~J., {Smith},
  K., \& {Wright}, D. 2015, IAU General Assembly, 22, 2258303

\bibitem[{{Hung} {et~al.}(2017){Hung}, {Gezari}, {Blagorodnova}, {Roth},
  {Cenko}, {Kulkarni}, {Horesh}, {Arcavi}, {McCully}, {Yan}, {Lunnan},
  {Fremling}, {Cao}, {Nugent}, \& {Wozniak}}]{hung17}
{Hung}, T., {et~al.} 2017, \apj, 842, 29

\bibitem[{{Hung} {et~al.}(2019){Hung}, {Cenko}, {Roth}, {Gezari}, {Veilleux},
  {Van Velzen}, {Gaskell}, {Foley}, {Blagorodnova}, {Yan}, {Graham}, {Brown},
  {Siebert}, {Frederick}, {Ward}, {Gatkine}, {Gal-yam}, {Yang}, {Schulze},
  {Dimitriadis}, {Kupfer}, {Shupe}, {Rusholme}, {Masci}, {Riddle}, {Soumagnac},
  {Van Roestel}, \& {Dekany}}]{hung19}
---. 2019, arXiv e-prints

\bibitem[{{Inserra} {et~al.}(2018){Inserra}, {Smartt}, {Gall}, {Leloudas},
  {Chen}, {Schulze}, {Jerkstrand}, {Nicholl}, {Anderson}, {Arcavi}, {Benetti},
  {Cartier}, {Childress}, {Della Valle}, {Flewelling}, {Fraser}, {Gal-Yam},
  {Guti{\'e}rrez}, {Hosseinzadeh}, {Howell}, {Huber}, {Kankare}, {Kr{\"u}hler},
  {Magnier}, {Maguire}, {McCully}, {Prajs}, {Primak}, {Scalzo}, {Schmidt},
  {Smith}, {Smith}, {Tucker}, {Valenti}, {Wilman}, {Young}, \&
  {Yuan}}]{inserra18}
{Inserra}, C., {et~al.} 2018, \mnras, 475, 1046

\bibitem[{{Kalberla} {et~al.}(2005){Kalberla}, {Burton}, {Hartmann}, {Arnal},
  {Bajaja}, {Morras}, \& {P{\"o}ppel}}]{kalberla05}
{Kalberla}, P.~M.~W., {Burton}, W.~B., {Hartmann}, D., {Arnal}, E.~M.,
  {Bajaja}, E., {Morras}, R., \& {P{\"o}ppel}, W.~G.~L. 2005, \aap, 440, 775

\bibitem[{{Kochanek}(1994)}]{kochanek94}
{Kochanek}, C.~S. 1994, \apj, 422, 508

\bibitem[{{Kochanek}(2016)}]{kochanek16}
---. 2016, \mnras, 461, 371

\bibitem[{{Kochanek} {et~al.}(2017){Kochanek}, {Shappee}, {Stanek}, {Holoien},
  {Thompson}, {Prieto}, {Dong}, {Shields}, {Will}, {Britt}, {Perzanowski}, \&
  {Pojma{\'n}ski}}]{kochanek17}
{Kochanek}, C.~S., {et~al.} 2017, \pasp, 129, 104502

\bibitem[{{Kriek} {et~al.}(2009){Kriek}, {van Dokkum}, {Labb{\'e}}, {Franx},
  {Illingworth}, {Marchesini}, \& {Quadri}}]{kriek09}
{Kriek}, M., {van Dokkum}, P.~G., {Labb{\'e}}, I., {Franx}, M., {Illingworth},
  G.~D., {Marchesini}, D., \& {Quadri}, R.~F. 2009, \apj, 700, 221

\bibitem[{{La Mura} {et~al.}(2009){La Mura}, {Di Mille}, {Ciroi},
  {Popovi\'{c}}, \& {Rafanelli}}]{lamura09}
{La Mura}, G., {Di Mille}, F., {Ciroi}, S., {Popovi\'{c}}, L.~v., \&
  {Rafanelli}, P. 2009, \apj, 693, 1437

\bibitem[{{Lacy} {et~al.}(1982){Lacy}, {Townes}, \& {Hollenbach}}]{lacy82}
{Lacy}, J.~H., {Townes}, C.~H., \& {Hollenbach}, D.~J. 1982, \apj, 262, 120

\bibitem[{{Lantz} {et~al.}(2004){Lantz}, {Aldering}, {Antilogus}, {Bonnaud},
  {Capoani}, {Castera}, {Copin}, {Dubet}, {Gangler}, {Henault}, {Lemonnier},
  {Pain}, {Pecontal}, {Pecontal}, \& {Smadja}}]{lantz04}
{Lantz}, B., {et~al.} 2004, in \procspie, Vol. 5249, Optical Design and
  Engineering, ed. L.~{Mazuray}, P.~J. {Rogers}, \& R.~{Wartmann}, 146--155

\bibitem[{{Leloudas} {et~al.}(2016){Leloudas}, {Fraser}, {Stone}, {van Velzen},
  {Jonker}, {Arcavi}, {Fremling}, {Maund}, {Smartt}, {Kr{\`i}hler},
  {Miller-Jones}, {Vreeswijk}, {Gal-Yam}, {Mazzali}, {De Cia}, {Howell},
  {Inserra}, {Patat}, {de Ugarte Postigo}, {Yaron}, {Ashall}, {Bar},
  {Campbell}, {Chen}, {Childress}, {Elias-Rosa}, {Harmanen}, {Hosseinzadeh},
  {Johansson}, {Kangas}, {Kankare}, {Kim}, {Kuncarayakti}, {Lyman}, {Magee},
  {Maguire}, {Malesani}, {Mattila}, {McCully}, {Nicholl}, {Prentice},
  {Romero-Ca{\~n}izales}, {Schulze}, {Smith}, {Sollerman}, {Sullivan},
  {Tucker}, {Valenti}, {Wheeler}, \& {Young}}]{leloudas16}
{Leloudas}, G., {et~al.} 2016, Nature Astronomy, 1, 0002

\bibitem[{{Liu} {et~al.}(2017{\natexlab{a}}){Liu}, {Zhou}, {Cao}, {Ho}, \&
  {Komossa}}]{liu17b}
{Liu}, F.~K., {Zhou}, Z.~Q., {Cao}, R., {Ho}, L.~C., \& {Komossa}, S.
  2017{\natexlab{a}}, \mnras, 472, L99

\bibitem[{{Liu} {et~al.}(2017{\natexlab{b}}){Liu}, {Tozzi}, {Wang}, {Brandt},
  {Vignali}, {Xue}, {Schneider}, {Comastri}, {Yang}, {Bauer}, {Paolillo},
  {Luo}, {Gilli}, {Wang}, {Giavalisco}, {Ji}, {Alexander}, {Mainieri},
  {Shemmer}, {Koekemoer}, \& {Risaliti}}]{liu17}
{Liu}, T., {et~al.} 2017{\natexlab{b}}, \apjs, 232, 8

\bibitem[{{Lodato} \& {Rossi}(2011)}]{lodato11}
{Lodato}, G., \& {Rossi}, E.~M. 2011, \mnras, 410, 359

\bibitem[{{Lupton}(2005)}]{lupton05}
{Lupton}, R. 2005, http://www.sdss.org/dr5/algorithms/sdssUBVRITransform.html

\bibitem[{{MacLeod} {et~al.}(2012){MacLeod}, {Ivezi{\'c}}, {Sesar}, {de Vries},
  {Kochanek}, {Kelly}, {Becker}, {Lupton}, {Hall}, {Richards}, {Anderson}, \&
  {Schneider}}]{macleod12}
{MacLeod}, C.~L., {et~al.} 2012, \apj, 753, 106

\bibitem[{{Magnier} {et~al.}(2013){Magnier}, {Schlafly}, {Finkbeiner}, {Juric},
  {Tonry}, {Burgett}, {Chambers}, {Flewelling}, {Kaiser}, {Kudritzki},
  {Morgan}, {Price}, {Sweeney}, \& {Stubbs}}]{magnier13}
{Magnier}, E.~A., {et~al.} 2013, \apjs, 205, 20

\bibitem[{{Marchesi} {et~al.}(2016){Marchesi}, {Lanzuisi}, {Civano}, {Iwasawa},
  {Suh}, {Comastri}, {Zamorani}, {Allevato}, {Griffiths}, {Miyaji}, {Ranalli},
  {Salvato}, {Schawinski}, {Silverman}, {Treister}, {Urry}, \&
  {Vignali}}]{marchesi16}
{Marchesi}, S., {et~al.} 2016, \apj, 830, 100

\bibitem[{{McConnell} \& {Ma}(2013)}]{mcconnell13}
{McConnell}, N.~J., \& {Ma}, C.-P. 2013, \apj, 764, 184

\bibitem[{{McCrum} {et~al.}(2014){McCrum}, {Smartt}, {Kotak}, {Rest},
  {Jerkstrand}, {Inserra}, {Rodney}, {Chen}, {Howell}, {Huber}, {Pastorello},
  {Tonry}, {Bresolin}, {Kudritzki}, {Chornock}, {Berger}, {Smith},
  {Botticella}, {Foley}, {Fraser}, {Milisavljevic}, {Nicholl}, {Riess},
  {Stubbs}, {Valenti}, {Wood-Vasey}, {Wright}, {Young}, {Drout}, {Czekala},
  {Burgett}, {Chambers}, {Draper}, {Flewelling}, {Hodapp}, {Kaiser}, {Magnier},
  {Metcalfe}, {Price}, {Sweeney}, \& {Wainscoat}}]{mccrum14}
{McCrum}, M., {et~al.} 2014, \mnras, 437, 656

\bibitem[{{McCrum} {et~al.}(2015){McCrum}, {Smartt}, {Rest}, {Smith}, {Kotak},
  {Rodney}, {Young}, {Chornock}, {Berger}, {Foley}, {Fraser}, {Wright},
  {Scolnic}, {Tonry}, {Urata}, {Huang}, {Pastorello}, {Botticella}, {Valenti},
  {Mattila}, {Kankare}, {Farrow}, {Huber}, {Stubbs}, {Kirshner}, {Bresolin},
  {Burgett}, {Chambers}, {Draper}, {Flewelling}, {Jedicke}, {Kaiser},
  {Magnier}, {Metcalfe}, {Morgan}, {Price}, {Sweeney}, {Wainscoat}, \&
  {Waters}}]{mccrum15}
---. 2015, \mnras, 448, 1206

\bibitem[{{Miller} {et~al.}(2009){Miller}, {Chornock}, {Perley},
  {Ganeshalingam}, {Li}, {Butler}, {Bloom}, {Smith}, {Modjaz}, {Poznanski},
  {Filippenko}, {Griffith}, {Shiode}, \& {Silverman}}]{miller09}
{Miller}, A.~A., {et~al.} 2009, \apj, 690, 1303

\bibitem[{{Mockler} {et~al.}(2018){Mockler}, {Guillochon}, \&
  {Ramirez-Ruiz}}]{mockler18}
{Mockler}, B., {Guillochon}, J., \& {Ramirez-Ruiz}, E. 2018, ArXiv e-prints

\bibitem[{{Moloney} \& {Shull}(2014)}]{moloney14}
{Moloney}, J., \& {Shull}, J.~M. 2014, \apj, 793, 100

\bibitem[{{Moretti} {et~al.}(2004){Moretti}, {Campana}, {Tagliaferri}, {Abbey},
  {Ambrosi}, {Angelini}, {Beardmore}, {Br{\"a}uninger}, {Burkert}, {Burrows},
  {Capalbi}, {Chincarini}, {Citterio}, {Cusumano}, {Freyberg}, {Giommi},
  {Hartner}, {Hill}, {Mori}, {Morris}, {Mukerjee}, {Nousek}, {Osborne},
  {Short}, {Tamburelli}, {Watson}, \& {Wells}}]{moretti04}
{Moretti}, A., {et~al.} 2004, in \procspie, Vol. 5165, X-Ray and Gamma-Ray
  Instrumentation for Astronomy XIII, ed. K.~A. {Flanagan} \& O.~H.~W.
  {Siegmund}, 232--240

\bibitem[{{Murray} \& {Chiang}(1997)}]{murray97}
{Murray}, N., \& {Chiang}, J. 1997, \apj, 474, 91

\bibitem[{{Murray} {et~al.}(1995){Murray}, {Chiang}, {Grossman}, \&
  {Voit}}]{murray95a}
{Murray}, N., {Chiang}, J., {Grossman}, S.~A., \& {Voit}, G.~M. 1995, \apj,
  451, 498

\bibitem[{{Oke} {et~al.}(1995){Oke}, {Cohen}, {Carr}, {Cromer}, {Dingizian},
  {Harris}, {Labrecque}, {Lucinio}, {Schaal}, {Epps}, \& {Miller}}]{oke95}
{Oke}, J.~B., {et~al.} 1995, \pasp, 107, 375

\bibitem[{{Phinney}(1989)}]{phinney89}
{Phinney}, E.~S. 1989, \nat, 340, 595

\bibitem[{{Piran} {et~al.}(2015){Piran}, {Svirski}, {Krolik}, {Cheng}, \&
  {Shiokawa}}]{piran15}
{Piran}, T., {Svirski}, G., {Krolik}, J., {Cheng}, R.~M., \& {Shiokawa}, H.
  2015, \apj, 806, 164

\bibitem[{{Pogge} {et~al.}(2010){Pogge}, {Atwood}, {Brewer}, {Byard},
  {Derwent}, {Gonzalez}, {Martini}, {Mason}, {O'Brien}, {Osmer}, {Pappalardo},
  {Steinbrecher}, {Teiga}, \& {Zhelem}}]{Pogge2010}
{Pogge}, R.~W., {et~al.} 2010, in Society of Photo-Optical Instrumentation
  Engineers (SPIE) Conference Series, Vol. 7735, Society of Photo-Optical
  Instrumentation Engineers (SPIE) Conference Series

\bibitem[{{Poole} {et~al.}(2008){Poole}, {Breeveld}, {Page}, {Landsman},
  {Holland}, {Roming}, {Kuin}, {Brown}, {Gronwall}, {Hunsberger}, {Koch},
  {Mason}, {Schady}, {vanden Berk}, {Blustin}, {Boyd}, {Broos}, {Carter},
  {Chester}, {Cucchiara}, {Hancock}, {Huckle}, {Immler}, {Ivanushkina},
  {Kennedy}, {Marshall}, {Morgan}, {Pandey}, {de Pasquale}, {Smith}, \&
  {Still}}]{poole08}
{Poole}, T.~S., {et~al.} 2008, \mnras, 383, 627

\bibitem[{{Popovi\'{c}} {et~al.}(2004){Popovi\'{c}}, {Mediavilla}, {Bon}, \&
  {Ili{\'c}}}]{popovic04}
{Popovi\'{c}}, L.~{\v{C}}., {Mediavilla}, E., {Bon}, E., \& {Ili{\'c}}, D.
  2004, \aap, 423, 909

\bibitem[{{Prieto} {et~al.}(2016){Prieto}, {Kr{\"u}hler}, {Anderson},
  {Galbany}, {Kochanek}, {Aquino}, {Brown}, {Dong}, {F{\"o}rster}, {Holoien},
  {Kuncarayakti}, {Maureira}, {Rosales-Ortega}, {S{\'a}nchez}, {Shappee}, \&
  {Stanek}}]{prieto16}
{Prieto}, J.~L., {et~al.} 2016, \apjl, 830, L32

\bibitem[{{Rees}(1988)}]{rees88}
{Rees}, M.~J. 1988, \nat, 333, 523

\bibitem[{{Ricci} {et~al.}(2017){Ricci}, {Trakhtenbrot}, {Koss}, {Ueda}, {Del
  Vecchio}, {Treister}, {Schawinski}, {Paltani}, {Oh}, {Lamperti}, {Berney},
  {Gandhi}, {Ichikawa}, {Bauer}, {Ho}, {Asmus}, {Beckmann}, {Soldi},
  {Balokovi{\'c}}, {Gehrels}, \& {Markwardt}}]{ricci17}
{Ricci}, C., {et~al.} 2017, \apjs, 233, 17

\bibitem[{{Romero-Ca{\~n}izales} {et~al.}(2016){Romero-Ca{\~n}izales},
  {Prieto}, {Chen}, {Kochanek}, {Dong}, {Holoien}, {Stanek}, \&
  {Liu}}]{romero16}
{Romero-Ca{\~n}izales}, C., {Prieto}, J.~L., {Chen}, X., {Kochanek}, C.~S.,
  {Dong}, S., {Holoien}, T.~W.-S., {Stanek}, K.~Z., \& {Liu}, F. 2016, \apjl,
  832, L10

\bibitem[{{Roming} {et~al.}(2005){Roming}, {Kennedy}, {Mason}, {Nousek}, {Ahr},
  {Bingham}, {Broos}, {Carter}, {Hancock}, {Huckle}, {Hunsberger}, {Kawakami},
  {Killough}, {Koch}, {McLelland}, {Smith}, {Smith}, {Soto}, {Boyd},
  {Breeveld}, {Holland}, {Ivanushkina}, {Pryzby}, {Still}, \&
  {Stock}}]{roming05}
{Roming}, P.~W.~A., {et~al.} 2005, SSR, 120, 95

\bibitem[{{Roth} \& {Kasen}(2018)}]{roth18}
{Roth}, N., \& {Kasen}, D. 2018, \apj, 855, 54

\bibitem[{{Roth} {et~al.}(2016){Roth}, {Kasen}, {Guillochon}, \&
  {Ramirez-Ruiz}}]{roth16}
{Roth}, N., {Kasen}, D., {Guillochon}, J., \& {Ramirez-Ruiz}, E. 2016, \apj,
  827, 3

\bibitem[{{Schlafly} \& {Finkbeiner}(2011)}]{schlafly11}
{Schlafly}, E.~F., \& {Finkbeiner}, D.~P. 2011, \apj, 737, 103

\bibitem[{{Schlafly} {et~al.}(2012){Schlafly}, {Finkbeiner}, {Juri{\'c}},
  {Magnier}, {Burgett}, {Chambers}, {Grav}, {Hodapp}, {Kaiser}, {Kudritzki},
  {Martin}, {Morgan}, {Price}, {Rix}, {Stubbs}, {Tonry}, \&
  {Wainscoat}}]{schlafly12}
{Schlafly}, E.~F., {et~al.} 2012, \apj, 756, 158

\bibitem[{{Shappee} {et~al.}(2014){Shappee}, {Prieto}, {Grupe}, {Kochanek},
  {Stanek}, {De Rosa}, {Mathur}, {Zu}, {Peterson}, {Pogge}, {Komossa}, {Im},
  {Jencson}, {Holoien}, {Basu}, {Beacom}, {Szczygie{\l}}, {Brimacombe},
  {Adams}, {Campillay}, {Choi}, {Contreras}, {Dietrich}, {Dubberley},
  {Elphick}, {Foale}, {Giustini}, {Gonzalez}, {Hawkins}, {Howell}, {Hsiao},
  {Koss}, {Leighly}, {Morrell}, {Mudd}, {Mullins}, {Nugent}, {Parrent},
  {Phillips}, {Pojmanski}, {Rosing}, {Ross}, {Sand}, {Terndrup}, {Valenti},
  {Walker}, \& {Yoon}}]{shappee14}
{Shappee}, B.~J., {et~al.} 2014, \apj, 788, 48

\bibitem[{{Shiokawa} {et~al.}(2015){Shiokawa}, {Krolik}, {Cheng}, {Piran}, \&
  {Noble}}]{shiokawa15}
{Shiokawa}, H., {Krolik}, J.~H., {Cheng}, R.~M., {Piran}, T., \& {Noble}, S.~C.
  2015, \apj, 804, 85

\bibitem[{{Smartt} {et~al.}(2016){Smartt}, {Chambers}, {Smith}, {Huber},
  {Young}, {Cappellaro}, {Wright}, {Coughlin}, {Schultz}, {Denneau},
  {Flewelling}, {Heinze}, {Magnier}, {Primak}, {Rest}, {Sherstyuk}, {Stalder},
  {Stubbs}, {Tonry}, {Waters}, {Willman}, {Anderson}, {Baltay}, {Botticella},
  {Campbell}, {Dennefeld}, {Chen}, {Della Valle}, {Elias-Rosa}, {Fraser},
  {Inserra}, {Kankare}, {Kotak}, {Kupfer}, {Harmanen}, {Galbany}, {Gal-Yam},
  {Le Guillou}, {Lyman}, {Maguire}, {Mitra}, {Nicholl}, {Olivares E},
  {Rabinowitz}, {Razza}, {Sollerman}, {Smith}, {Terreran}, {Valenti}, {Gibson},
  \& {Goggia}}]{smartt16}
{Smartt}, S.~J., {et~al.} 2016, \mnras, 462, 4094

\bibitem[{{Steele} {et~al.}(2004){Steele}, {Smith}, {Rees}, {Baker}, {Bates},
  {Bode}, {Bowman}, {Carter}, {Etherton}, {Ford}, {Fraser}, {Gomboc}, {Lett},
  {Mansfield}, {Marchant}, {Medrano-Cerda}, {Mottram}, {Raback}, {Scott},
  {Tomlinson}, \& {Zamanov}}]{steele04}
{Steele}, I.~A., {et~al.} 2004, in Society of Photo-Optical Instrumentation
  Engineers (SPIE) Conference Series, Vol. 5489, Ground-based Telescopes, ed.
  J.~M. {Oschmann}, Jr., 679--692

\bibitem[{{Storchi-Bergmann} {et~al.}(2017){Storchi-Bergmann}, {Schimoia},
  {Peterson}, {Elvis}, {Denney}, {Eracleous}, \& {Nemmen}}]{storchi17}
{Storchi-Bergmann}, T., {Schimoia}, J.~S., {Peterson}, B.~M., {Elvis}, M.,
  {Denney}, K.~D., {Eracleous}, M., \& {Nemmen}, R.~S. 2017, \apj, 835, 236

\bibitem[{{Storchi-Bergmann} {et~al.}(2003){Storchi-Bergmann}, {Nemmen da
  Silva}, {Eracleous}, {Halpern}, {Wilson}, {Filippenko}, {Ruiz}, {Smith}, \&
  {Nagar}}]{storchi03}
{Storchi-Bergmann}, T., {et~al.} 2003, \apj, 598, 956

\bibitem[{{Storey} \& {Hummer}(1995)}]{storey95}
{Storey}, P.~J., \& {Hummer}, D.~G. 1995, \mnras, 272, 41

\bibitem[{{Strubbe} \& {Murray}(2015)}]{strubbe15}
{Strubbe}, L.~E., \& {Murray}, N. 2015, \mnras, 454, 2321

\bibitem[{{Strubbe} \& {Quataert}(2009)}]{strubbe09}
{Strubbe}, L.~E., \& {Quataert}, E. 2009, \mnras, 400, 2070

\bibitem[{{Tonry} {et~al.}(2012){Tonry}, {Stubbs}, {Lykke}, {Doherty},
  {Shivvers}, {Burgett}, {Chambers}, {Hodapp}, {Kaiser}, {Kudritzki},
  {Magnier}, {Morgan}, {Price}, \& {Wainscoat}}]{tonry12}
{Tonry}, J.~L., {et~al.} 2012, \apj, 750, 99

\bibitem[{{Tonry} {et~al.}(2018){Tonry}, {Denneau}, {Heinze}, {Stalder},
  {Smith}, {Smartt}, {Stubbs}, {Weiland}, \& {Rest}}]{tonry18}
---. 2018, \pasp, 130, 064505

\bibitem[{{Tozzi} {et~al.}(2006){Tozzi}, {Gilli}, {Mainieri}, {Norman},
  {Risaliti}, {Rosati}, {Bergeron}, {Borgani}, {Giacconi}, {Hasinger},
  {Nonino}, {Streblyanska}, {Szokoly}, {Wang}, \& {Zheng}}]{tozzi06}
{Tozzi}, P., {et~al.} 2006, \aap, 451, 457

\bibitem[{{Tucker} {et~al.}(2018{\natexlab{a}}){Tucker}, {Huber}, {Shappee},
  {Dong}, {Bose}, \& {Chen}}]{SCATref}
{Tucker}, M.~A., {Huber}, M., {Shappee}, B.~J., {Dong}, S., {Bose}, S., \&
  {Chen}, P. 2018{\natexlab{a}}, The Astronomer's Telegram, 11444, 1

\bibitem[{{Tucker} {et~al.}(2018{\natexlab{b}}){Tucker}, {Huber}, {Shappee},
  {Prieto}, {Holoien}, {Dong}, {Bose}, {Chen}, {Falco}, {Calkins}, {Chambers},
  {Flewelling}, {Magnier}, {Schultz}, {Waters}, {Wainscoat}, {Wilman}, {Smith},
  {Smartt}, {Young}, \& {Wright}}]{ps18kh_spec_atel}
{Tucker}, M.~A., {et~al.} 2018{\natexlab{b}}, The Astronomer's Telegram, 11473

\bibitem[{{van Velzen} {et~al.}(2011){van Velzen}, {Farrar}, {Gezari},
  {Morrell}, {Zaritsky}, {{\"O}stman}, {Smith}, {Gelfand}, \&
  {Drake}}]{velzen11}
{van Velzen}, S., {et~al.} 2011, \apj, 741, 73

\bibitem[{{Vink{\'o}} {et~al.}(2015){Vink{\'o}}, {Yuan}, {Quimby}, {Wheeler},
  {Ramirez-Ruiz}, {Guillochon}, {Chatzopoulos}, {Marion}, \&
  {Akerlof}}]{vinko15}
{Vink{\'o}}, J., {et~al.} 2015, \apj, 798, 12

\bibitem[{{Voges} {et~al.}(1999){Voges}, {Aschenbach}, {Boller},
  {Br{\"a}uninger}, {Briel}, {Burkert}, {Dennerl}, {Englhauser}, {Gruber},
  {Haberl}, {Hartner}, {Hasinger}, {K{\"u}rster}, {Pfeffermann}, {Pietsch},
  {Predehl}, {Rosso}, {Schmitt}, {Tr{\"u}mper}, \& {Zimmermann}}]{voges99}
{Voges}, W., {et~al.} 1999, \aap, 349, 389

\bibitem[{{Waters} {et~al.}(2016){Waters}, {Magnier}, {Price}, {Chambers},
  {Burgett}, {Draper}, {Flewelling}, {Hodapp}, {Huber}, {Jedicke}, {Kaiser},
  {Kudritzki}, {Lupton}, {Metcalfe}, {Rest}, {Sweeney}, {Tonry}, {Wainscoat},
  {Wood-Vasey}, \& {Builders}}]{waters16}
{Waters}, C.~Z., {et~al.} 2016, ArXiv e-prints

\bibitem[{{Wevers} {et~al.}(2017){Wevers}, {van Velzen}, {Jonker}, {Stone},
  {Hung}, {Onori}, {Gezari}, \& {Blagorodnova}}]{wevers17}
{Wevers}, T., {van Velzen}, S., {Jonker}, P.~G., {Stone}, N.~C., {Hung}, T.,
  {Onori}, F., {Gezari}, S., \& {Blagorodnova}, N. 2017, \mnras, 471, 1694

\bibitem[{{Wright} {et~al.}(2015){Wright}, {Smartt}, {Smith}, {Miller},
  {Kotak}, {Rest}, {Burgett}, {Chambers}, {Flewelling}, {Hodapp}, {Huber},
  {Jedicke}, {Kaiser}, {Metcalfe}, {Price}, {Tonry}, {Wainscoat}, \&
  {Waters}}]{wright15}
{Wright}, D.~E., {et~al.} 2015, \mnras, 449, 451

\bibitem[{{Wright} {et~al.}(2010){Wright}, {Eisenhardt}, {Mainzer}, {Ressler},
  {Cutri}, {Jarrett}, {Kirkpatrick}, {Padgett}, {McMillan}, {Skrutskie},
  {Stanford}, {Cohen}, {Walker}, {Mather}, {Leisawitz}, {Gautier}, {McLean},
  {Benford}, {Lonsdale}, {Blain}, {Mendez}, {Irace}, {Duval}, {Liu}, {Royer},
  {Heinrichsen}, {Howard}, {Shannon}, {Kendall}, {Walsh}, {Larsen}, {Cardon},
  {Schick}, {Schwalm}, {Abid}, {Fabinsky}, {Naes}, \& {Tsai}}]{wright10}
{Wright}, E.~L., {et~al.} 2010, \aj, 140, 1868

\end{thebibliography}
\bibliographystyle{apj}


\begin{deluxetable}{cccc}[h!]
\tabletypesize{\footnotesize}
\tablecaption{Spectroscopic Observations of PS18kh}
\tablehead{
\colhead{Date} &
\colhead{Telescope} &
\colhead{Instrument} &
\colhead{Exposure Time}}
\startdata
2018 March 07 & University of Hawaii 88-inch & SNIFS & 1x1200s \\
2018 March 18 & University of Hawaii 88-inch & SNIFS & 1x2000s \\
2018 March 20 & Fred L. Whipple Observatory Tillinghast 60-inch & FAST & 1x1800s \\
2018 March 20 & du Pont 100-inch & WFCCD & 2x1200s, 1x900s \\
2018 March 25 & Magellan Baade 6.5-m & IMACS & 1x1200s \\
2018 March 31 & University of Hawaii 88-inch & SNIFS & 3x1800s \\
2018 April 01 & Gemini North 8.2-m & GMOS & 1x900s \\
2018 April 06 & Liverpool Telescope 2-m & SPRAT & 1x900s \\
2018 April 07 & Liverpool Telescope 2-m & SPRAT & 2x900s \\
2018 April 11 & Gemini North 8.2-m & GMOS & 1x900s \\
2018 April 13 & Liverpool Telescope 2-m & SPRAT & 2x900s \\
2018 April 13 & Keck I 10-m & LRIS & 1x2200s \\
2018 April 16 & Liverpool Telescope 2-m & SPRAT & 2x900s \\
2018 April 25 & University of Hawaii 88-inch & SNIFS & 3x1800s \\
2018 April 25 & Gemini North 8.2-m & GMOS & 3x900s \\
2018 April 27 & Liverpool Telescope 2-m & SPRAT & 2x900s \\
2018 April 27 & University of Hawaii 88-inch & SNIFS & 2x1800s, 1x1200s \\
2018 April 29 & Liverpool Telescope 2-m & SPRAT & 2x900s \\
2018 May 04 & Liverpool Telescope 2-m & SPRAT & 2x900s \\
2018 May 11 & University of Hawaii 88-inch & SNIFS & 1x1800s \\
2018 May 12 & University of Hawaii 88-inch & SNIFS & 1x1800s \\
2018 May 14 & Keck I 10-m & LRIS & 1x1200s \\
2018 May 15 & University of Hawaii 88-inch & SNIFS & 1x1800s \\
2018 May 17 & University of Hawaii 88-inch & SNIFS & 1x1800s, 1x1200s \\
2018 May 18 & University of Hawaii 88-inch & SNIFS & 1x1800s \\
2018 May 19 & University of Hawaii 88-inch & SNIFS & 1x1800s \\
2018 May 21 & Large Binocular Telescope 8.2-m & MODS & 3x1200s \\
\enddata 
\tablecomments{Date, telescope, instrument, and exposure time for each of the spectroscopic observations obtained of PS18kh for the initial classification of the transient and as part of our follow-up campaign.} 
\label{tab:spec_details} 
\end{deluxetable}

\begin{deluxetable}{ccc}[h!]
\tabletypesize{\footnotesize}
\tablecaption{Measured \halpha{} and \hbeta{} line luminosities}
\tablehead{
\colhead{Rest-Frame Days Relative to Peak} &
\colhead{\halpha{} Luminosity} &
\colhead{\hbeta{} Luminosity}}
\startdata
-0.09 & ($1.06\pm0.32$)$\times10^{41}$ & --- \\ 
1.77 & ($1.33\pm0.4$)$\times10^{41}$ & --- \\ 
1.77 & ($0.77\pm0.23$)$\times10^{41}$ & --- \\ 
6.44 & ($2.96\pm0.89$)$\times10^{41}$ & ($1.46\pm0.44$)$\times10^{41}$ \\ 
11.11 & ($3.26\pm0.98$)$\times10^{41}$ & ($2.51\pm0.75$)$\times10^{41}$ \\ 
12.04 & ($2.68\pm0.8$)$\times10^{41}$ & ($3.35\pm1.0$)$\times10^{41}$ \\ 
12.98 & ($3.06\pm0.92$)$\times10^{41}$ & ($1.16\pm0.35$)$\times10^{41}$ \\ 
18.58 & ($3.21\pm0.96$)$\times10^{41}$ & ($1.2\pm0.36$)$\times10^{41}$ \\ 
22.32 & ($5.32\pm1.59$)$\times10^{41}$ & ($0.99\pm0.3$)$\times10^{41}$ \\ 
24.18 & ($4.45\pm1.34$)$\times10^{41}$ & ($1.25\pm0.37$)$\times10^{41}$ \\ 
24.18 & ($4.93\pm1.48$)$\times10^{41}$ & ($1.89\pm0.57$)$\times10^{41}$ \\ 
26.98 & ($3.61\pm1.08$)$\times10^{41}$ & ($1.51\pm0.45$)$\times10^{41}$ \\ 
35.39 & ($6.26\pm1.88$)$\times10^{41}$ & ($1.28\pm0.39$)$\times10^{41}$ \\ 
37.25 & ($6.18\pm1.86$)$\times10^{41}$ & --- \\ 
39.12 & ($4.71\pm1.41$)$\times10^{41}$ & ($2.32\pm0.69$)$\times10^{41}$ \\ 
43.79 & ($2.88\pm0.86$)$\times10^{41}$ & ($1.08\pm0.32$)$\times10^{41}$ \\ 
49.39 & ($5.05\pm1.52$)$\times10^{41}$ & ($1.18\pm0.35$)$\times10^{41}$ \\ 
51.26 & ($4.84\pm1.45$)$\times10^{41}$ & --- \\ 
54.06 & ($6.02\pm1.81$)$\times10^{41}$ & ($2.39\pm0.72$)$\times10^{41}$ \\ 
57.8 & ($4.68\pm1.4$)$\times10^{41}$ & --- \\ 
59.66 & ($3.94\pm1.18$)$\times10^{41}$ & ($1.66\pm0.5$)$\times10^{41}$ \\ 
\enddata 
\tablecomments{\halpha{} and \hbeta{} line luminosities measured from the follow-up spectra of PS18kh. In some epochs \hbeta{} was not measurable. The uncertainties shown are 30\% uncertainties on the measured fluxes.} 
\label{tab:line_lum} 
\end{deluxetable}

\end{document}